\documentclass[prd,aps,twocolumn,superscriptaddress,showpacs,preprintnumbers,
               amsmath,amssymb,floatfix]{revtex4}

\usepackage{graphicx}
\usepackage{dcolumn}
\usepackage{bm}
\usepackage{epsfig}

\def\insitu{{\em in situ}}
\def\eg{{{e.g.}}}
\def\ie{{{i.e.}}}
\def\calif252{$^{252}$Cf}
\def\am241{$^{241}$Am}
\def\c14{$^{14}$C}
\def\co60{$^{60}$Co}
\def\pb210{$^{210}$Pb}
\def\asi{$\alpha$-Si}

\def\er{E_{\mathrm{R}}}

\def\ystar{Y^\ast}

\def\gev{GeV~c$^{-2}$}

\def\cm2{cm$^2$}
\def\persec{s$^{-1}$}

\def\perm2{m$^{-2}$}
\def\percmsq{cm$^{-2}$}
\def\percm3{cm$^{-3}$}
\def\perkev{keV$^{-1}$}
\def\perday{d$^{-1}$}
\def\perkg{kg$^{-1}$}
\def\gpercm3{g~\percm3}
\def\percmsqpersec{\percmsq~\persec}
\def\dru{\perkev~\perkg~\perday}
\def\iru{\perkg~\perday}

\def\nvrthz{$\mathrm{nV/\sqrt{Hz}}$}

\def\micros{$\mu$s}
\def\micron{$\mu$m}
\def\chisq{$\chi^2$}

\newcommand{\bea}{\begin{eqnarray}}
\newcommand{\eea}{\end{eqnarray}}
\newcommand{\scinot}[2]{$#1~\times~10^{#2}$}

\begin{document}

\preprint{CWRU-P4-02/UCSB-HEP-02-02}

\title{Exclusion limits on the WIMP-nucleon cross-section from \\
the Cryogenic Dark Matter Search}

\affiliation{Department of Physics, Brown University,   
		  Providence, RI 02912, USA} 
\affiliation{Department of Physics, Case Western Reserve University,   
		  Cleveland, OH 44106, USA   }
\affiliation{Fermi National Accelerator Laboratory,  
                  Batavia, IL 60510, USA   }  
\affiliation{Lawrence Berkeley National Laboratory,  
                  Berkeley, CA 94720, USA   } 
\affiliation{National Institute of Standards and Technology,  
                  Boulder, CO 80303, USA   } 
\affiliation{Department of Physics, Princeton University,  
                  Princeton, NJ 08544, USA   }
\affiliation{Department of Physics, Santa Clara University,  
	          Santa Clara, CA 95053, USA}
\affiliation{Department of Physics, Stanford University,  
                  Stanford, CA 94305, USA   }
\affiliation{Department of Physics, 
		  University of California, Berkeley,  
                  Berkeley, CA 94720, USA   }
\affiliation{Department of Physics, 
                  University of California, Santa Barbara,   
	          Santa Barbara, CA 93106, USA   } 
\affiliation{Department of Physics, University of Colorado,   
		 Denver, CO 80217, USA  }
\author{D.~Abrams}
\affiliation{Department of Physics, Stanford University,  
                  Stanford, CA 94305, USA   }
\author{D.S.~Akerib}
\affiliation{Department of Physics, Case Western Reserve University,   
		  Cleveland, OH 44106, USA   }
\author{M.S.~Armel-Funkhouser}
\affiliation{Department of Physics, 
		  University of California, Berkeley,  
                  Berkeley, CA 94720, USA   }
\author{L.~Baudis}
\affiliation{Department of Physics, Stanford University,  
                  Stanford, CA 94305, USA   }
\author{D.A.~Bauer}
\affiliation{Department of Physics, 
                  University of California, Santa Barbara,   
	          Santa Barbara, CA 93106, USA   } 
\author{A.~Bolozdynya}
\affiliation{Department of Physics, Case Western Reserve University,   
		  Cleveland, OH 44106, USA   }
\author{P.L.~Brink}
\affiliation{Department of Physics, Stanford University,  
                  Stanford, CA 94305, USA   }
\author{R.~Bunker}
\affiliation{Department of Physics, 
                  University of California, Santa Barbara,   
	          Santa Barbara, CA 93106, USA   } 
\author{B.~Cabrera}
\affiliation{Department of Physics, Stanford University,  
                  Stanford, CA 94305, USA   }
\author{D.O.~Caldwell}
\affiliation{Department of Physics, 
                  University of California, Santa Barbara,   
	          Santa Barbara, CA 93106, USA   } 
\author{J.P.~Castle}
\affiliation{Department of Physics, Stanford University,  
                  Stanford, CA 94305, USA   }
\author{C.L.~Chang}
\affiliation{Department of Physics, Stanford University,  
                  Stanford, CA 94305, USA   }
\author{R.M.~Clarke}
\affiliation{Department of Physics, Stanford University,  
                  Stanford, CA 94305, USA   }
\author{M.B.~Crisler}
\affiliation{Fermi National Accelerator Laboratory,  
                  Batavia, IL 60510, USA   }  
\author{R.~Dixon}
\affiliation{Fermi National Accelerator Laboratory,  
                  Batavia, IL 60510, USA   }  
\author{D.~Driscoll}
\affiliation{Department of Physics, Case Western Reserve University,   
		  Cleveland, OH 44106, USA   }
\author{S.~Eichblatt}
\affiliation{Fermi National Accelerator Laboratory,  
                  Batavia, IL 60510, USA   }  
\author{R.J.~Gaitskell}
\affiliation{Department of Physics, Brown University,   
		  Providence, RI 02912, USA} 
\author{S.R.~Golwala}
\affiliation{Department of Physics, 
		  University of California, Berkeley,  
                  Berkeley, CA 94720, USA   }
\author{E.E.~Haller}
\affiliation{Lawrence Berkeley National Laboratory,  
                  Berkeley, CA 94720, USA   } 
\author{J.~Hellmig}
\affiliation{Department of Physics, 
		  University of California, Berkeley,  
                  Berkeley, CA 94720, USA   }
\author{D.~Holmgren}
\affiliation{Fermi National Accelerator Laboratory,  
                  Batavia, IL 60510, USA   }  
\author{M.E.~Huber}
\affiliation{Department of Physics, University of Colorado,   
		 Denver, CO 80217, USA  }
\author{S.~Kamat}
\affiliation{Department of Physics, Case Western Reserve University,   
		  Cleveland, OH 44106, USA   }
\author{A.~Lu}
\affiliation{Department of Physics, 
                  University of California, Santa Barbara,   
	          Santa Barbara, CA 93106, USA   }
\author{V.~Mandic}
\affiliation{Department of Physics, 
		  University of California, Berkeley,  
                  Berkeley, CA 94720, USA   }
\author{J.M.~Martinis}
\affiliation{National Institute of Standards and Technology,  
                  Boulder, CO 80303, USA   }
\author{P.~Meunier}
\affiliation{Department of Physics, 
		  University of California, Berkeley,  
                  Berkeley, CA 94720, USA   }
\author{S.W.~Nam}
\affiliation{National Institute of Standards and Technology,  
                  Boulder, CO 80303, USA   } 
\author{H.~Nelson}
\affiliation{Department of Physics, 
                  University of California, Santa Barbara,   
	          Santa Barbara, CA 93106, USA   }
\author{T.A.~Perera}
\affiliation{Department of Physics, Case Western Reserve University,   
		  Cleveland, OH 44106, USA   }
\author{M.C.~Perillo~Isaac}
\affiliation{Department of Physics, 
		  University of California, Berkeley,  
                  Berkeley, CA 94720, USA   }
\author{W.~Rau}
\affiliation{Department of Physics, 
		  University of California, Berkeley,  
                  Berkeley, CA 94720, USA   }
\author{R.R.~Ross}
\affiliation{Lawrence Berkeley National Laboratory,  
                  Berkeley, CA 94720, USA   }
\affiliation{Department of Physics, 
		  University of California, Berkeley,  
                  Berkeley, CA 94720, USA   } 
\author{T.~Saab}
\affiliation{Department of Physics, Stanford University,  
                  Stanford, CA 94305, USA   }
\author{B.~Sadoulet}
\affiliation{Lawrence Berkeley National Laboratory,  
                  Berkeley, CA 94720, USA   }
\affiliation{Department of Physics, 
		  University of California, Berkeley,  
                  Berkeley, CA 94720, USA   }
\author{J.~Sander}
\affiliation{Department of Physics, 
                  University of California, Santa Barbara,   
	          Santa Barbara, CA 93106, USA   }
\author{R.W.~Schnee}
\email[Corresponding author, email address:]{schnee@po.cwru.edu}
\affiliation{Department of Physics, Case Western Reserve University,   
		  Cleveland, OH 44106, USA   }
\author{T.~Shutt}
\affiliation{Department of Physics, Princeton University,  
                  Princeton, NJ 08544, USA   }
\author{A.~Smith}
\affiliation{Lawrence Berkeley National Laboratory,  
                  Berkeley, CA 94720, USA   }
\author{A.H.~Sonnenschein}
\affiliation{Department of Physics, 
                  University of California, Santa Barbara,   
	          Santa Barbara, CA 93106, USA   }
\author{A.L.~Spadafora}
\affiliation{Department of Physics, 
		  University of California, Berkeley,  
                  Berkeley, CA 94720, USA   }
\author{G.~Wang}
\affiliation{Department of Physics, Case Western Reserve University,   
		  Cleveland, OH 44106, USA   }
\author{S.~Yellin}
\affiliation{Department of Physics, 
                  University of California, Santa Barbara,   
	          Santa Barbara, CA 93106, USA   }
\author{B.A.~Young}
\affiliation{Department of Physics, Santa Clara University,  
	          Santa Clara, CA 95053, USA}

\collaboration{CDMS Collaboration}
\noaffiliation

\date{\today}

\begin{abstract}
The Cryogenic Dark Matter Search (CDMS) employs low-temperature
Ge and Si detectors to
search for Weakly Interacting Massive Particles
(WIMPs) via their elastic-scattering interactions with nuclei
while discriminating against interactions of background particles.
For recoil energies above
10~keV, events due to background photons are rejected with $>$~99.9\%
efficiency, and surface events are rejected with $>$~95\%
efficiency.
The estimate of the  background due to neutrons is based primarily on the 
observation of multiple-scatter events that should all be neutrons.
Data selection is determined 
primarily by examining calibration data and vetoed events.
Resulting efficiencies should be accurate to $\sim$10\%.
Results of CDMS data from 1998 and 1999 with a relaxed fiducial-volume
cut (resulting in 15.8 kg-days exposure on Ge)
are consistent with an earlier analysis with a more restrictive 
fiducial-volume cut.
Twenty-three WIMP candidate events are observed, 
but 
these events are consistent with a background from neutrons in all ways 
tested.
Resulting limits on the
spin-independent WIMP-nucleon elastic-scattering cross-section 
exclude unexplored parameter space 
for WIMPs with masses between 10--70~GeV~c$^{-2}$.
These limits border, but do not exclude,
parameter space allowed by supersymmetry 
models and accelerator constraints.
Results are compatible with some regions 
reported as allowed at $3\sigma$ by the annual-modulation measurement 
of the DAMA collaboration.
However, under the assumptions of standard WIMP interactions and a standard halo,
the results are incompatible with the DAMA most likely value
at $> 99.9$\%~CL, 
and are 
incompatible with the model-independent 
annual-modulation signal of DAMA at 99.99\%~CL in the
asymptotic limit.  
\end{abstract}


\maketitle

\section{Introduction}

This paper presents details of a new search for matter in the universe 
that is nonluminous, or ``dark.''
Extensive observational evidence indicates that this dark matter comprises
a large fraction of the matter in
the universe~\cite{bergstrom}.
However, the nature and
quantity of the dark matter in the universe remain unknown, 
providing a central problem for astronomy and
cosmology~\cite{kolbturner,peebles}. 
Recent measurements of the cosmic microwave background 
radiation~\cite{boomerang,dasi,maxima}, as well as arguments based on Big Bang 
Nucleosynthesis and the growth of structure in the universe~\cite{srednicki},  
suggest that dark matter consists predominantly of
non-baryonic particles outside the standard model of particle
physics.
Supersymmetric particle
physics models provide a natural candidate for dark matter: the lightest
superpartner, usually
taken to be a neutralino with typical mass 
about 100~GeV~c$^{-2}$~\cite{jkg,ellis,gondolo,bottino}; 
experimental bounds 
from LEP give a lower limit of 46~GeV~c$^{-2}$~\cite{ellis00lep}.

More
generically, one can consider a class of Weakly Interacting Massive
Particles, or
WIMPs~\cite{lee}, which were once in thermal equilibrium with the early
universe, but were ``cold,'' \ie\ moving
non-relativistically at the time of structure formation. 
Their density today
is then determined roughly by their annihilation rate,
with weak-scale interactions if the dark matter is mainly composed of WIMPs. 
WIMPs are expected to
have collapsed into a roughly isothermal, spherical halo within which
the visible portion of our galaxy resides,
consistent with measurements of spiral
galaxy rotation curves~\cite{salucci}. 

The best possibility for direct detection of WIMPS lies in elastic
scattering from nuclei~\cite{goodman,primack}. 
Calculations of the fundamental WIMP-quark cross sections require 
a model, usually the Minimal Supersymmetric Standard Model 
(MSSM)~\cite{jkg}.  This interaction, summed over the quarks
present in a nucleon, gives an effective WIMP-nucleon cross section.
In the low momentum-transfer limit, the contributions of individual
nucleons are summed coherently 
to yield a
WIMP-nucleus cross section; these are typically smaller than 
$10^{-42}$~cm$^{2}$. 
The nuclear-recoil energy is typically a few keV~\cite{lewin},
since 
WIMPs should have velocities typical for Galactic objects.

Due to the extremely small WIMP scattering rate 
and the 
small energy of the recoiling nucleus, 
a direct-detection experiment must have a low 
energy threshold and very low
backgrounds from radioactivity and cosmic rays (or be able to reject such
backgrounds). The sensitivity of such an experiment improves 
linearly with detector mass, $M$, and exposure time, $T$, 
if there is no background.
If there is a background of known size, 
the sensitivity can improve as $\propto \sqrt{MT}$.

The Cryogenic Dark Matter Search (CDMS) is
an experiment designed to measure the nuclear recoils generated by Galactic
WIMPs using cryogenic Ge and Si detectors operating within a carefully
shielded environment.  CDMS detectors provide active rejection of
backgrounds that would otherwise swamp any signal.  
Consequently, the assessment of detector
performance, rejection efficiency, and known backgrounds constitutes a
substantial component of our analysis effort.

This paper presents a new analysis of the data obtained by
the CDMS collaboration in its 1998 and 1999 experimental runs.
The original analysis of these data and the associated exclusion limit
on the WIMP-nucleon elastic-scattering cross-section appeared
in a Letter~\cite{r19prl}.  
Significant changes introduced in this new analysis include a relaxed 
fiducial volume cut, resulting in a $\sim$40\% larger
exposure, as well as detailed treatment of possible systematic errors.

The organization of this article is as follows.
Section~II describes the CDMS experimental apparatus
including the detectors, hardware, cryogenics, electronics,
facilities, and data acquisition systems.  
Section~\ref{r19datapipe} summarizes the methods by
which the data are reduced and calibrated.  
Section~\ref{sect:Ge} presents the data
obtained with the Ge detectors and 
details the application of cuts to the data.
Because the measurements analyzed in this report were made in a shallow 
facility, there is a significant unrejectable neutron background.
Determination of this background is described in Sec.~\ref{sect:n}.  
Section~\ref{ConRegion} explains the procedure by which the limits 
on cross-sections are calculated.  
Section~\ref{sect:results} contains the results of the new analysis
including new limits on the WIMP-nucleon elastic-scattering cross-section.

\section{The Experiment}

The first stage of the Cryogenic Dark
Matter Search (CDMS~I) operates at the Stanford Underground Facility, a tunnel 
10.6~m beneath the Stanford University campus. The 
experiment consists of a 2-meter, nearly cubic, layered shield 
(with an active-scintillator muon veto) 
surrounding a cold volume which houses the Ge and Si detectors. The 
cold volume is connected via a horizontal stem to a dilution refrigerator 
and via a separate stem to a vacuum bulkhead where detector signals 
are brought out to front-end electronics. The amplified signals are coupled to 
a data acquisition system approximately 20~m away, where a trigger is 
formed and the signals are recorded. The Ge and Si detectors are  
cooled to sub-Kelvin temperatures so that the 
phonons produced by particle interactions are detectable above the 
ambient thermal phonon population.
Simultaneous determination of the ionization energy and the phonon energy
deposited in these Ge or Si
crystals makes it possible to distinguish between a nuclear-recoil event
produced by a WIMP (or a neutron) 
and an electron-recoil
event 
due to the otherwise dominant background from radioactive decay products 
(mainly $\alpha$-particles, electrons, and photons).
Such discrimination is possible 
because nuclear recoils dissipate a significantly smaller fraction
of their
energy into electron-hole pairs than do electron
recoils~\cite{tomprl2}.

\subsection{Detectors}
\label{blip}

The data discussed here were
obtained with two types of detectors, Berkeley Large Ionization- and
Phonon-mediated (BLIP)~\cite{tomprl1,tomprl2,blipltd7}
and Z-sensitive Ionization- and Phonon-mediated (ZIP) detectors
\cite{irwin2,alexltd7,alexsheffield,clarkethesis,clarkeapl}.
One early-design ZIP detector was operated in 1998, 
and four BLIP detectors were operated during a data run mostly in 1999.

\begin{figure*}  
\psfig{figure=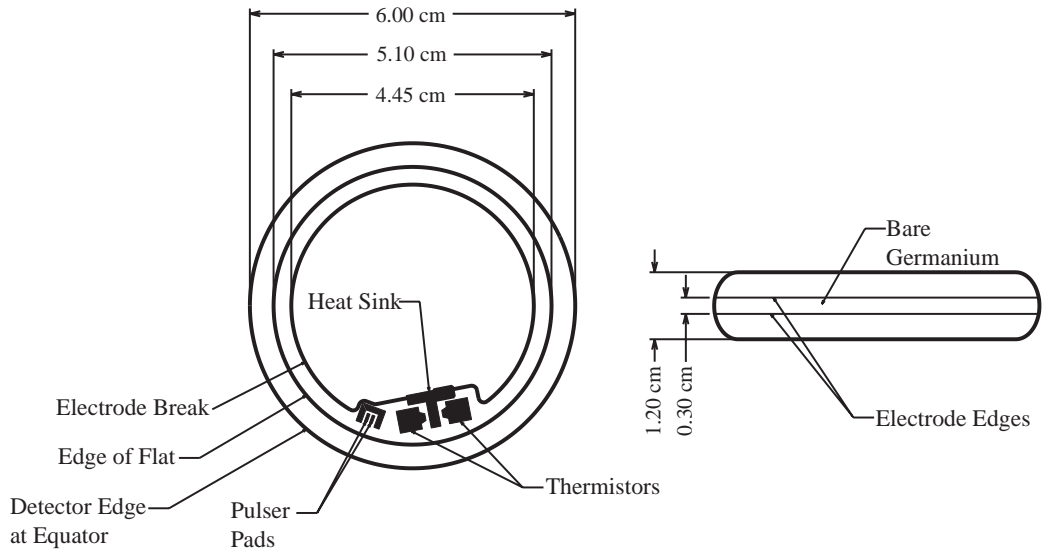,width=6in, 
        bbllx=40,bblly=220,bburx=620,bbury=560,clip=}
\caption{\label{blipfig}A BLIP detector.  
The ionization-electrode breaks are
indicated.  The NTD thermistors are not shown in the side view; they are
0.26-cm high.}
\end{figure*}

Each BLIP detector consists of a cylindrical crystal of high-purity, undoped, 
p-type, single-crystal Ge with rounded 
edges, as shown schematically in Fig.~\ref{blipfig}.
The BLIP substrates are 165~g in
mass, 6~cm in diameter, and 1.2-cm thick.
Phonon
production is determined from the detector's calorimetric temperature
change, as measured with two
neutron-transmutation-doped (NTD) Ge thermistors (each approximately
$3.1\times 3.1\times 2.6$~mm$^3$) eutectically
bonded to the crystal~\cite{hallereutectic}.
Charge-collection electrodes on the top and
bottom faces of each BLIP detector define the ionization drift field 
and provide electrical contact to the
ionization bias circuits and amplifier~\cite{tomltd8}.  
For the 1999 data run, the four BLIP detectors (numbered 3--6 from top to bottom) 
were stacked 3~mm apart with no
intervening material.  This close packing 
helped shield the detectors from
low-energy electron sources on surrounding surfaces.  
The close-packing arrangement also increased the 
probability that a 
background event in one detector would multiple-scatter 
into another detector.
Division of the electrodes into an annular outer electrode and a 
disk-shaped inner electrode helped define an inner fiducial region 
that was further shielded from low-energy electron sources.

\begin{figure*}
\psfig{figure=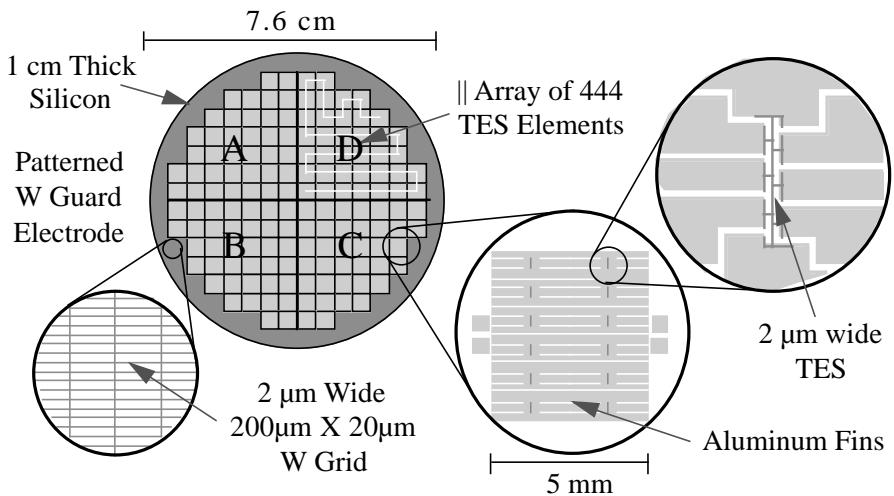,width=12cm}
\caption[]{\label{AlexFig}
A diagram of the phonon sensors for the 100-g Si ZIP
detector run in 1998. The central item depicts the basic layout 
with each phonon sensor occupying a detector quadrant.  Each sensor is
divided into 37 units each 5\,mm square (magnified to the right) which 
themselves contain 12 individual transition-edge-sensor (TES) elements 
(far right) connected in parallel.
Aluminum quasiparticle-collector fins cover 82\,$\%$ of the top surface
of the Si and also provide the ground electrode for the ionization 
measurement.  On the far left is shown the W outer ionization electrode
that is patterned (10\,\% area coverage) to minimize athermal-phonon absorption.
}
\end{figure*}

In ZIP detectors, athermal phonons are collected to determine
both the phonon production and xy-position of each event.
The ZIP detector operated in 1998 is a high-purity, single-crystal
cylinder of Si, 
100~g in mass, 7.6~cm in diameter, and 1-cm thick.
The detector has two concentric charge-collection electrodes.
One side of the detector is patterned with an active aluminum/tungsten 
film that defines four independent phonon sensors (see Fig.~\ref{AlexFig}).
Around the perimeter of the phonon-sensor region is a passive 
tungsten grid, which 
provides 10\% area coverage and is used in the ionization measurement. 

The energy deposited in the detector by an interacting particle
is called ``recoil energy'' $E_{\mathrm{R}}$.
If the particle interacts with an electron or electrons
(\eg\ by Compton scattering, K-capture, etc.),
the event is called an electron recoil;
if the particle interacts with a nucleus
(\eg\ by WIMP-nucleus or neutron-nucleus elastic scattering),
the event is a nuclear recoil.
Most of the recoil energy is converted
almost immediately into phonons, while the rest is dissipated via
ionization losses in the creation of electron-hole pairs.  By the
time the calorimetric temperature rise is detected, the electron-hole
pairs have recombined in the electrodes, releasing the energy
initially dissipated in their creation.  Thus, all of the recoil
energy has been converted to phonons and is detected.
In principle, a small fraction of the recoil energy can be lost
to permanent crystal damage, to trapped charges,
or to direct thermal conduction of high-energy, recombination phonons 
through a detector's electrodes.
Comparisons of the collected phonon energy to kinematic energy 
measurements indicate at most a few percent of the recoil energy is 
lost~\cite{shuttthesis,tomprl1,tomprl2}.


Depending on the material and the type of recoil,
between about one sixteenth and one third of the recoil energy is 
dissipated via ionization before subsequent conversion to phonons.
On average, one electron-hole pair is produced for every
$\epsilon\approx3.0$~eV (3.8~eV) of energy 
from an electron recoil in Ge (in Si). 
The ``ionization energy'' $E_{\mathrm{Q}}$ is defined for convenience
as the recoil energy inferred from the detected
number of charge pairs $N_{\mathrm{Q}}$ 
by assuming that the event is an electron recoil 
with 100\% charge-collection efficiency:
\begin{equation}
E_{\mathrm{Q}} \equiv N_{\mathrm{Q}} \times \epsilon . 
\end{equation}
Ionization energy is usually reported in units such as ``keVee,''
or keV of the equivalent electron recoil.
The ionization yield $Y \equiv E_{\mathrm{Q}} / E_{\mathrm{R}}$,
so $Y\approx1$ for electron recoils with complete charge collection.

\begin{figure}
\psfig{figure=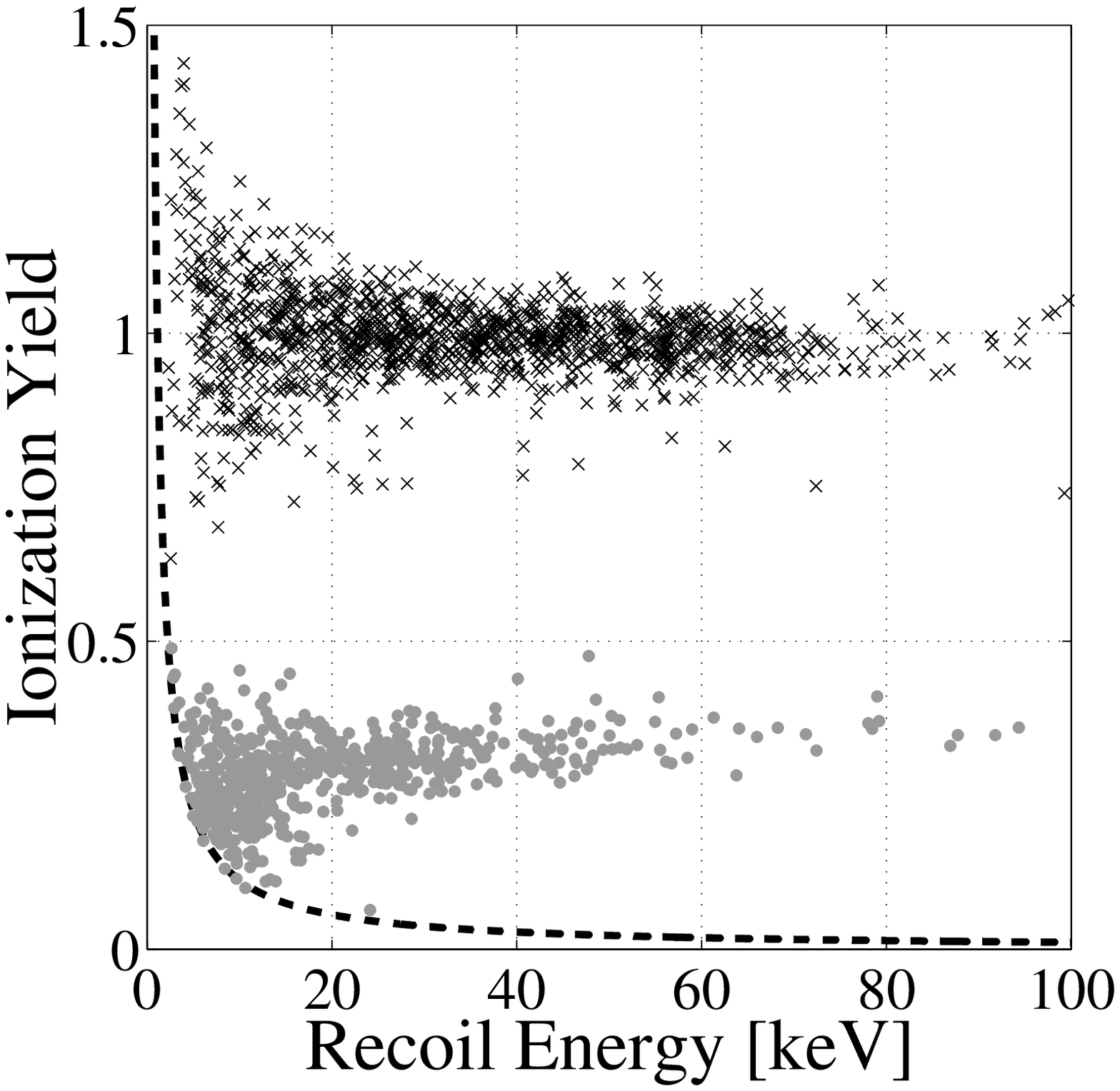,width=3.2in}
\caption[]{\label{Ydemo}
Ionization yield $Y$ versus recoil energy $E_{\mathrm{R}}$
for 1334 electron-recoil events due to photons 
from an external \co60\ source ($\times$'s)
and for 616 nuclear-recoil events due to neutrons
from a separate calibration with an external 
\calif252\ source (grey dots) for a Ge BLIP detector.  
These \insitu\ external-source calibrations are described below 
in Sec.~\ref{sect:cal}.
The dashed curve (at $E_{\mathrm{Q}}=1.1$~keV) indicates the 
ionization-search threshold (described below in Sec.~\ref{sect:thresh}) 
for the neutron-calibration data.
}
\end{figure}

Nuclear recoils produce fewer charge pairs,
and hence less ionization energy $E_{\mathrm{Q}}$,
than electron recoils of the same recoil energy do.
The ionization yield $Y$ for nuclear-recoil events
depends on both the material and the recoil energy,
with $Y \sim 0.3$ ($Y \sim 0.25$) in Ge (in Si) 
for $E_{\mathrm{R}} \agt 20 $~keV, as shown in Fig.~\ref{Ydemo} for Ge.

Energy is dissipated in the drifting of charges in the electric field,
increasing phonon production by an amount equal to the work done by the
electric field.  These ``Neganov-Luke'' phonons contribute to the 
total observed phonon signal, yielding
\begin{equation}
E_{\mathrm{P}} = E_{\mathrm{R}} + e V_{\mathrm{b}} N_{\mathrm{Q}} 
               = E_{\mathrm{R}} + \frac{e V_{\mathrm{b}}}{\epsilon}
                 E_{\mathrm{Q}} ,
\end{equation}
where $V_{\mathrm{b}}$ is the bias voltage across the 
detector~\cite{neganov,luke}.
Because the ionization measurement effectively weights 
the number of charge pairs by their drift distances 
(see Sec.~\ref{deadlayer}),
this equation is valid even for events with incomplete charge collection 
(due, for example, to trapping or recombination in the wrong electrode).
Since $E_{\mathrm{Q}} = E_{\mathrm{R}}$ for electron recoils
with full charge collection,
$E_{\mathrm{P}} = \left( 1 + \frac{e V_{\mathrm{b}}}{\epsilon} \right)
E_{\mathrm{R}}$
for these events.
Calibration of the detectors at several bias voltages using photon 
sources confirms that $\epsilon \approx 3$~eV (3.8~eV) in Ge (in Si).
For electron recoils with full charge collection in Ge at 6~V bias
(the bias voltage for most of the data described here),
$E_{\mathrm{P}} = 3E_{\mathrm{R}}$.
In practice, the recoil energy $E_{\mathrm{R}}$ of an event is
inferred from measurements of the phonon and ionization energies:
\begin{equation}
E_{\mathrm{R}} = E_{\mathrm{P}} - \frac{e V_{\mathrm{b}}}{\epsilon}
                 E_{\mathrm{Q}}  .		 
\end{equation}

\subsubsection{\label{deadlayer}The ionization measurement}

Charge-collection electrodes deposited on the two
faces of each disk-shaped  
detector are maintained at different
voltages to supply an electric field, so that electrons drift toward
one face and holes to the other.
However, because the electrons and holes generated by an interaction
are created ``hot'' and are not in local thermodynamic equilibrium 
with the crystal,
some may diffuse before the drift field has
a significant effect upon their motion.  
The charge cloud produced by a
recoiling particle may also shield itself because the separating
electron-hole pairs have dipole fields that counter the drift field.
As a result,
charges produced near a surface of the detector can diffuse 
against the applied electric field into the nearby electrode, 
causing a fraction of the event ionization to be ``lost.''  
The surface region in which ionization is lost 
is termed the detector's ``dead layer''~\cite{shuttthesis}.

\begin{figure}
\psfig{figure=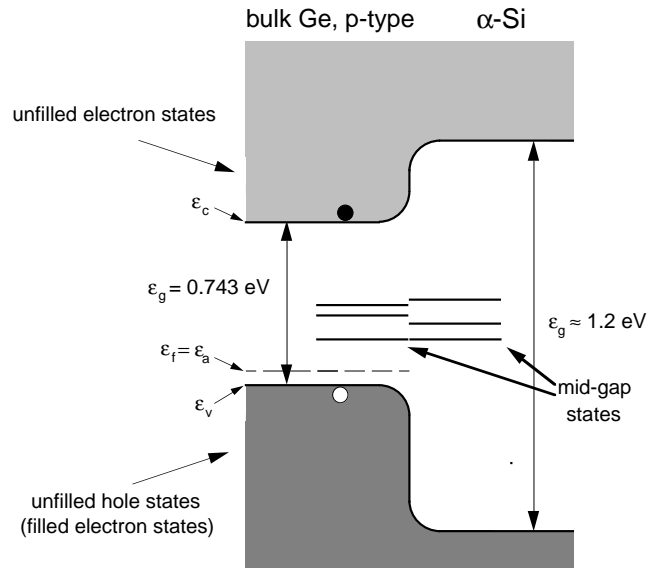,width=3.4in}
\caption{\label{asicontact}
Schematic illustration of bulk-Ge/\asi\ interface, indicating  
qualitative misalignment suggested by data from test devices.
Mid-gap states that may serve to
define the alignment are schematically indicated.}
\end{figure}

In order to reduce the loss of ionization near detector surfaces,
the BLIP detectors used in 1999 were made with
hydrogenated, amorphous-silicon (\asi)
contacts~\cite{tomltd8}.  
Amorphous Si possesses a
bandgap $\varepsilon_{\mathrm{g}}= 1.2$~eV, 
almost twice as large as that of bulk Ge. 
So long as the bands of the bulk Ge and the deposited layer of
\asi\ are nearly centered on each other, the \asi\ can block diffusion of
charges of both polarities.
See Fig.~\ref{asicontact} for a schematic illustration of this effect.  
Data taken with test devices indicates that using \asi\ contacts 
dramatically reduces the dead-layer problem~\cite{tomltd8,tomltd9}. 

The dead layer is a problem particularly for electrons
incident on the surface of a detector,
since electrons
have a very small penetration depth. 
The 90\% stopping length, or practical range, in Ge (in Si)
is 0.5~\micron\ (0.7~\micron) at 10~keV, and is 10~\micron\
(23~\micron) at 60~keV. 
Although most low-energy electrons suffer incomplete ionization collection 
even with our \asi\ electrodes,
only a small fraction of the electrons produce an ionization yield  
indistinguishable from that characteristic of nuclear recoils.

As described below in Sec.~\ref{sect:Ge},
we have measured the efficiencies of our detectors for discriminating 
between nuclear recoils, bulk electron recoils, and surface electron 
recoils using conventional radioactive sources of neutrons, photons, and 
electrons. 
Above
10~keV, BLIP detectors reject bulk electron recoils with $>$~99.9\%
efficiency and surface events with $>$~95\%
efficiency. 
ZIP detectors provide further surface-event rejection based on the
differing phonon pulse shapes 
of bulk and surface events~\cite{clarkethesis,clarkeapl}.  
This phonon-based surface-event rejection alone is $>$~99.7\% efficient
above 20~keV while retaining 40\% of the nuclear-recoil events.
Because the ZIP detector run in 1998 did not have \asi\ electrodes,
rejection of surface events in this detector was provided primarily by 
phonon pulse-shape analysis.

The ionization measurement depends on the drifting of charges to the 
detector's electrodes.
The p-type Ge has many more acceptor sites than donor 
sites, $N_{\mathrm{A}}\gg N_{\mathrm{D}}$, with number density
$n_{\mathrm{A}}-n_{\mathrm{D}} \approx 6 \times 10^{10}$~cm$^{-3}$,
and the dominant acceptor levels at 
$\epsilon_{\mathrm{a}} \approx 12$~meV above the valence band.
Because the detectors are cooled to $\sim$20~mK, 
the number of free charges is
Boltzmann suppressed by a factor $\exp(-\epsilon_a/kT )\sim e^{-5800}$ --- 
\ie\ there is no free charge.  
It is energetically favorable for
the $N_{\mathrm{D}}$ electrons to fall onto acceptor sites 
rather than to bind to
the $N_{\mathrm{D}}$ donor sites.
If left alone, the resulting $N_{\mathrm{D}}$ ionized donor sites
and $N_{\mathrm{D}}$ ionized acceptor sites
would trap charges generated by events. 
Trapping is minimized, however, 
by neutralizing the ionized impurity sites once the detectors
have been cooled, by exposing them to photons emitted by an LED
while the detectors' electrodes 
are grounded~\cite{shuttthesis}.  
Photons from the LED produce electron-hole pairs in the detector;
the absence of a drift field allows these free
charges to either recombine or be trapped on ionized impurities. 
When the detector is in the resulting neutralized state, charge-collection 
efficiency is 100\%.
The neutralized state degrades with time,
presumably due to the liberation of trapped charges as 
drifting charges scatter off the trapping sites.   
Restoration of the 
neutralized state is accomplished by  
grounding the electrodes for a 
brief period; particle interactions 
(or additional flashes of light from an LED) 
create the necessary free charge to refill the traps.  
During the 
CDMS run in 1999, the BLIPs showed no signs of degraded ionization 
collection when used with a 50-minute-biased/5-minute-grounded 
neutralization cycle.  Slightly more conservative cycles were used in the 
1998 run for the Si ZIP detector, with comparable results. 

\begin{figure*}
\begin{center}
\psfig{figure=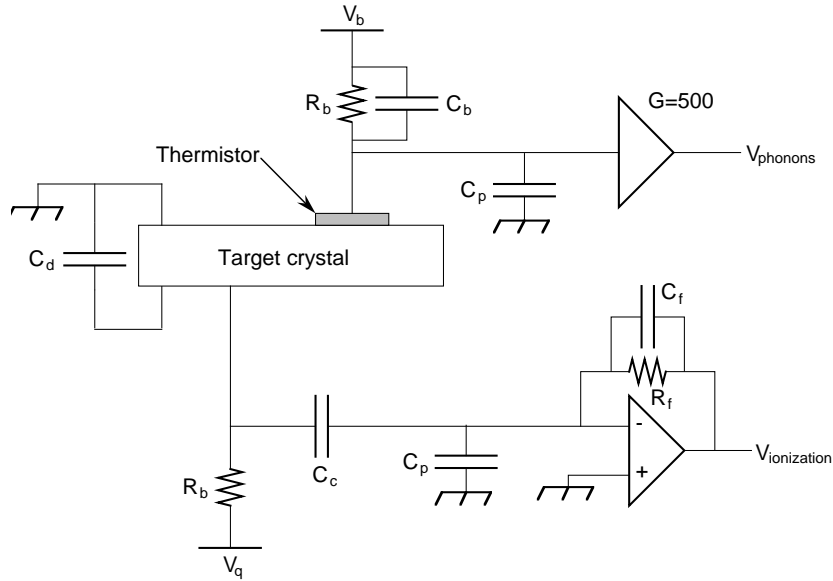,width=4.5in,
        bbllx=70,bblly=400,bburx=470,bbury=680,clip=}
\caption[Combined ionization- and phonon-readout circuit]
{Ionization-readout circuit used for 
both BLIPs and ZIPs, 
together with the BLIP phonon-readout circuit.
The ionization amplifier
connects to the biased side via a coupling capacitor with
$C_{\mathrm{c}}\approx 330$~pF.  
The detector capacitance $C_{\mathrm{d}} \approx 40$~pF.  The
ionization-bias resistor $R_{\mathrm{b}} = 40\,\mathrm{M}\Omega$.  
The parasitic capacitance $C_{\mathrm{p}} \approx 50$~pF 
is dominated by FET capacitance.
Figure taken from~\protect\cite{dasilvathesis}.
}
\label{truechargeelec}
\end{center}
\end{figure*}

The readout circuit for the CDMS detectors is shown schematically in 
Fig.~\ref{truechargeelec}.
Because the phonon circuit necessitates
establishing a true ground on one side of the detector,
the ionization amplifier is connected to the
biased side through a coupling capacitor. 
The ionization amplifier operates
as a current integrator; the signal observed is the voltage drop
across the feedback capacitor, 
which collects a charge corresponding to the product of the number
of electron-hole pairs created and the distance they drift across the
detector.  For complete charge collection, the total drift distance
for a given pair is the the detector thickness, so the integrated
charge simply gives the number of pairs created.  When trapping
occurs during drift, the integrated signal for a trapped charge is
decreased to the fraction of the detector thickness across which it
drifts before trapping.
More details on the ionization- and phonon-readout electronics can be 
found in~\cite{clarkethesis,golwalathesis,coldelect}.

\subsubsection{The BLIP phonon measurement}

The BLIP detectors rely on the fact that
the heat capacity of an insulating crystal drops as $T^3$ at low
temperatures. 
Thus, very small depositions can
cause large temperature rises.  
For a 165-g BLIP operated at 20~mK, a 10-keV deposition results in a
measurable temperature rise of 2.4~$\mu$K.

The detector's coupling to the refrigerator is via 
a gold wirebond connecting the detector mount to
a gold heat-sink pad deposited on the detector.
The dominant thermal impedance is the 
area-dependent acoustic-mismatch
resistance between the crystal substrate and the heat-sink pad.
Thermal impedances within the heat-sink pad and the wirebond
are negligible in comparison because these systems are metallic.  
Bias power dissipated in the thermistor heats the electron system in 
the thermistor
and, to a lesser extent, the crystal to a few~mK above the 
refrigerator temperature.

\begin{figure}
\psfig{figure=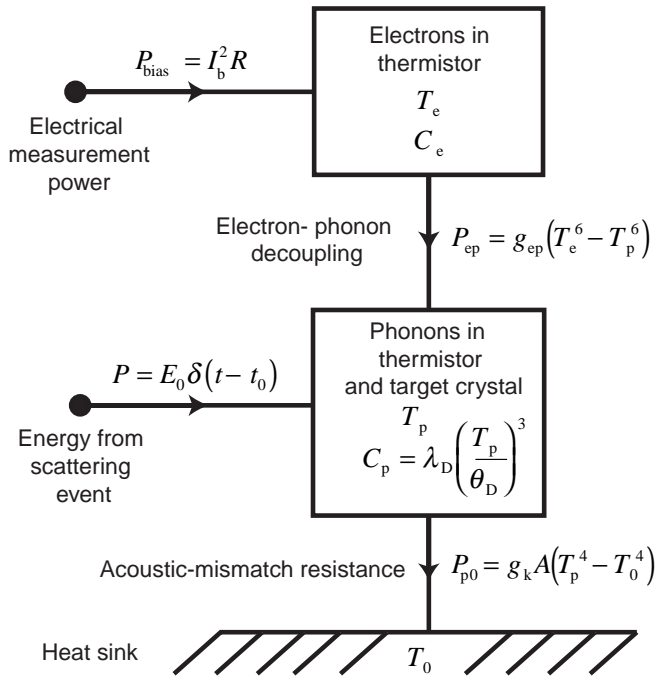,width=3.4in}
\caption{\label{blipthermal}
BLIP thermal model.  The top box is the electrons and the bottom the
crystal/thermistor phonons.  The heat sink is shown at the bottom.
The power flows are described
in the text.  Figure taken from~\cite{andrewthesis}.}
\end{figure}

A simplified thermal model for BLIP detectors, including only one
thermistor, is shown in Fig.~\ref{blipthermal}.  
One system in this model includes the phonons in
the crystal substrate and in the thermistors
since the eutectic bond is transparent to phonons.  
The other system includes the thermistor's electrons, which can be
taken to be separate from the phonon system because of the low-temperature
phenomenon of electron-phonon decoupling. 
At these low temperatures, electron-phonon interaction rates
are so low that the time needed for the
electron and phonon systems of the thermistor to 
equilibrate with each other is significant compared to the 
internal thermalization times of the 
individual phonon and electron systems within the thermistor.  
Moreover, because a significant DC power is deposited
into the electron system  of a 
thermistor (in order to bias it),
and the thermistor is heat-sunk via its phonons, a
large steady-state temperature difference arises
between 
electrons and phonons in the thermistor,
as described in~\cite{nwthesis}.

Schematically, the power flows are as follows.  A thermistor-bias
current $I_{\mathrm{b}}$ produces a measurable voltage $I_{\mathrm{b}} R$.  
This dissipates
power $I_{\mathrm{b}}^2 R$ in the thermistor.  (A current bias is needed to
prevent thermal runaway because $dR/dT < 0$.)  This power flows to the
heat sink via the phonon system.  An interaction in the crystal 
produces a
$\delta$-function energy deposition in the {\em phonon} system.  The
phonons heat up, warming the electrons via the electron-phonon
coupling and yielding a measurable change in resistance.  The energy
flows out of the system via the connection to the heat sink.  The
couplings are chosen so the electron system senses the phonon-system
temperature rise before the energy can leave the detector.

Two thermistors are used to provide rejection of interactions in
the thermistors.  Use of two thermistors also decreases the phonon
readout noise by $1/\sqrt{2}$.  For crystal interactions and assuming
the two thermistors are identical, the temperature-evolution solutions
have the same form as a one-thermistor system: the two thermistors can
be treated thermally and electrically as a single thermistor.  For
interactions within a single thermistor, 
the symmetry is broken and the results
become more complicated,
altering the signal shapes in the two thermistors.  

\begin{figure}
\psfig{figure=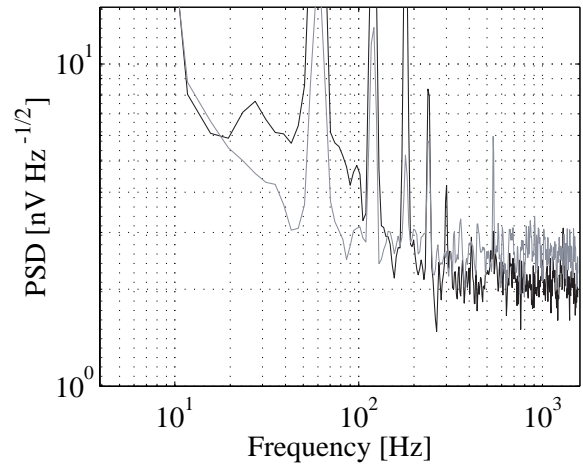,width=3in}
\caption{\label{pnoisenolockinlog}
Phonon-channel noise spectra without lockin, logarithmic scales.
Dark: phonon sensor 1.  Light: phonon sensor 2.  The continuum noise,
above about 100~Hz, is dominated by thermistor Johnson noise --- the
FET contributes $\sim 1$~\nvrthz.  The spectral lines are 60~Hz and
harmonics.  The significant increase in ``smooth'' noise and in 60~Hz
and harmonics at low frequencies motivates the use of an
AC modulation/demodulation 
technique: the fall and rise times of the phonon pulses
correspond to $\sim 3$~Hz and $\sim 30$~Hz, so essentially all of
the phonon signal is below 30~Hz.   
}
\end{figure}

The thermistor signal is a negative-going voltage pulse
given by the product of the fixed bias current and the resistance
decrease arising from an energy deposition.  A low-noise voltage amplifier 
is used to measure this signal.  The time
constants are slow enough that a significant component of the signal
lies at low frequencies.  The rise and fall times of the BLIP phonon
signals are $\sim 5$~ms and $\sim 50$~ms, 
corresponding to poles in the pulse
frequency spectrum at $\sim 30$~Hz and $\sim 3$~Hz.  Below 500~Hz,
$1/f$ noise in the JFET, thermistor, or electrical connections, and
spurious 60~Hz noise become significant; see
Fig.~\ref{pnoisenolockinlog}.  
We have found it advantageous to use an AC modulation/demodulation
technique for the BLIP phonon measurement.
To take advantage of the very clean noise environment around 1~kHz,
the DC current bias is replaced by a 1-kHz
sine-wave bias~\cite{lockin,golwalathesis,coldelect}.

\subsubsection{The BLIP pulsers}
\label{pulser}

\begin{figure}
\psfig{figure=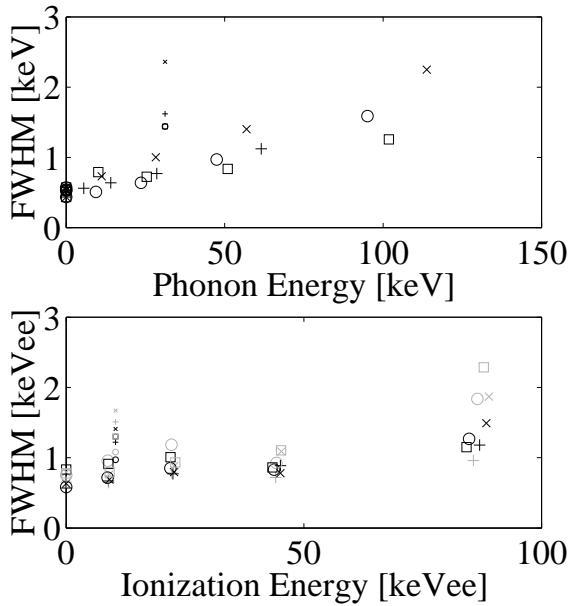,width=3in}
\caption{\label{eres}
Phonon energy resolutions and ionization 
electron-equivalent energy resolutions (full-width, half maximum) 
as functions of energy, for 
BLIP3 (crosses), BLIP4 ($\times$'s), BLIP5 (circles), and BLIP6 (squares),
as measured using the pulsers, or the 10.4~keV (31.2~keV phonon energy)
background line from gallium (small symbols).  
Resolutions of both the inner (black) and the outer (grey) ionization 
electrodes are shown.  
The apparent resolutions as determined by the widths of the 10.4~keV
background line are likely worsened by the existence of another line 
at 9.65~keV.
Phonon energy resolutions are worsened further by the 
effect of long-term drifts.
}
\end{figure}

In order to help calibrate each BLIP detector, 
a small resistive heater ($\sim 100\,\Omega$) on
the detector surface is used to produce heat pulses.
Additionally, pulser capacitors placed at the gates 
of the ionization-amplifier
FETs allow 
$\delta$-function current pulses to be sent to the ionization 
amplifiers~\cite{golwalathesis}.  
These pulsers produce signals of fixed amplitude at known times,
allowing measurement of the ionization and phonon energy resolutions
as functions of energy (see Fig.~\ref{eres}).
Every few hours during our normal data-acquisition process, 
a series of phonon-pulser events was taken.
This data allows calibration of the effect of detector temperature
on pulse height,
allowing real-time corrections for small drifts in refrigerator
temperature, as described in Sec.~\ref{dcref}.
For most of the run, ionization pulses were triggered by an asynchronous 
process, allowing independent measurement of the experiment 
live time and cut efficiencies.

\subsubsection{The ZIP phonon measurement}

In contrast to the relatively slow, calorimetric measurement 
of phonon energy with the BLIP detectors, 
ZIP detectors rapidly detect athermal phonons before 
significant thermalization occurs, using 
quasiparticle-trap-assisted electrothermal-feedback 
transition-edge sensors\cite{irwin2}.
These  phonon  sensors 
consist of 
photolithographically patterned, overlapping thin films of superconducting 
aluminum and tungsten,
divided  into  4  independent  channels  (see Fig.~\ref{AlexFig}).  
Each channel contains a parallel array of 444 tungsten 
transition-edge 
sensors (TESs) each coupled to 6 aluminum 
phonon-collection pads.  

Energy deposited in the bulk detector leads via anharmonic decay
to generation primarily of 
high-frequency $\sim$THz ($\sim$4~meV), 
quasi-diffusive phonons~\cite{tamura}.
These athermal phonons 
propagate to the detector
surface, where most of them have enough energy 
($>2\Delta_{\mathrm{Al}} \approx 0.34$~meV) 
to be absorbed in 100-nm thick, superconducting aluminum pads 
which cover 82\%
of the detector's surface~\cite{alexltd7,clarkethesis}.
Quasiparticles generated in the aluminum
by the phonons breaking Cooper pairs
diffuse in $\sim$10~\micros\ through the aluminum 
to the detector's tungsten TES,
where they become trapped.
Through electron-electron interactions, these 
quasiparticles rapidly lose their potential energy by heating the 
conduction electrons in the tungsten, which has no gap since the 
tungsten film is biased in the middle of its superconducting-to-normal 
transition.  
The net result is that a few percent 
of the energy in athermal phonons from an event in the
detector substrate is 
measured in the tungsten TES.
For the ZIP detector run in 1998, this collection efficiency was 
$\sim$2\%.

The TESs are voltage biased, and the current through them is monitored by a 
high-bandwidth HYPRES SQUID array~\cite{squid1,squid2}.  
The phonons released in the tungsten raise the temperature of the
film, increasing its resistance and reducing the current.    
To ensure operation in the extreme feedback limit, 
the substrate is kept much colder ($T<50$~mK) than the transition
temperature of the tungsten sensor ($T_{\mathrm{c}} \sim 80 $~mK).  
The tungsten is maintained stably within the transition
by electrothermal feedback based on Joule self-heating:
if the sensor
were hotter, the resistance would increase, decreasing the current and the
Joule heating;
an analogous argument applies if the sensor were cooler.  
The interaction
energy deposited in the tungsten as phonons is entirely removed by the
reduction in Joule heating caused by the current drop. 
Therefore,
in the limit of very sharp transitions,
the energy absorbed by the tungsten 
is just the integral of the current drop times the bias voltage:  
\begin{equation}
E = V_{\mathrm{b}}\int \delta I dt .
\end{equation}

The tungsten sensors are
intrinsically very fast, with pulse rise times electronics bandwidth
limited (at $\sim$100~ns), and fall times governed by the electrothermal
feedback time (20--40~\micros). 
The actual pulse shapes measured from ZIP phonon sensors
are dependent on both the phonon propagation in the detector
substrate, and the quasiparticle diffusion in the Al collection fins. 
The pulses typically have rise times in the range 5--15~\micros, 
and fall times $\sim$100~\micros,
dominated by the phonon collection.
Comparison of phonon-pulse arrival times in the four independent channels 
allows localization in the xy-plane of a ZIP detector. 
In addition, energy deposited near detector surfaces apparently gives rise 
to slightly lower-frequency phonons, 
which undergo less scattering and hence travel 
ballistically~\cite{clarkeapl}.
The shorter rise times of the resulting phonon pulses allow rejection 
of such surface events.

\subsection{Cryogenics}

The detectors are located inside a large cold
volume~\cite{icebox,pdbthesis}. The nested cans of the cryostat,
each of which corresponds to a thermal stage 
in our modified Oxford Instruments S-400 dilution
refrigerator, serve as both thermal radiation shields and heat sinks for
detector wiring and support
structures. The cryostat is connected to the dilution refrigerator via a copper
coldfinger and a set of
coaxial copper tubes.
Each tube connects one can to the
corresponding thermal stage in the refrigerator, with the copper coldfinger
connecting the innermost
can directly to the mixing chamber. 
The nominal temperatures of the cryostat cans (and refrigerator 
thermal stages) are 
10~mK, 50~mK, 600~mK, 4~K, 77~K, and 300~K. 
The cryostat itself contains no cryogenic liquid; all
cooling power is generated in the refrigerator, and the cryostat is cooled via
conduction.
The innermost can is 30~cm in diameter and 30~cm high, 
providing approximately 21~liters of experimental space at $\sim20$~mK
base temperature. 
Access to this space is
obtained by removing the can lids.

A cryogenic detector readout package addresses the
unusual combination of requirements in CDMS --- low noise, low
background, high channel count, and low
temperature~\cite{coldelect}. The anchor for the system is a
multi-temperature-stage modular coaxial wiring package, or ``tower.''
Directly below the tower are mounted up
to six detector holders with modular coaxial wiring assemblies.
Mounted on top of the tower are cold electronics
cards that carry either four FETs (for a BLIP detector),
or four DC superconducting
quantum interference device (SQUID) arrays and two FETs
(for a ZIP detector).
Because of the susceptibility to microphonic
pickup for the gate wires of the FET, a vacuum coaxial geometry is
used in which the wires are tensioned and attached to printed circuit board at
the ends of covered copper channels. The absence of a dielectric near 
the gate wires
minimizes the presence of static charge, thereby reducing microphonic
pickup. The printed circuit boards also serve to heatsink the wires to
the various temperature stages. The electrical connections from the
FET/SQUID cards at 4~K to the room-temperature vacuum bulkhead 
feedthroughs
are made through a 3-meter-long shielded copper-kapton flex circuit, or
``stripline.''
The tower and detector packaging is constructed so that
infrared radiation from room temperature and the 130~K FETs is
efficiently blocked and absorbed at each layer.
Except for the warm end of the stripline, which is
outside the radioactive shielding, all of the components of the
towers, stripline, electronics cards and detector packages are made
from materials 
that have been prescreened for U/Th isotopes, with the
goal of having $<0.1$ ppb of the mass of the material surrounding the
detector package, or approximately $<1 \; {\rm \mu Bq/g}$. 
One such material is  a custom-made low-activity
solder~\cite{cdms_solder}.

\subsection{\label{facility}The Stanford Underground Facility}

Due to the cryogenic technology and continuing development of our Ge and Si
detectors,  
the initial dark matter search has been conducted at a local site. 
The Stanford Underground Facility (SUF) is a tunnel
10.6~meters below ground level
in the Hansen
Experimental Physics Laboratory,  
on the Stanford University campus.
The tunnel housing the experiment is a clean area supplied 
with cooled, filtered air from the surface to suppress radon.
The earth above SUF absorbs the hadronic component of cosmic-ray showers 
which would otherwise produce a large background rate 
and activate materials near the detectors. 
The overburden also reduces the muon flux by a factor of 5; 
the muon flux measurements indicate that the overburden is 
equivalent to $\sim16$~meters of water.
A substantial vertical muon flux
($29~$m$^{-2}$~s$^{-1}$~sr$^{-1}$)
is still present
in the SUF tunnel due to the relatively shallow depth. 
The muon-induced neutron flux, 
and the ambient photons and neutrons from radioactivity in the tunnel walls,
dictate that a passive shield and an active veto surround the detectors.

\subsection{Shielding and muon veto}

The goal of shielding is to minimize the rate
of interactions arising from external particle
sources that can mimic nuclear
recoils in the cryogenic detectors. These external sources
include photons and neutrons from radioactivity in the surrounding
environment,
photons and neutrons produced by cosmic-ray muons, and electrons
from radioactivity on surfaces.
The external sources are primarily from the $^{238}$U and $^{232}$Th
decay chains, with
photon energies up to 2.6~MeV, and from $^{40}$K,
which emits a 1.46~MeV photon.
Passive shielding consisting of lead, polyethylene, and copper reduces the flux
from radioactive contamination,
while
active shielding efficiently vetoes 
the flux produced by muons from cosmic rays.

\begin{figure}
\psfig{figure=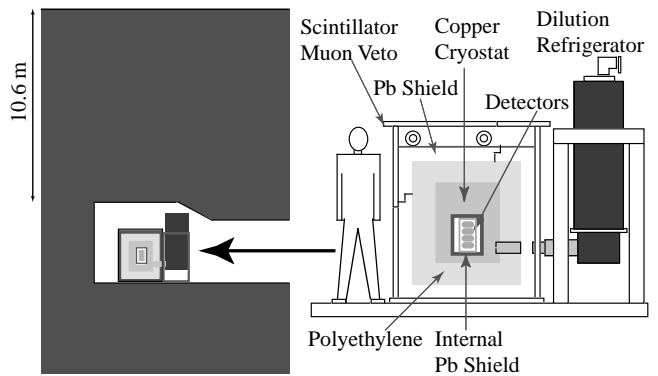,width=3.375in}
\caption{\label{fig:cdmsshield}
Layout of the CDMS I shielding at the Stanford Underground Facility.}
\end{figure}

The concentric shields around the WIMP detectors at SUF are
shown schematically
in Fig.~\ref{fig:cdmsshield}. 
Outermost is the active
veto~\cite{andrewthesis}, fashioned of
NE-110 plastic scintillator 
with waveshifter bars coupled to 2"
RCA 8575 photomultiplier tubes (PMTs). 
Each scintillator is coupled to 1--4 PMTs,
depending on its size and shape. 
The PMT signals are summed together for each
scintillator, then presented to LeCroy NIM discriminators. 
The discriminator thresholds are
set to be sensitive to (minimum-ionizing) cosmic-ray muons, 
which deposit about 8~MeV in the
4.1-cm-thick scintillator,
and insensitive to the vast majority of
photons from radioactivity, whose spectrum ends at 2.6~MeV. To
reject events in the detectors that occur close in time
with the passage of a muon, we record the times of all veto hits above
threshold in a $\pm 10$~ms window about each detector trigger and use
a $\sim$25~\micros\ window to establish correlations. 
The total veto-trigger rate during 
normal operation is approximately 6~kHz, leading to $\sim$15\% dead time 
due to accidental correlations. 
To monitor possible
changes in veto performance, analog-to-digital converters 
read out the pulse heights
from all six sides of the veto for each event.

\begin{figure}
\psfig{figure=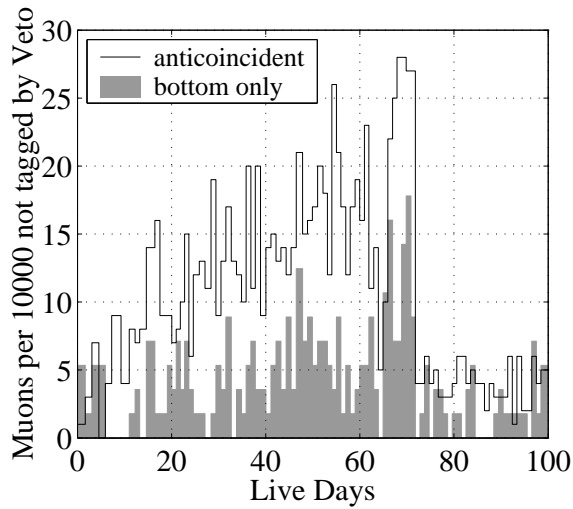,width=3in}
\caption{\label{fig:vetoeff}
Veto inefficiency for detector-tagged muons 
during the 1999 Ge data run described in Sec.~\ref{sect:Ge}.
The dark, unfilled histogram indicates the number of 
detector-through-going muons anticoincident with the muon veto, per
10000 detector-through-going muons detected.  
The grey, shaded histogram shows the fraction of muons passing through
both a detector and the bottom layer of the veto that were not tagged 
by one of the other sides of the veto.
The performance of the veto slowly 
degraded over the course of the run.  It was improved briefly on 
June~20 (live day~64).  It was improved more permanently on July~30 
(live day~72).  See Fig.~\ref{ltvsdayall} for the dates corresponding to the 
integral live days into the run. 
}
\end{figure}

A thorough mapping of the veto with an
x-ray source documented a few areas of relatively poor light collection
in late 1998, just before the start of the 1999 Ge data run described 
in Sec.~\ref{sect:Ge}.
To compensate, high
voltages and thresholds for all veto counters were tuned to ensure that muons
passing through these
areas would not be missed (at the expense of reduced livetime due to a
higher rate of vetoing by
environmental photons passing through the areas of the counter with better
light collection). 
The
efficiency of the veto for detecting muons can be measured using muons
identified by their
large energy depositions in the Ge detectors. The average measured
efficiency of this veto for
muons during the 1999 Ge data run described in Sec.~\ref{sect:Ge} was 99.9\%, 
with time variation 
shown in Fig.~\ref{fig:vetoeff}. 
The rejection inefficiency for cosmic-induced
neutrons generated in material surrounded by the veto should be 
$\sim3\times$ worse ($\sim$0.3\%);
this rejection efficiency is sufficient to reduce the
background from these neutrons 
to a level comparable to 
the background from neutrons produced
outside the veto.
The measured efficiency of the veto for muons during the 
1998 data run is even higher, 99.995\%.

The veto surrounds a lead 
shield of 15 cm thickness, which attenuates
the external photon flux by a factor of 1000.
The inner 5~cm of this
lead shell is made from
Glover lead, which has substantially less of the long-lived (22-year
half-life) $^{210}$Pb isotope which is
present at some measurable level in all sources of recently manufactured
lead~\cite{dasilvapb210}.  Decays of $^{210}$Pb yield a bremsstrahlung
spectrum (from $^{210}$Bi with 1.16~MeV endpoint), which
results in background photons that interact in the detectors.
Inside the lead, a 25-cm thickness of polyethylene surrounds the
cryostat. The
polyethylene moderates
and attenuates neutrons from the material surrounding the tunnel
and from the interaction of cosmic-ray muons with the lead shield.
Previous studies at this depth indicate that thicker polyethylene 
would increase the neutron flux at the detectors due to neutron production 
in the polyethylene itself.
The cryostat and detector-wiring assembly constitute an 
average thickness  of about 3~cm of copper. 
The most important contribution of the veto is to reject events
from neutrons produced by cosmic-ray muons entering this copper.
Samples of
all construction materials were screened to ensure low
radioactive contamination. 
A 1-cm-thick ``internal'' shield made of ancient Pb, which has very little 
$^{210}$Pb, immediately surrounds
the detectors in order to further reduce the photon background~\cite{nantes}.
The layers of the shield outside the
cryostat can be partially lifted and
rolled away for easy access to the detector volume.
None of the shielding is hermetic because copper tubes providing 
cooling or electrical connections must 
penetrate the shields; however, shielding inside these copper tubes 
helps reduce the external photon flux.

\subsection{\label{expBackgrounds}Expected backgrounds}

The shielding was designed in conjunction with Monte Carlo
simulations and measurements of particle fluxes at 
SUF~\cite{dasilvathesis,dasilvaneutron}.
The measured event rate between 10--100~keV in Ge detectors due
to photons is roughly 60~keV$^{-1}$~kg$^{-1}$~d$^{-1}$ overall and
2~keV$^{-1}$~kg$^{-1}$~d$^{-1}$ anticoincident with the veto.
These anticoincident photons are presumably due to residual radioactivity 
in and around
the inner shielding and detector package. 
Detector discrimination of $99.9\%$ 
should reduce the photon background 
to $\approx$ \scinot{5}{-4}~events~keV$^{-1}$~kg$^{-1}$~d$^{-1}$, 
negligible compared to other expected backgrounds.
The non-muon-induced low-energy-electron background is more 
difficult to predict, as it depends critically on the level of 
radioactive contamination on parts immediately next to the detectors.
This background is also potentially more troubling
because of the CDMS detectors' ionization dead layer.
Discussion of the measured low-energy-electron background is
described in Sec.~\ref{sect:Ge}.

The rate of
neutrons from natural radioactivity of materials inside the shield is
negligible because of the careful choice of construction materials.
Neutrons from natural radioactivity in the tunnel walls and outer lead
can also be ignored; because their spectrum is softer than that of neutrons
produced by muons, they are well moderated by the
polyethylene. 
Neutrons with energies capable of producing keV nuclear recoils in 
the detectors are
produced by muons interacting inside or outside the veto
(``internal'' or ``external'' neutrons, respectively).  
The dominant,
low-energy ($< 50$~MeV) component of these neutrons is moderated well by the
polyethylene~\cite{dasilvaneutron}.  
Essentially all remaining internal neutrons are tagged as 
muon-coincident by the scintillator veto.
However, relatively rare, high-energy external
neutrons may ``punch through'' the polyethylene and yield secondary neutrons
that produce keV nuclear recoils.  
A large fraction of the events induced by high-energy external
neutrons are vetoed: $\sim$40\% due to neutron-scintillator
interactions, and an unknown fraction due to hadronic
showers associated with the primary muon.  
This unknown fraction, combined with a factor of 4
uncertainty in their production rate, makes it difficult to accurately
predict the absolute flux of unvetoed external neutrons.  

Two methods are used to {\em measure} this flux 
of unvetoed external neutrons.
The first method involves comparing the rate of nuclear-recoil events 
in the Ge detectors with the rate in the Si detector,  
since Ge is more sensitive to WIMPS and Si is more sensitive
to neutrons.
The second method is to count the number of events 
consisting of nuclear recoils in two or more detectors.  
Since WIMPs interact too weakly to multiple scatter,
these events must be due to neutrons, thereby 
providing a clean measurement of the neutron background.
Predictions from Monte Carlo simulations of the expected ratio
of single-detector scatters to multiple-detector scatters
are then used to determine the  
expected rate of
neutron single-scatter events.
Neutron backgrounds are simulated 
using the MICAP~\cite{MICAP} and FLUKA~\cite{FLUKA} extensions
to the GEANT~\cite{GEANT} particle-physics simulation package. 
The MICAP and FLUKA packages track neutrons above and below 20~MeV,
respectively.  For this work, no attempt is made to simulate the
production of the neutrons. 
Instead, production rates and spectra 
from~\cite{khalchukov1983}
 are used, and
only the propagation of the neutrons and their interactions in the
detectors are simulated. These simulations will be discussed further 
in Sec.~\ref{sect:n}. 

\subsection{Data acquisition}

\begin{figure*}
\psfig{figure=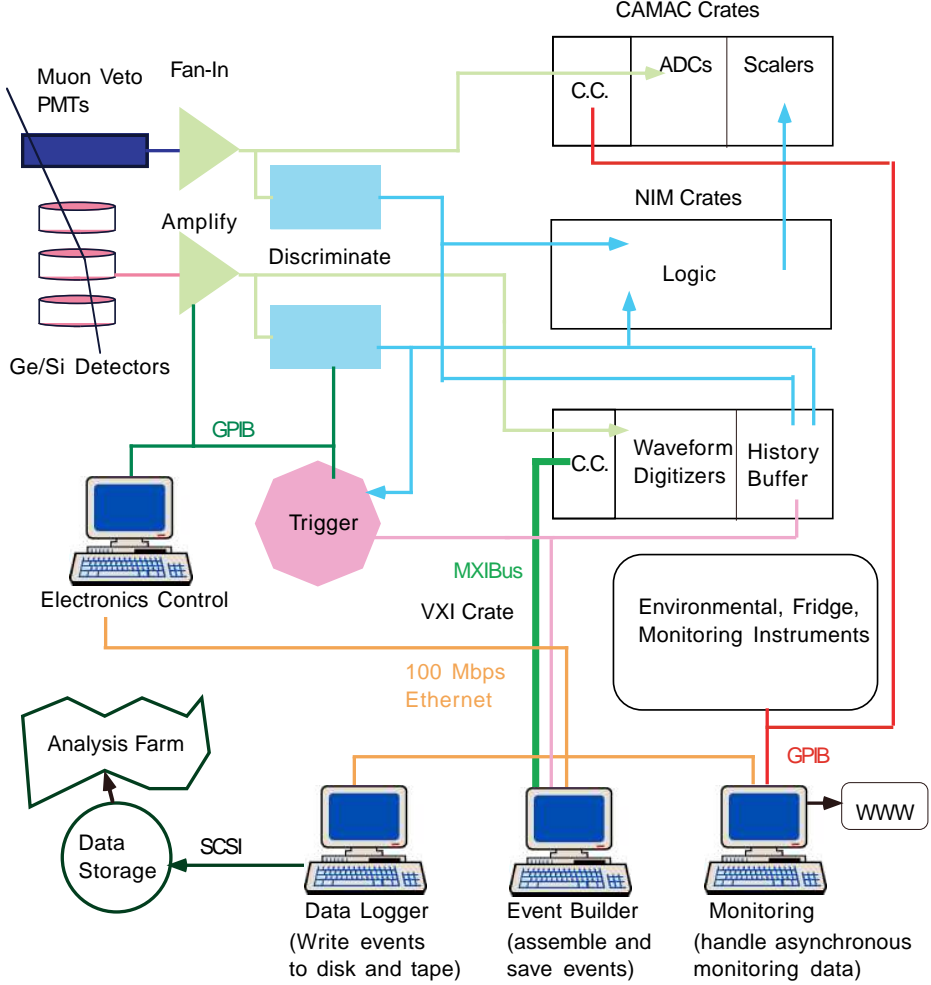,width=5in,
bbllx=30,bblly=170,bburx=583,bbury=762,clip=}
\caption{\label{fig:daq}
(Color). Block diagram of the CDMS data acquisition system.}
\end{figure*}

The purpose of the data acquisition system for CDMS (shown as the 
block diagram in Fig.~\ref{fig:daq}) is to generate an 
experimental trigger and faithfully record all detector and veto 
activity within 
a specified time interval about that trigger.
Detector signals from the front-end electronics are received, 
conditioned, and anti-alias-filtered 
in custom 9U electronics boards.
These boards also contain discriminators which provide low-threshold 
ionization-trigger and 
phonon-trigger signals, as well as high-threshold trigger signals for 
vetoing high-energy events during calibrations.
The trigger signals are combined in a separate 9U board which generates 
a global trigger
signal to inform the data acquisition computer that an event has 
occurred. The individual trigger signals are also stored in a history 
buffer (VXI Technology 1602,  clocked at 1~MHz), which preserves a 
triggering history for up to 10~ms 
before and after each global trigger. 
Trigger thresholds and logic are configured via a backplane digital bus 
that is interfaced to GPIB.

The filtered detector pulses are routed to VME waveform digitizers 
(Omnibyte Comet and Joerger VTR1012)
situated in a VXI mainframe, which provides better ambient noise rejection 
than VME crates. These 12-bit, 5-10 MHz digitizers record
the entire waveform, or trace, for each detector channel, including the 
pre-trigger baselines.
This information is
crucial for extracting the best signal-to-noise from the detectors, 
and for rejecting artifacts such as pulse pile-up,
at a cost of large event sizes (typically 50--100 kB).

The muon-veto PMT signals are processed by NIM discriminators and logic,
then recorded
in a VXI history buffer (VXI Technology 1602) which is clocked at 1~MHz.
A buffer extending on average from 15~ms before trigger to 5~ms after 
trigger
is read out on every trigger,
allowing 
correlations with cosmic-ray muons to be made strictly in software.

Monitoring information is provided by GPIB and CAMAC instruments.
The dilution
refrigerator and cryostat temperatures and pressures are sampled every 30
minutes, while detector temperatures,
trigger/veto rates, and veto high voltages are measured once a minute.
This information is
constantly on display at SUF and is remotely accessible from any WWW browser.
E-mail and phone
alarms warn of serious problems.

The online data acquisition software is written in 
LabVIEW~\cite{labview} and runs
on a cluster of Power Macintoshes. The system is
modular, in that the main event-builder program runs on one computer which
communicates over a high-speed
link to the VXI crate, while all front-end control and
environmental monitoring runs on separate computers. 
A VME I/O module (HP 1330B) synchronizes the software 
 to the trigger
hardware and provides the path for a random (software) trigger
to be recognized by the hardware. 
The online acquisition system is capable of running
with better than 85\% livetime for up to six detectors
at the typical total low-background trigger rate of $\sim$0.4~Hz. 
Data is
written over the local Fast Ethernet (100 Mbps) network to fast SCSI disks,
where it
is promptly analyzed via
a Matlab/C analysis system running on Unix/Linux workstations. Both raw
data and summary information
are written to DLT tapes.

\section{\label{r19datapipe}G\lowercase{e} BLIP Data Reduction}

\begin{figure}
\psfig{figure=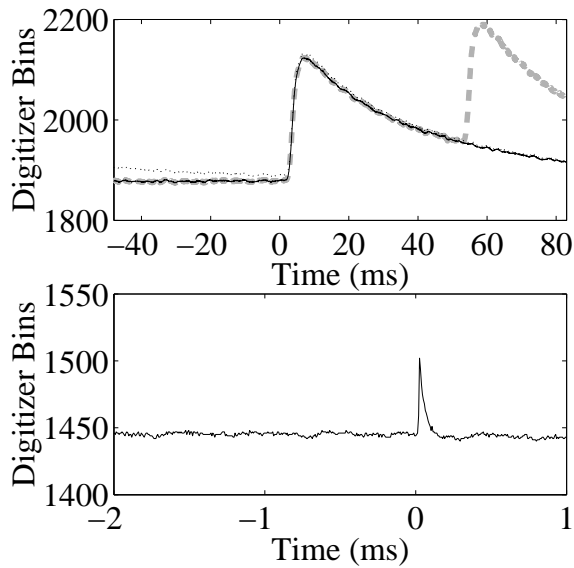,width=3in}
\caption{\label{pulses}
Typical BLIP phonon-channel (top) and ionization-channel (bottom) pulse shapes,
with times shown relative to the trigger time.
Overlaid on the phonon pulse shape (solid) are examples of how the pulse
might look with pretrigger pile-up (dots) or post-trigger pile-up 
(grey dashes).
Traces shown are from BLIP4 phonon sensor~1 and inner-electrode 
ionization channel,
for a neutron-calibration event with $E_{\mathrm P}=199$~keV and
inner-electrode ionization energy
$E_{\mathrm{QI}}=23$~keV.
The full downloaded phonon trace is shown, but the ionization trace actually 
extends from 9.8~ms before trigger to 3.3~ms after trigger.
}
\end{figure}

Automated analysis reduces the detector pulses 
(see Fig.~\ref{pulses})
to quantities 
describing the energies, times, and quality of various fits performed.
First, it is necessary to determine the
event ``delay'' --- the position of the global trigger time relative
to the particle interaction, 
as determined using the detector that gave rise to the
global trigger.  In the vast majority of events, any
multiple scattering occurs on  
timescales much shorter than
the pulse rise times, so it is reasonable to speak of a single
particle-interaction time.  
Once this delay is determined (see Sec.~\ref{sect:delay}), 
the pulse energy is fit using templates, as described in 
Sec.~\ref{sect:energyfit}.
These energies are calibrated daily, as described in 
Sec.~\ref{dcref}.

\subsection{\label{sect:delay}Determination of the event delay}

Calculation of the delay is done using optimal (Weiner) filtering on the
triggering detector~\cite{golwalathesis,numrecipes}.  
If a trace 
baseline is below the digitizer range, 
the event is not fitted.  For a trace with its peak 
above the digitizer range, 
a simplified
delay algorithm, which takes advantage of the fact that the start of a 
large pulse is easy to find, is employed.  

If the event's global trigger is an ionization
trigger, the calculation is done on the ionization pulse summed over
both electrodes,
and the trigger time is used for correlating with the veto.
If the event is a phonon trigger, first the delay of the 
average of the two phonon channel pulses is calculated,
using a time-domain convolution.  
Because the phonon pulses 
have a 5-ms rise time, this delay does not provide a sufficiently
precise time-offset estimate to allow correlation with the veto ---
the veto-trigger rate is $\sim$6~kHz, making accidental
coincidences too frequent.  
Instead, the optimal-filter
convolution is performed on the ionization traces over a 
search window restricted by this phonon delay.  
If no above-threshold pulse
exists, the search finds a noise excursion.  In the case of a phonon
trigger, the widths of the search windows for the phonon and ionization
signals are 14.4~ms and 1.6~ms, respectively, large enough that 
pulses above noise are not found near the window edges.

The delay determined in the above way is used as the time offset
in the fitting algorithm for the pulses in all the detectors.  It is
also used to determine the nearest veto hit.  Phonon-trigger events are 
characterized as veto-anticoincident if there is no veto hit within 
25~\micros\ of
the time of this inferred delay.
Ionization-trigger events are veto-anticoincident if there is no veto 
hit in the 25~\micros\ before the event trigger. 

\subsection{\label{sect:energyfit}Pulse-energy fitting}

Once the delay is determined, the pulse energy is fit
using templates.
For each channel, a template is built by averaging a number of
ionization-triggered pulses.  Pulses with energies of 100--200~keV are
used to ensure high signal-to-noise while being low enough in
energy to be unaffected by pulse-shape variations with energy.
To form templates for the shape of the ionization crosstalk, events 
with energy only in a single electrode are used. 
It is necessary to build different templates for each
detector and channel because of pulse-shape variations.  
In the phonon channels, 
variations are caused by small differences in
thermistor properties and detector heat sinking.  
Variations in the
ionization pulse shape occur because of differences in
feedback-component values and amplifier open-loop gains.

For the phonon pulses, linear template fits are performed, 
minimizing the \chisq\ defined by
\begin{equation}
\chi^2 = \sum_{i = 1}^{N} \frac{| V_i - V_0\, s_i |^2}{\sigma^2}
\end{equation}
where $V_i$ are the ($N = 2048$) digitized data samples, 
$s_i$ is the pulse-shape template,
$V_0$ is the fitted pulse amplitude, and $\sigma$ is the rms noise per
sample.  
In practice, additional linear terms are included
(a baseline
offset and an arbitrarily normalized exponential with time
constant fixed to the known pulse fall time 
to fit the tail of a possible previous pulse), but this simplified
description well summarizes the method.
Minimization with respect to $V_0$ yields
\begin{equation}
V_0 = \frac{\sum_{i=1}^N \frac{V_i s_i}{\sigma^2} } 
           {\sum_{i=1}^N \frac{s_i^2}{\sigma^2}}  \label{linearfit}
\end{equation}
The
\chisq\ of the fit is incorrectly normalized because 
correlations in the noise
between time samples are not taken into account.
Cuts based on the \chisq\ values are therefore formed empirically, 
ignoring the overall normalization.

For the ionization traces, it is advantageous to use optimal filtering
to calculate the fit energy because of the significant frequency structure
of the noise of the ionization
channels (due to FET $1/f$ noise, 60~Hz pickup, and pickup of 1~kHz and
harmonics from the thermistor bias).  
Optimal filtering calculates
the pulse fit in frequency space, where frequency components with low
signal-to-noise are deweighted to minimize their effect on the fit.
The optimal time-offset and energy estimators are given by the time and
the value of the peak of the 
convolution of the optimal filter with the trace.
The time offset provides
the phase factor to apply to the template in frequency space to allow
calculation of the \chisq\ in frequency space, where it can be
correctly normalized because noise components at different frequencies
are uncorrelated.
A complication arises because of cross-talk between the inner
and outer ionization channels of a single detector.  
Each ionization channel's trace
is the sum of its own pulse and a cross-talk component whose amplitude
is proportional to that of the pulse in the other channel.  There is 
an analogous matrix equation for the
\chisq\ in this case, which fits both 
ionization channels at once~\cite{golwalathesis}.

\subsection{\label{dcref}Energy calibrations}

Due to drifts in both refrigerator base temperature and the
electronics, the phonon energies fit by the above procedure exhibit 
slow drifts with time.  
Although the ionization energies do not drift with time, 
discrete events such as cycling of power on the front-end
electronics crate can cause changes in the ionization calibration.  
It is
necessary to perform an absolute, time-dependent calibration to
correct these changes.

\begin{figure}
\psfig{figure=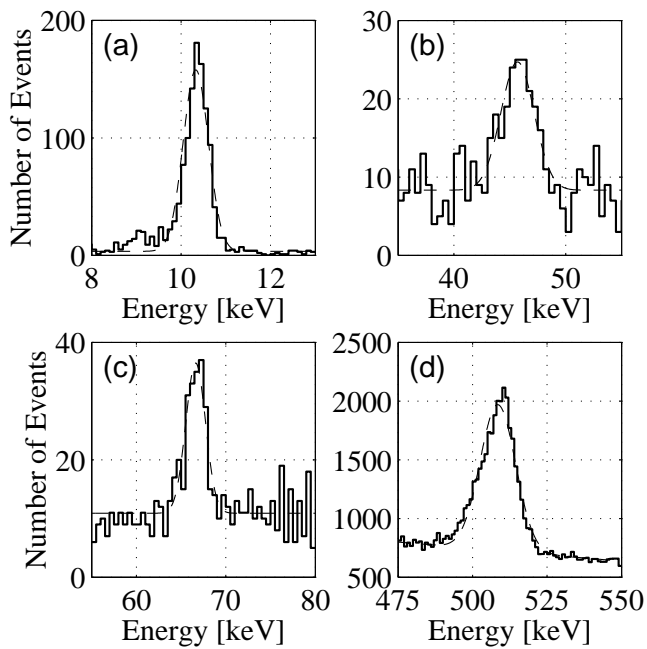,width=3.4in}
\caption{\label{Elines}
     Spectral lines visible during low-background running, 
     in recoil energy $E_{\mathrm{R}}$, summed
     over all four Ge detectors.  Gaussian fits are shown as dashed curves.  
     (a) Line at 10.4~keV from internal Ga, using phonon sensors.  
     (b) Line at 46.5~keV from \pb210, evident in events 
         with energy in the outer electrode only, using the phonon sensors.  
     (c) Line at 66.7~keV from $^{73m}$Ge, using phonon sensors.  
     (d) Line at 511 keV from positron annihilation, using ionization sensors.
}
\end{figure}

The energy $E_{\mathrm{Q}}$ of the ionization channels is 
calibrated for large blocks of time (days to weeks) using the 511~keV 
positron-annihilation line, which appears during normal low-background running.  
To account for phonon drifts on scales longer than a day, 
the overall energy scale of each phonon sensor is calibrated  
against ionization 
using the prominent bulk
electron-recoil band and the relation $E_{\mathrm{P}} = 
\left(1+\frac{eV_{\mathrm{b}}}{\epsilon} \right) E_{\mathrm{Q}}$.
To account for phonon drifts due to temperature drifts over shorter 
timescales, a simple linear correction is made to the phonon pulse height
based on the phonon-lockin DC-reference measurement of each
thermistor's average resistance, made every ten seconds.
To first order, the phonon pulse height is
linear in deviations of the thermistor resistance due to thermal
drifts.  
The correction is calibrated using
phonon-pulser events of known energy.  
Occasionally, large temperature excursions drive a phonon sensor 
out of the range for which the correction is calibrated; the detector 
is considered to be dead during such periods.
Success of the energy calibration is demonstrated by the appearance of
low-energy spectral lines (see Fig.~\ref{Elines})
in the low-background data set described below.

\section{\label{sect:Ge}G\lowercase{e} BLIP Data Set}

\begin{figure}
\psfig{figure=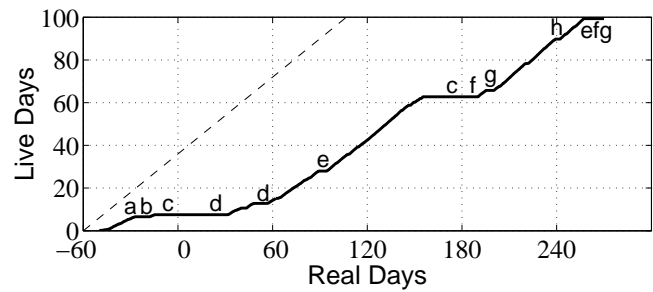,width=3.4in}
\caption{\label{ltvsdayall}
Cumulative time waiting for a trigger.  The
        dashed line has a slope of 0.6, the
        maximum observed slope during stable running.  
       The origin of the
       horizontal axis is January 1, 1999.
      Labeled periods of significant deadtime were due primarily to 
      a)~computer 
      problems and work, b)~slow pulses (see~\cite{golwalathesis} for 
      details), 
      c)~refrigerator warm-ups, d)~electronics work, e)~neutron 
      calibrations f)~low-bias studies g)~photon 
      calibrations, and h)~pump failure.}
\end{figure}

Between November, 1998, and September, 1999, 99.4 raw live-days of 
low-background data were
obtained using 3 of 4 165~g Ge BLIP detectors.  
Raw live-days denotes the live time of the data-acquisition (DAQ) system, 
before any
cuts are made, excepting periods when the raw data are 
discarded due to obvious problems.  
Figure~\ref{ltvsdayall}
shows the integrated 
live time for which the DAQ was taking low-background data
(\ie, excluding grounding and calibrations).  
The largest slope is $\sim$ 0.6 live-day/real day; 
periods of significant dead time are labeled in the figure.
During stable low-background running, the dead time consists of time 
for cryogen transfers ($\sim$10\%), detector grounding ($\sim$10\%), 
phonon pulser calibrations ($\sim$5\%), and DAQ deadtime ($\sim$15\%).

\subsection{\label{sect:cal}Calibrations}

As shown in Fig.~\ref{ltvsdayall}, \insitu\ detector 
calibrations with external photon and neutron sources were performed during 
the 1999 Ge data run.
These calibrations are used to help determine cut efficiencies, as 
described in Sec.~\ref{sectCuts},
and to estimate particle-misidentification rates
and other possible systematic errors in the 
analysis of the low-background data.

\subsubsection{Neutron calibrations}

In order to provide nuclear-recoil events that mimic WIMP 
interactions, a
\calif252-fission neutron source is placed on
the top face of the scintillator veto.  
Because the neutrons emitted by this source
have such low energies (see \eg\ \cite{knoll}), 
the top layers of polyethylene inside the
shield are removed to permit the neutrons to penetrate to the
cryostat.  
With the source and shielding in this configuration, 
the data set is dominated by neutrons, making 
the total event rate about 3 times higher than during low-background
data-taking.  
In all other ways, the data-taking conditions
are as usual.  
The source activity is known to $\sim$5\% accuracy, so the absolute
normalization of the spectrum is well determined.  
The overall cut
efficiency, determined by the methods discussed in 
Sec.~\ref{sectCuts},
is smaller than for the low-background data because the higher event 
rate significantly increases the amount of event pileup.  

\subsubsection{Photon calibrations}

The photon calibration is performed by inserting a \co60 source through a
small, pluggable hole in the lead shield.  \co60 emits two high-energy photons,
at 1173~keV and 1332~keV.  
These photons Compton scatter in the material surrounding the 
detectors, resulting in a secondary photon spectrum similar to the 
expected radioactive backgrounds.
The photons yield a large sample of bulk electron
recoils with $\sim$3\% surface electron recoils.
Although some surface events arise from electrons ejected from surrounding
materials,
simulations indicate that most low-energy surface events 
are due to electrons kicked
through the dead layer (and then out of the detector)
by high-energy photons Compton-scattering {\em inside} a
detector. 

Because the calibration results in many high-energy events,
whereas the WIMP search uses only low-energy events,
a hardware trigger veto rejects events with recoil energy 
$E_{\mathrm{R}}\agt 100$~keV
during the photon calibration.  
The calibration data are
analyzed in the same way as the normal data stream.  
As with the neutron-calibration data, a larger fraction
of events are cut due to pileup.
This larger fraction is not a concern because the photon
misidentification is determined by beginning with a set of events that
pass all data-quality cuts and then calculating the fraction that 
also pass the nuclear-recoil-acceptance cut. 
The efficiency of the data-quality cuts has no effect, 
since no data-quality cuts depend on ionization yield.

\subsubsection{Electron calibrations}
\label{eleccal}

\begin{figure}
\psfig{figure=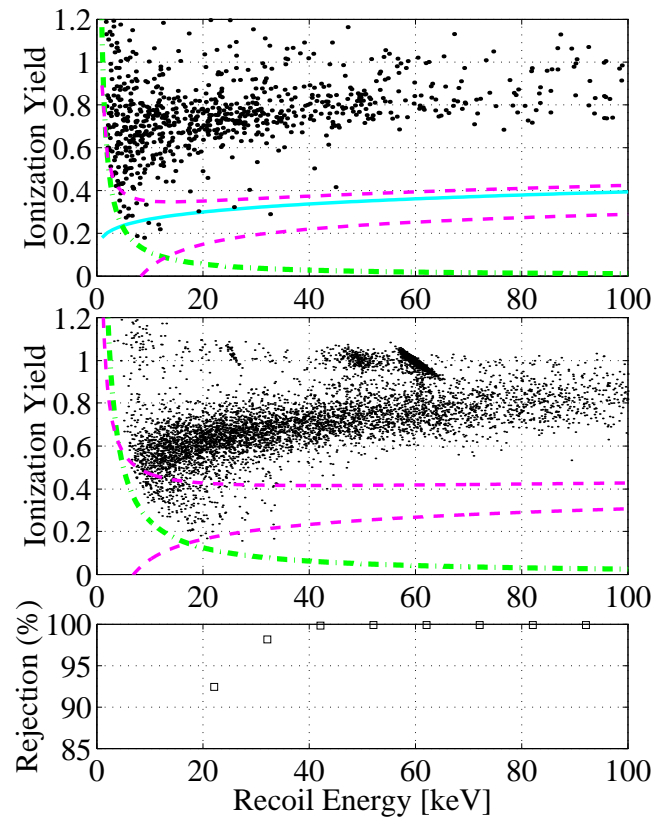,width=3.4in}
\caption{\label{electroncal}
Electron calibration data.
Hyperbolic dot-dashed lines: mean ionization-search thresholds.
Solid curves: mean centers of nuclear-recoil bands.
Dashed curves: mean nuclear-recoil-acceptance regions.
Top: 1999 run electron-calibration set consisting of 407
veto-anticoincident events tagged as multiple scatters in BLIP3 and 
BLIP4.
Middle:
Data from external \c14 source data taken with test device ABL1 with 
source-side electrode at positive bias. 
Bottom: Rejection efficiency for the test device.}
\end{figure}

Unfortunately, \insitu\ calibrations with external electron 
sources are not practical
because of the substantial material forming the cold volume.
Furthermore, 
BLIPs 3--6 were never tested with an external electron source
in the lab.  
Small devices prepared with variants of the electrode have been
tested with an electron source (see Fig.~\ref{electroncal}), 
but no laboratory electron
calibration was performed with the exact electrode structure used on
the detectors.

The photon calibration contains a very small fraction of electrons,
$\sim$0.7\% in the 10-to-100-keV range according to Monte Carlo
simulations.  The typical number of events
observed in this energy range during the calibration is $\sim$9000 
per detector.
Therefore, only $\sim$60 electrons are expected per detector,
insufficient for placing a useful limit on electron
misidentification.  

The veto-anticoincident data provide an electron calibration because BLIP3
appears to be heavily contaminated with an electron source that
results in clear electron bands in BLIPs~3 and 4.  The contamination
likely consists of \c14 atoms from a leaking \c14 source 
to which the detector was 
exposed during an attempted laboratory calibration.
Low-energy (10-100~keV) veto-anticoincident multiple-scatter events 
between BLIP3 and BLIP4 appear to be dominated by this electron 
``source'' on the surface of BLIP3.
Figure~\ref{electroncal} shows ionization yield vs. recoil
energy in the two detectors for the calibration data set.  The surface
events form a clear band in ionization yield, similar to that seen in 
a test device with \asi\ contacts. 
The bulk of the
events are concentrated at low recoil energy, so this data set probes
energies where electron misidentification is worst.

\subsection{\label{sect:thresh}Hardware and analysis thresholds}

\begin{figure}
\psfig{figure=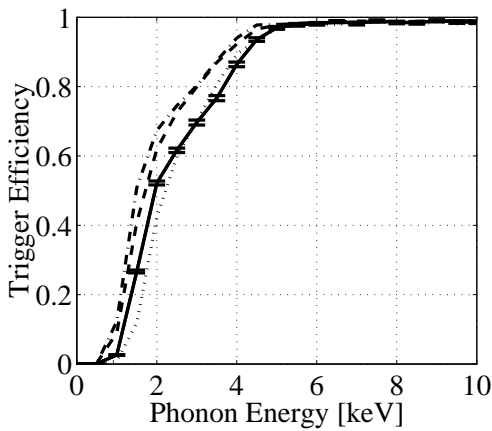,width=2.6in} 
\caption{\label{ptrigeff}
Efficiency of hardware phonon trigger vs. phonon energy 
$E_{\mathrm{P}}$, for BLIP3 
(solid), BLIP4 (dashes), BLIP5 (dot-dashes), and BLIP6 (dots).  
Statistical uncertainty (1$\sigma$), shown for BLIP3, is similar in the 
other detectors.  These results are averaged over the entire data set;
the slight residual trigger inefficiency above 5~keV is 
dominated by a four-week period with slightly worse trigger filters.
}
\end{figure}

For all
events, every detector channel is digitized and trace fits done.
The hardware-trigger efficiency for each detector can be measured using
events in which any of the other detectors was the first to trigger.  The
trigger efficiency for a given detector as a function of energy is defined
as the fraction of such events for which that detector's trigger is found
in the post-trigger history.
This analysis
is done separately for the phonon trigger as a function of phonon
energy and for the ionization trigger as a function of ionization energy.
To ensure good energy estimates, 
this calculation is done on the set of events passing all 
data-quality cuts (note data-quality cuts do not require that events 
are single scatters; see Sec.~\ref{sectCuts}).  
Figure~\ref{ptrigeff} shows the
phonon-trigger efficiency as a function of phonon energy.

\begin{figure}
\psfig{figure=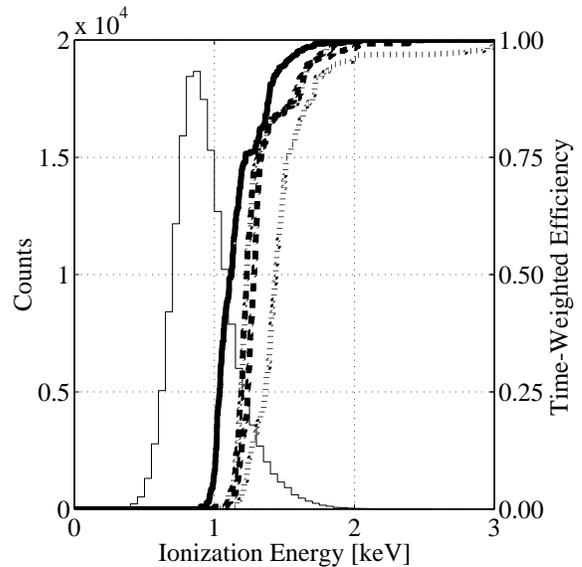,width=3.0in}
\caption{\label{slidingnoiseqsum}
Solid: Distribution of summed ionization energy in BLIP6 for random triggers
as determined by the ``sliding'' noise fit. Also shown are the 
data-averaged ionization search threshold efficiency curves for BLIP3 
(solid), BLIP4 (dashes), BLIP5 (dot-dashes), and BLIP6 (dots).}
\end{figure}

For phonon-trigger events, it must be determined whether the 
ionization signal is due to amplifier noise or to real ionization.
Because the phonon pulses have $\sim$5~ms rise times, 
for phonon-trigger events
we search for ionization pulses inside a 1.6-ms-wide time window. 
An optimal-filter algorithm picks out the largest peak in the window.
Random-trigger events are used to determine, on a day-by-day basis, the 
ionization search threshold above which the ionization is 
unlikely to be just noise.
The standard optimal-filter algorithm finds the delay and energy
for the random-trigger events. 
The resulting energy distribution is approximately
Gaussian but is offset positively from zero, is narrower than the
zero-delay noise distribution, and has a non-Gaussian tail to high
energy:  
\begin{equation}
P(E) = M \left[ {\mathrm{erf}}( E,\sigma_E) \right]^{M-1}
                       \frac{1}{\sigma_E \sqrt{2 \pi}} 
                       \exp\left( - \frac{E^2}{2 \sigma_E^2} \right) 
                       \label{slidingnoise}
\end{equation}
where $M$ is the number of samples in the search window and
$\sigma_E$ is the width of the zero-delay noise 
distribution~\cite{golwalathesis}.  
A histogram of energies yielded by the sliding noise fit for random triggers is
shown in Fig.~\ref{slidingnoiseqsum}, together with the 
data-averaged ionization search threshold efficiencies for each of the four 
detectors.  Events with no real ionization are called ``ionization-noise'' 
events.

Only events above the ionization-search threshold are included in the 
analysis because two classes of events otherwise could mimic WIMP 
events.
Muon-induced events without a clear ionization pulse cannot be vetoed 
because the slow phonon timing information is too poor to allow 
correlations with the muon veto.
Thermal events, such as detector displacement in its support,
yield phonon energy but no ionization,
and hence could also be mistaken for WIMP events were no ionization threshold 
applied.

Although the phonon-trigger efficiency is $\sim$100\% for phonon energies 
$E_{\mathrm{P}}>5$~keV, 
an analysis threshold is placed at recoil energy $E_{\mathrm{R}}=10$~keV 
for two reasons.  
First, for energies $E_{\mathrm{R}}\alt10$~keV
the efficiency for identifying nuclear
recoils decreases precipitously as energy decreases
because of the fraction of 
nuclear-recoil events producing less ionization
than the ionization-search threshold.
Below 10~keV, the uncertainty in our determination of this
efficiency would make interpretation of the number of identified 
nuclear-recoil events unreliable. 
Second, at these same energies, the expected contamination of the nuclear-recoil 
band with electron-recoil events appears to be non-negligible.   

Analysis is further restricted to events below 100~keV 
because the nuclear-recoil efficiency above 100~keV is not well
determined.  This uncertainty arises simply because 
there are so few neutron-calibration interactions above 100~keV 
that the position of
the nuclear-recoil band cannot be determined.  This restriction does not
significantly degrade the detectors' sensitivity to WIMPs or to 
background neutrons 
because both types of particles produce recoil-energy spectra that are
approximately exponential with $\langle E_{\mathrm{R}} \rangle
\alt 30$~keV.

\subsection{Software cuts}
\label{sectCuts}

To prepare the data for a search for WIMP-induced nuclear recoils, a
number of data-quality cuts are made, as described in 
Sec.~\ref{sectPretriggerCuts}--\ref{sect:chisq}.  
The goals of these cuts are to
remove pileup, to remove periods of high noise or trace-baseline wandering, and
to select only those events where the pulse fits are of sufficient
quality to ensure the accuracy of the energy estimate
and hence the ability to reject electron-recoil background events.  
Additional, ``physics'' cuts preferentially reject background events,
as described in Sec.~\ref{sect:vetocut}--\ref{sect:NRcut}.
All cuts other than the nuclear-recoil cut were set after initial 
examination of the data.
In order to minimize the potential for introducing bias,
these cuts were set without regard to the number of events passing the 
nuclear-recoil cut, as described below.
In particular, the data-quality cuts were set using a random 10\% of 
the data with no other cuts applied.  The veto-anticoincidence cut 
(see Sec.~\ref{sect:vetocut}) was set from a random 10\% of the 
data with only the data-quality cuts applied.

\subsubsection{Pretrigger-trace-quality cuts}
\label{sectPretriggerCuts}

A number of cuts are made using information not about the 
events, but only on the quality of the 
set-up prior to the event trigger.  
Periods of known poor energy resolution are discarded.  
For
the early part of this run, 
problems with the detectors'
electronics were the dominant cause of such cuts.
Detectors failing these cuts are discarded for the periods in 
question, but events in other detectors during these periods are not 
cut.  These cuts remove 5--10\% 
of the low-background data for each detector,
slightly decreasing the expected fraction of neutron-induced events 
that multiply scatter between detectors.
A detector is considered to be ``live'' for the events for which it 
passes these cuts.

Additional cuts are made on
pretrigger-trace quantities to ensure the traces are
free of pileup, the pulses are within the digitizer window, and the noise
environment is reasonable.  
First, the mean pretrigger baselines of 
all channels are
required to lie in a range so that an event of interest ($<100$~keV) 
would not saturate the digitizers.
Second, the standard deviations of the pretrigger baselines are 
required not to be too large. These cuts 
remove events with pretrigger pileup, high phonon noise, or
low-level baseline wandering that increases the baseline noise.
Any of these problems may compromise the energy measurement.
Third, the detector temperatures, as measured by the phonon-lockin DC 
reference voltages, are required to be in the range for which the 
linear ``DC-reference correction'' discussed above 
(Sec.~\ref{dcref}) is calibrated.
For an event to be accepted,
all live detectors must pass all these cuts.

The calculation of the efficiency of these combined pretrigger cuts
is straightforward
because the cuts have no dependence on the event characteristics.  
The efficiency is given simply by 
the fraction of ionization-pulser events passing the cuts (see 
Sec.~\ref{pulser}).
Furthermore, both lower and upper bounds 
on the pretrigger-cut efficiency 
may be calculated easily from the data itself.
The livetime of an event is defined as the time waiting for the trigger
after the trigger is armed.
An upper bound on the pretrigger-cut efficiency
is given by the ratio of the sum of the livetime of the events 
passing the cut set to the sum of the livetime of all events.
If the experiment were live for all the livetime preceding events 
that pass the pretrigger cuts, then this ratio would yield the cut 
efficiency.
Since the experiment may actually be dead for part of this time (\eg\ 
time recovering from  a high-energy deposition in one or more 
detectors),
this method yields an upper bound on the efficiency.
A lower bound on the pretrigger-cut efficiency is given by the 
fraction of events passing the cuts.
If the trigger rate were constant over the entire run,
then the fraction of events passing the cut would naturally yield the 
cut efficiency.
Because more triggers occur during periods when events are more 
likely to fail the pretrigger cut (\eg\ due to periods of high noise, 
which can induce triggers),
this estimate yields a lower bound on the efficiency.
Table~\ref{pretrigeff} displays the efficiencies 
together with these bounds
for the final all-detector pre-trigger 
trace-quality cuts.  

\begin{table}[ht]
\begin{ruledtabular}
\begin{tabular}{lrrrr} 
Pretrigger Cut  Efficiency  &   \multicolumn{1}{c}{BLIP3}
       &   \multicolumn{1}{c}{BLIP4}
       &   \multicolumn{1}{c}{BLIP5}
       &   \multicolumn{1}{c}{BLIP6}  \\ \hline
Fraction of data live time   & 0.79   &  0.77   &   0.82  &   0.83  \\ 
Fraction of pulser events    & 0.76   &  0.73   &   0.78  &   0.78  \\  
Fraction of data events      & 0.71   &  0.69   &   0.75  &   0.75  \\ 
\end{tabular}
\end{ruledtabular}
\caption{\label{pretrigeff}
Pretrigger-trace-quality cut efficiencies for the four detectors,
as measured by three different methods.  
The total live time before any cuts is 99.4 live-days.  
As noted in the text,
the fraction of pulser events passing pretrigger cuts accurately 
measures the efficiency,
while
the estimates based on fractions of events should be systematically low,
and the estimates based on fractions of live time should be 
systematically high.}
\end{table}

\subsubsection{Post-trigger pile-up cuts}
\label{sect:PosttriggerCuts}

Because the phonon pulses for the BLIP detectors are considerably slower 
than the ionization pulses, events with 
accidental additional hits on the $\sim$80-msec 
time scale of the 
phonon pulse could result in additional phonon energy without 
additional ionization energy on the shorter timescale of the 
ionization pulse, potentially mimicking the 
signature of nuclear recoils.
To avoid contamination by these events, 
additional care is taken to reject detectors with 
evidence of pile-up.
Events with discernible pulses in the post-trigger phonon digitization window 
(as evidenced by a second peak in the pulse larger than the triggering peak) 
are rejected.
To reject accidental pile-up with small delays ($<10$~ms)
that may not result in 
a distinguishable second phonon pulse, we also reject
detectors with additional accidental ionization 
triggers more than 50~\micros\ before or more than 300~\micros\ after 
the primary trigger (additional triggers very near the primary 
trigger may be due to double triggering in the electronics or
multiple-scattering).
Further cuts (described in Sec.~\ref{sect:chisq}) remove the 
remaining events that are contaminated with pile-up.
All these cuts remove only the detector(s) whose events are
contaminated with pile-up; events in detectors without pile-up are not 
cut.

The efficiency $\epsilon_{\mathrm{p}}$ of the pile-up cut can be
calculated directly from the trigger rate 
by assuming that the occurrence of a second event of any
energy causes an event to fail the cut.  This estimate is a good
one at low energies --- if the first event is below 100~keV, the
second event is likely to be more energetic simply because most of
the trigger rate comes from events above 100~keV.  
This efficiency $\epsilon_{\mathrm{p}}$ is
given by the accidental rate for a second event to appear in the 
10~ms 
pretrigger dead period or in the 83~ms phonon
post-trigger period, which is 
\begin{equation}
1 - \epsilon_{\mathrm{p}} = 0.093\mbox{s} \times\ R
\end{equation}
where $R$ is the measured single-detector trigger rate. 
The typical single-detector trigger rate is 0.33~Hz, so 
$\epsilon_{\mathrm{p}} \approx 0.97$.  
This result agrees well with the fractions of events that pass the cut, 
$0.96 < \epsilon_{\mathrm{p}} < 0.98$  
for the four detectors.

\subsubsection{Trace-quality cuts}
\label{sect:chisq}

\begin{figure}
\psfig{figure=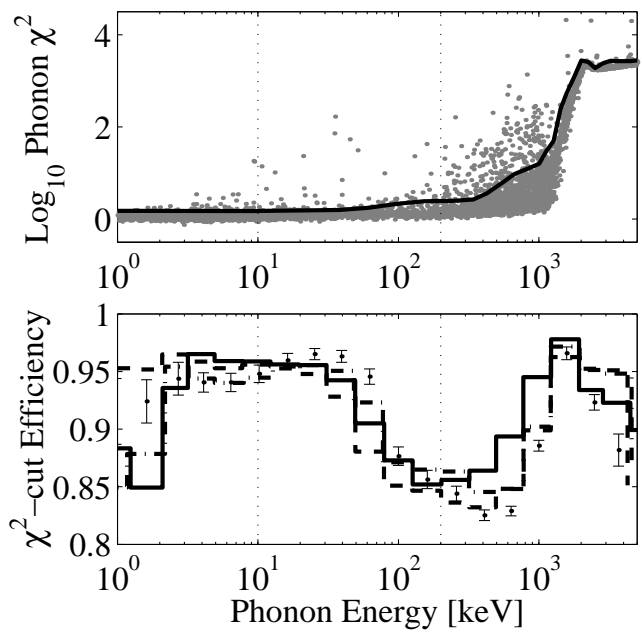,width=3.4in}
\caption{\label{pchisqexample}
Top: Typical phonon-pulse-fit \chisq\
vs. phonon energy.  The phonon \chisq\ is a reduced \chisq\ for approximately
2000 samples, but it is not properly
normalized.  The line on the plot indicates the position of the cut
calculated by the automated algorithm.
Bottom: Efficiency of phonon-\chisq\ cut vs. phonon energy for the four BLIP 
detectors.  Error bars are shown for BLIP3 data only.  Curves 
indicate data for BLIP4 (solid), BLIP5 (dashed), and BLIP6 
(dot-dashed).  For both plots, the vertical dotted lines indicate the 
approximate phonon energies corresponding to the 10--100~keV recoil-energy
analysis region.}
\end{figure}

In order to ensure rejection of all events with pile-up, and in order to 
discard pulses that may result in
misestimated energies, cuts are made on the pulse-shape \chisq\ values.
Pulse-shape templates are formed to match the shapes of low-energy 
pulses to ensure best energy resolution for such events.
At high energy, as shown in Fig.~\ref{pchisqexample}, 
pulse-shape changes result in
severe deviation of \chisq\ from its low-energy value.
The slow rise away from the
low-energy \chisq\ value is due to minor pulse-shape nonlinearity
as the energy is increased.  
The abrupt change at $\sim1$~MeV 
coincides with the beginning of digitizer saturation.  
Furthermore, the
\chisq\ distributions change on timescales of one to a few days,
as the phonon pulse shape changes
due to thermal drifts.  
An automated empirical approach is taken in
defining the phonon-\chisq\ cut as a function of energy
separately for each day of data\cite{golwalathesis}.
Figure~\ref{pchisqexample} shows a typical
cut determined by this automated technique.

The efficiency of the cut in each energy bin is estimated simply as 
the fraction of events that pass it.  
Although the cut efficiency varies over time,
the efficiency calculated from
the data set as a whole should correctly incorporate the variations.  
For example, a period with a low cut efficiency is
weighted according to the total number of events in the set before the
\chisq\ cut, which is proportional to the live time of the
period, providing the correct weighting.  The prior cuts remove
extraordinary periods, so this procedure is valid.  Furthermore, the
assumption is conservative in that it can only underestimate the efficiency.
For example, if a trigger outburst is left in the data set from which
the efficiency is calculated, then it is overweighted because it
has too many events.  The efficiency for such a period is
lower than is typical because of the higher noise.  Thus, the mean
efficiency is decreased by such a period.

The efficiency of the phonon-\chisq\ cut as a function of phonon
energy is shown in Fig.~\ref{pchisqexample}.  The efficiency has
structure that arises mainly from the fact that, at a few hundred~keV, 
the \chisq\ distribution broadens and exhibits a tail.
While the shape of the efficiency function may appear strange, it is
correct --- a more stringent cut is made at higher energy, giving a
lower efficiency.

Because the ionization \chisq\ is well behaved, a cut on ionization
\chisq\ is barely necessary.  A very liberal cut is made, accepting all events
that do not saturate the digitizers.

\begin{figure}
\psfig{figure=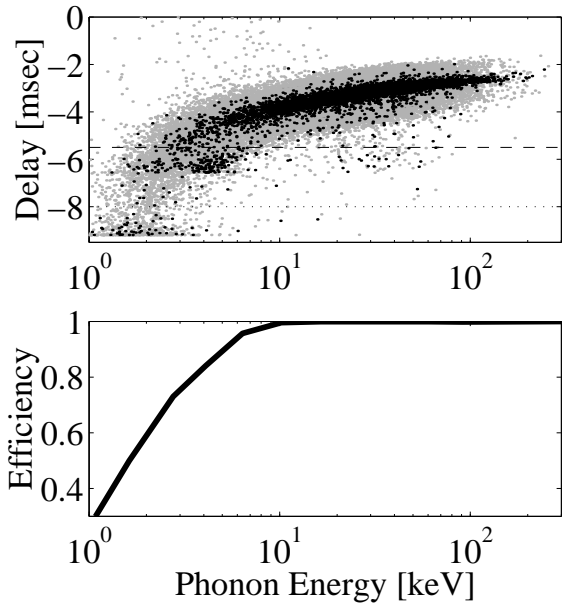,width=3.0in}
\caption{\label{qdelayvspheat}
Top: ``Ionization Delay''
vs. phonon ({\em not} 
recoil) energy for a random one tenth of the data, 
showing the timewalk of the phonon trigger.  
The ionization delay is the time of the ionization pulse
relative to trigger time, 
with negative values 
indicating the ionization pulse occurred before the trigger.
The dashed and dotted lines indicate the position of the ionization-delay cut; 
the cut at
$-5.5$~ms is used for data with 6~ms of pretrigger information and the
cut at $-8$~ms for data with 9~ms of pretrigger information. 
Dark (light) dots indicate events with ionization above (below)
the ionization-search threshold.
Bottom: Efficiency of the cut vs. phonon energy in the triggering
detector.
}
\end{figure}

An additional trace-quality cut is made because 
low-energy phonon-trigger events could in principle trigger so late 
that the ionization pulse lies before the downloaded section of the 
digitized trace.  Furthermore, for data from the first part of the run, 
the ionization-search algorithm 
was allowed to fit a pulse with falling edge at the very beginning of the digitization 
window, typically resulting in a poor energy estimation.  
Such events are rejected by cutting events with ionization-pulse 
start times too close to 
the beginning of the digitization window.
The length of the ionization pretrigger trace was
increased from about 6~ms to 9~ms midway through the data set; 
therefore, two cut values are used:
$-5.5$~ms for the 6~ms data and $-8$~ms for the 9~ms data.  These two cut
values are indicated in Fig.~\ref{qdelayvspheat}.  

As is seen in
Fig.~\ref{qdelayvspheat}, even with the cut at $-5.5$~ms, a 
significant number of ionization 
pulses should be missed only for phonon energies $E_{\mathrm{P}}< 8$~keV.
For this reason, although the efficiency of this cut is calculated, it 
has a small effect for the analysis, which considers only events with 
recoil energies $E_{\mathrm{R}}> 10$~keV.

\subsubsection{Veto-anticoincidence cut}
\label{sect:vetocut}

\begin{figure}
\psfig{figure=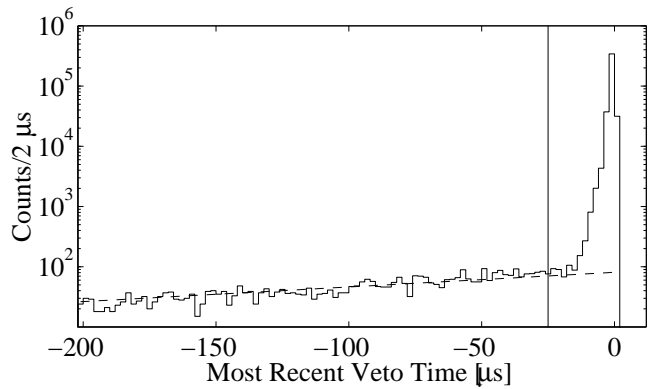,width=3.4in}
\caption{\label{vetopretime}
Distribution of last veto-trigger times for ionization-trigger
events, for a random 10\% of the data.  
The exponential background distribution has a slope
corresponding to $\tau = 150$~\micros\ (shown as dashes).  
The 25-\micros\ coincidence
window is indicated.}
\end{figure}

For dark-matter analysis, a cut is made to remove events
coincident with activity in the veto.  
Because of the high veto rate $R_{\mathrm{v}}\approx6$~kHz,
narrow veto windows in time must be used to minimize
the rate of accidental coincidences.
If an event's global trigger is an ionization trigger, 
the veto-coincidence window 
extends only before the trigger time, because an ionization trigger 
may occur only after the particle interaction that caused it.
An ionization-trigger event with any veto hits in the 25~\micros\
before the detector trigger is considered veto-coincident.
This window size was determined by choosing the point where the
distribution of last veto-trigger times deviates from the 
$\tau=150$~\micros\ background exponential 
(see Fig.~\ref{vetopretime}). This exponential is 
due to background photons emitted following
thermal-neutron capture on the polyethylene moderator.

\begin{figure}
\psfig{figure=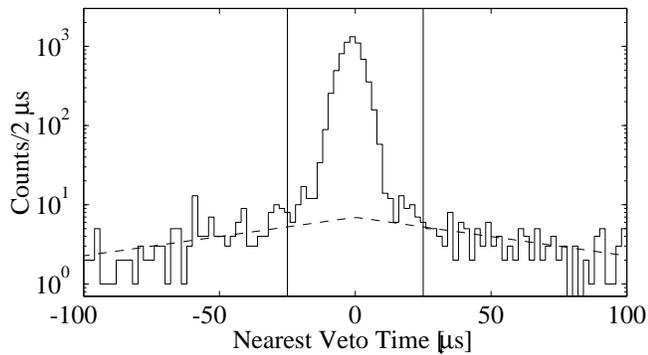,width=3.4in}
\caption{\label{vetoqneartime}
Distribution of nearest veto-trigger times for phonon-trigger
events, relative to the time of the ionization pulse,
for events above the ionization-search threshold.
The width of the peak is dominated by the uncertainty on the 
reconstructed time of the ionization pulse.
The exponential accidental distribution is shown as dashes.
The $\pm 25$-\micros\ coincidence window is indicated. 
}
\end{figure}

For an event with a phonon trigger but no ionization trigger, 
the veto-coincidence cut is different.  As described in
Sec.~\ref{sect:delay}, a search for a pulse in the ionization
trace is performed for phonon triggers.  If an ionization event is
found, its time can be compared to the veto-trigger history.  
The uncertainty on the time of the ionization pulse makes it 
necessary to search for the nearest veto hit not only before the 
inferred time of the pulse, but also after it.
The distribution of nearest veto-trigger times for phonon triggers with an
ionization pulse found is shown in Fig.~\ref{vetoqneartime}.  Based
on the points where the distribution deviates from an exponential
accidental distribution, a cut window of $\pm 25$~\micros\
is set.  
For phonon triggers without ionization, 
the uncertainty on the event time is comparable to the
average time between veto events, making vetoing useless.  Primarily 
for this reason, all events without ionization pulses are discarded.

The efficiencies of the veto-anticoincidence cuts 
are determined by the
fraction of random-triggered events that they reject averaged over the course of 
the run.
Using the random-triggered events 
accurately takes into account variations in veto rate over the course 
of the run.
The resulting efficiencies, 87\% for ionization triggers 
and 75\% for phonon triggers with ionization found,
agree with the measured 
average veto-trigger
rate $R_{\mathrm{v}}\approx6$~kHz.  
For ionization triggers, 
the probability that an
accidental coincidence occurs
is $1-\exp(-6~\mathrm{kHz}\,\times\,25~\mu\mathrm{s}) = 0.13$, 
yielding an efficiency of 0.87.  For
phonon triggers with ionization found, the window is $\pm 25$~\micros,
giving an efficiency of 0.75.

\subsubsection{Removal of thermistor-contained events}

\begin{figure}
\psfig{figure=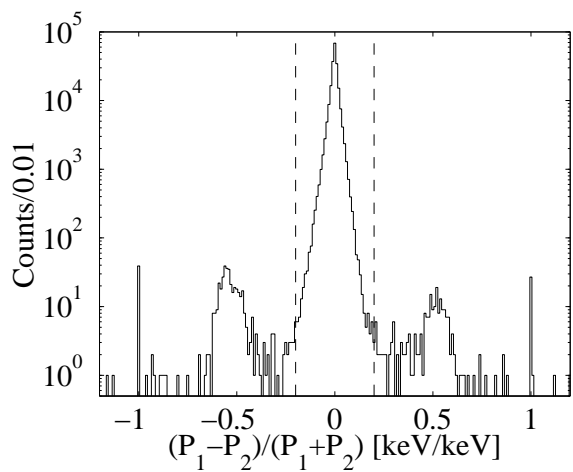,width=3.0in}
\caption{\label{pparthist}
Histogram of phonon partition.  The dashed lines indicate the acceptance
region; events failing the cut are dominated by interactions in the 
NTD thermistors. }
\end{figure}

Particle interactions may occur in the thermistors themselves, 
resulting in little or no ionization energy.  The
resulting phonon pulses in the two thermistors are very different from
crystal-interaction pulses.
When fitted with a
standard pulse template, such events result in extremely different pulse
heights $P_{1}$ and $P_{2}$ for the two thermistors. 
To reject detectors with interactions in one or the other thermistor,
a cut rejects detectors with events for which
$\left|{(P_{1} - P_{2})/(P_{1} + P_{2})}\right|>0.2$.
As shown in Fig.~\ref{pparthist}, this 
cut results in a negligible loss of efficiency for 
events in the crystal.

\subsubsection{Removal of BLIP3}

\begin{figure}
\psfig{figure=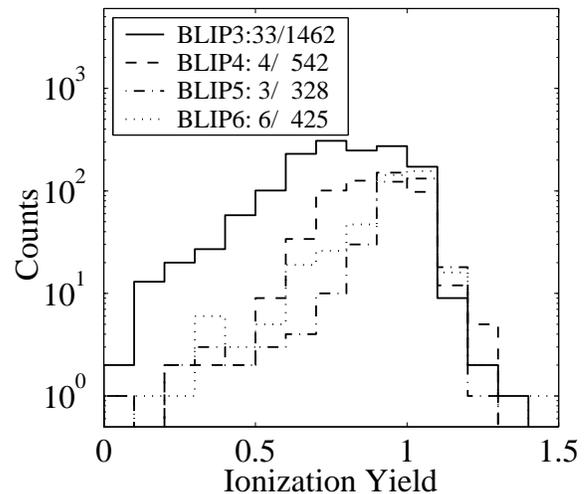,width=3.0in}
\caption{\label{qiantisingleyhist}
Distributions of ionization yield $Y$ for
veto-anticoincident 
single-scatter events with recoil energies between 10--100~keV,
fully contained in the inner electrode of
BLIP3 (solid), BLIP4 (dashed), BLIP5 (dot-dashed), or BLIP6 
(dotted).
BLIP3's high event rate, particularly for yields slightly too high to 
be nuclear recoils ($Y\approx 0.5$), indicates its 
contamination by a source of low-energy electrons.  
Although BLIP4 shows a high rate of events with 
$Y\approx 0.8$, its rate just above the nuclear-recoil 
acceptance region is similar to that of BLIP5 and BLIP6.
The legend lists the number of events that fall in the nuclear-recoil 
acceptance region 
for each detector as a fraction of the total number of events in that 
detector.}
\end{figure}

The rate of low-ionization-yield events in BLIP3,
the top detector of the 4-detector stack,
is significantly
higher than the rates in the other detectors
(230~kg$^{-1}$~d$^{-1}$ as compared to 50~kg$^{-1}$~d$^{-1}$ for the
other detectors).
BLIP3 was the prototype detector for these four BLIPs; 
it suffered repeated processing steps during
development of a new electrode-fabrication method~\cite{golwalathesis}, 
so its electrodes may have been damaged during processing. 
Moreover, exposure to an external \c14 source recently found to be 
leaky appears to have contaminated BLIP3's surface with \c14.
For this reason, BLIP3 is discarded for dark-matter analysis.
BLIP4 also shows an elevated rate of low-yield events contained in 
the inner electrode,
likely due to electrons emitted by the \c14
contaminant on BLIP3.  
As shown in Fig.~\ref{qiantisingleyhist},
there is good separation between BLIP4's low-yield
band and the nuclear-recoil-acceptance region.
Because of this good separation, BLIP4 is included in the experiment's 
fiducial volume along with BLIP5 and BLIP6.

\subsubsection{Fiducial-volume cut}

As described in Sec.~\ref{blip},
the detectors have 
radially segmented electrodes to allow rejection of events 
due to particles incident on 
the sides of the detectors, which are less shielded.
The two electrodes result in three categories of events. 
``Inner-electrode-contained'' events have inner-electrode signal $>4\sigma$ 
above the noise mean
and have outer-electrode signal within $\pm 2\sigma$ of the noise mean.
The strict requirement on the inner-electrode signal ensures that
events are not classified as inner-electrode-contained due to noise 
fluctuations.
``Outer-electrode-contained'' events have inner-electrode signal $<4\sigma$ 
above the noise mean and outer-electrode signal $>2\sigma$ above the noise mean.
Finally, ``shared-electrode'' events have inner-electrode signal $>4\sigma$ 
above the noise mean, 
and have outer-electrode signal $>2\sigma$ above the noise mean.
The shared-electrode events arise either due to interactions in the 
physical volume near the break between the inner and outer electrodes,
or due to multiple scatters under each electrode.
Here, the noise mean and standard deviation are given by the
noise parameters calculated from random-trigger events on a day-by-day basis.

The fraction of the detector volume accepted by the
three volume cuts is determined using
the relative numbers of calibration neutrons passing each cut 
at high energy, where thresholds have a reduced effect.  
The fractions averaged over 20--100~keV are 
47\%, 22\%, and 31\% (with $\pm2$\% statistical uncertainty) for 
the inner-electrode, shared-electrode, and outer-electrode volumes respectively.

Two straightforward corrections must be made.  First, according to 
Monte Carlo simulation of the neutron calibration data, 
9\% of neutrons yielding 20--100~keV recoil energy 
scatter once under each electrode of a given
detector, yielding a shared event.  
Second, the simulation shows that the
probability of a neutron interacting in the outer electrode is 14\% higher
than expected from the volume fraction, simply due to 
self-shielding~\cite{TAPthesis}
(WIMPs of course interact too weakly to show a shielding effect
or to multiple scatter).
The results for the inner-, shared-, and outer-electrode fractions 
are therefore 46\%, 19\%, and 35\%.  
The inner electrode nominally contains 56\% of the detector
volume, so these numbers are consistent with the shared volume being
geometrically equally divided between the inner and outer electrodes,
as expected.  Systematic uncertainty on the fiducial-volume fractions, 
due to possible inaccuracies in the Monte Carlo simulation, is 
estimated at 3\%~\cite{TAPthesis}. 
At low energies, the importance of thresholds makes the calculated 
fiducial volume more dependent on how ionization is shared between the 
two electrodes for events in the shared volume.
For this reason, at low energies the uncertainty on the efficiencies 
of the fiducial-volume cuts is $\sim10$\%.

\begin{figure}
\psfig{figure=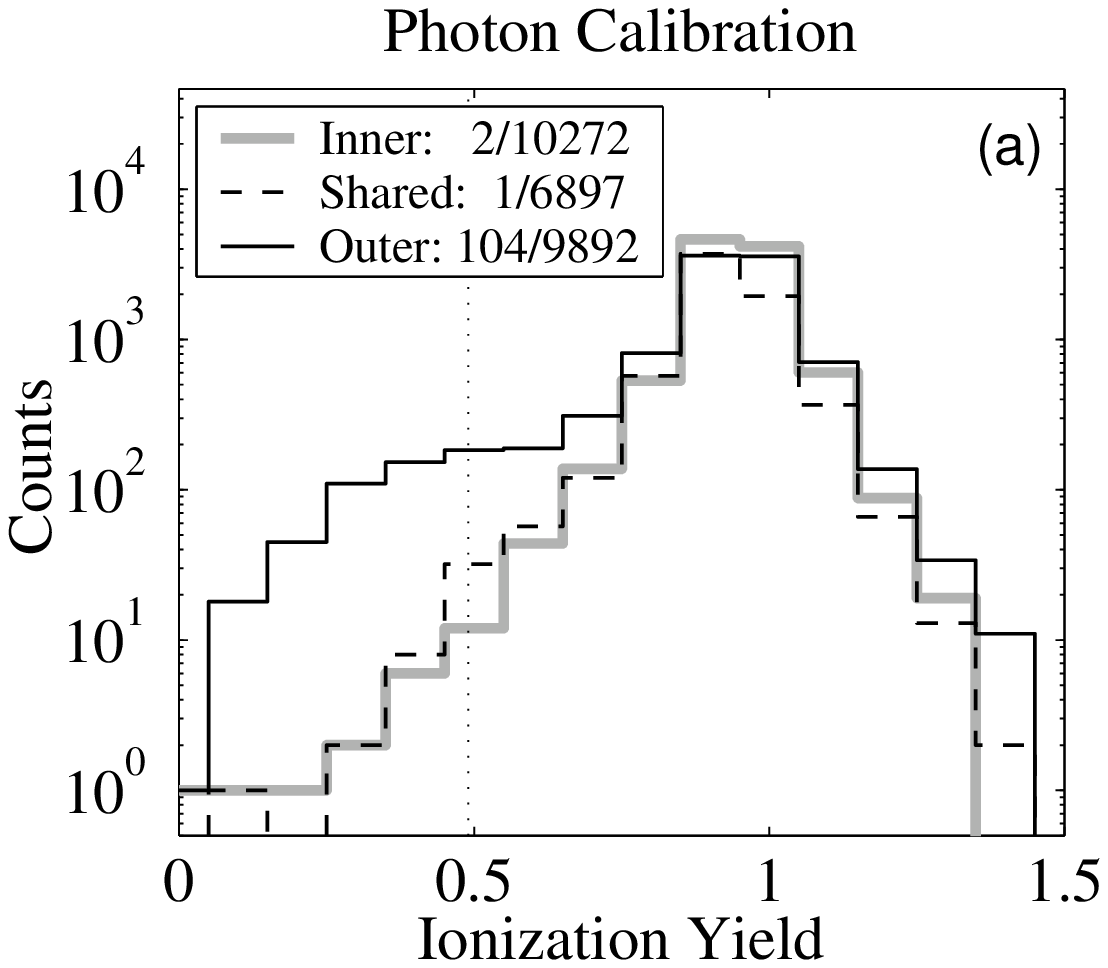,width=2.3in}\\
\psfig{figure=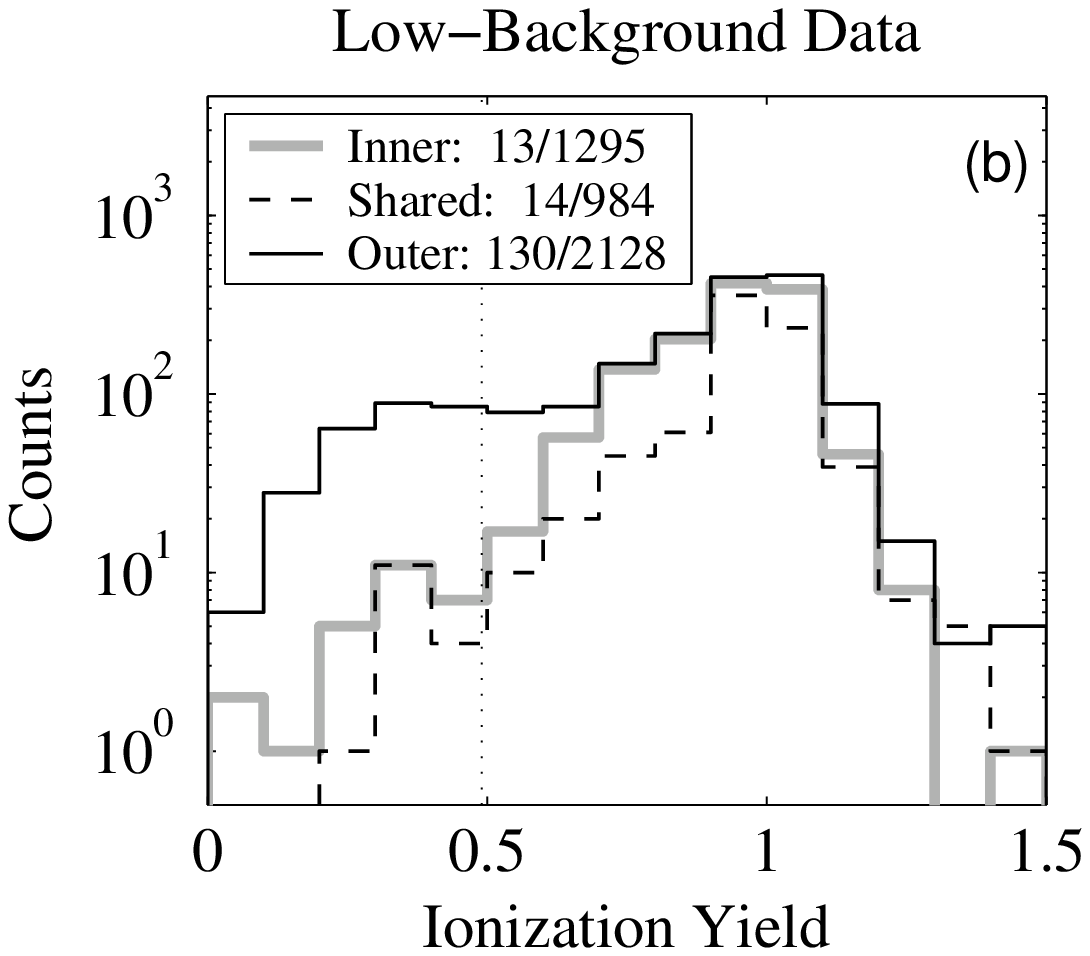,width=2.3in}
\caption{\label{ygammacalhist}
Histograms of ionization yield $Y$ 
for interactions with 10~keV$<E_{\mathrm{R}}<100$~keV in BLIP 4, 5 or 6
in (a) photon-calibration data  and  
(b) veto-anticoincident low-background data.  
The vertical lines indicate the maximum position of
the nuclear-recoil-acceptance region for any energy or detector.
The legend gives the number of events
in the nuclear-recoil-acceptance region as a fraction of
the total number of events; the former number is
determined using the fully energy-dependent acceptance region, not
just the line shown in the plots.
The high fraction of outer-electrode photon-calibration events
in the nuclear-recoil acceptance region, together with the high 
fraction of low-background 
events with yields slightly too high to 
be nuclear recoils ($Y\approx 0.5$), indicates the outer
electrode's poor discrimination against electron
contamination. 
Four (27) of the shared-electrode (outer-electrode) 
events in the nuclear-recoil acceptance region,
and 191 (310) of the events overall, occurred during the 4-V-bias
section of the data.}
\end{figure}

Calibration and low-background data are used 
in order to determine whether events in the outer electrode and
events shared between the two electrodes should be 
rejected.
Histograms
of ionization yield, shown in Fig.~\ref{ygammacalhist}, 
suggest that the outer-electrode events should be discarded. 
The photon calibration
indicates that the photon misidentification is $\sim$50~times 
higher for outer-electrode than for inner-electrode or shared events.
Beyond this, the much flatter
$Y$ distributions for the outer-electrode data indicate that, though
the outer-electrode electron rate is not significantly different from the
rates seen for the inner-electrode and shared cuts, the
electron-misidentification fraction is likely to be much worse.  

There appears to be no reason to discard the shared-electrode data 
from most of the run.
As shown in Sec.~\ref{bgnd}, 
the shared-electrode electron- and photon-background rates 
are not significantly higher than for the inner-electrode
data set.  The photon-calibration data set indicates that 
the photon- and electron-misidentification fractions for the shared region 
are
no worse than for the inner-electrode region.  
The $Y$ histograms
for the background data corroborate this point.  Because both the rates
and the misidentification fractions of photons and electrons are 
not too different for the two regions, the expected
rate of misidentified photons and electrons in the two regions should
be about the same.  

However, for a short part of the run, the charge electrodes were 
biased at 4~V, as opposed to the 6~V bias used for the rest of the 
run and for all the calibration data.
As shown in Sec.~\ref{effcheck} below, 
veto-coincident data indicate the possibility of worse contamination 
for the 4~V shared-electrode data than for the 6~V shared-electrode 
data.
For this reason, the 4~V shared-electrode data are discarded.

The original WIMP-search analysis of this data used only events 
with at least one detector hit 
fully contained in the inner electrode~\cite{r19prl}.  
For the current analysis, we include all events with any ionization 
energy in an inner electrode (both ``inner-electrode-contained'' and 
``shared'' events), excepting the 4~V shared-electrode data.  
We will call these events ``QIS'' events.
We will also show how the results 
would change if we enforced the stricter requirement that all events 
be ``QI'' events, fully contained in the inner electrode. 
We will use ``QS'' as a shorthand for the ``shared'' events.

\subsubsection{Nuclear-recoil cut}
\label{sect:NRcut}

To determine the position of the nuclear-recoil-acceptance region in
ionization yield as a function of recoil energy, two neutron calibrations
were performed during the 1999 run: one in April, approximately midway
through the run, and a second in September, at the end of the run. 

\begin{figure}
\psfig{figure=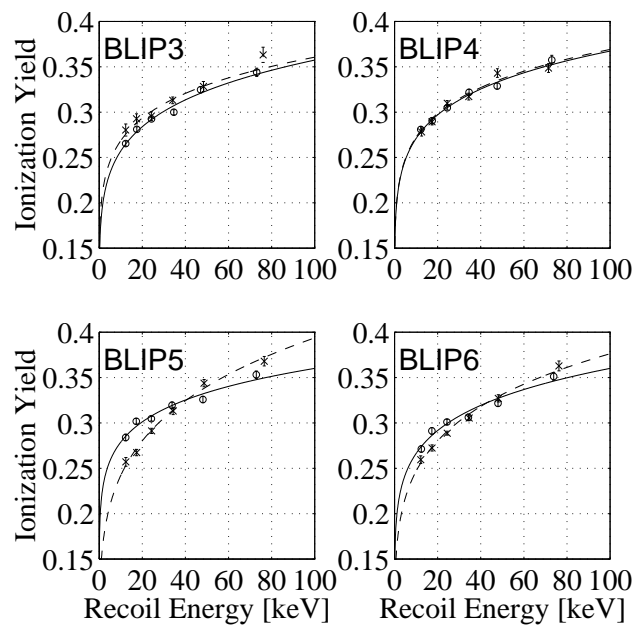,width=3.3in}
\caption{\label{ysumneutcalfits}
Nuclear-recoil-line data points
and fits for the April (circles and solid curves) and September 
($\times$'s 
and dashed curves) neutron calibrations.
For BLIP5 in particular, the two nuclear-recoil lines are clearly 
shifted relatively to each other.}
\end{figure}

The timing of the first neutron calibration was fortunate, as it 
occurred on April~2, one day before a 
Stanford-wide power outage that damaged the electronics chain,
introducing a nonlinearity in the ionization-energy response.  
An empirical linearization corrects the nonlinearity
using the well-defined
band of bulk electron recoils provided by 
the single-scatter veto-coincident photon data~\cite{golwalathesis}. 
In spite of this linearization, the nuclear-recoil acceptance region
shifts between the pre- and post-April 3 data sets.
This shift is apparent in both the veto-coincident-neutron data and
the second neutron calibration.  To account for this shift, the nuclear-recoil
band is defined separately for data before and after the power outage, 
based on the two neutron calibrations.  
Figure~\ref{ysumneutcalfits} shows the power-law functions
$Y_{\mathrm{NR}}= c E_{\mathrm{R}}^{d}$
that best fit the center of the nuclear-recoil band for the two neutron 
calibrations.
The observed one-standard-deviation width $\sigma_{\mathrm{NR}}$
of the nuclear-recoil band is also parameterized as 
a function of recoil energy:
$\sigma_{\mathrm{NR}} = a E_{\mathrm{R}}+b$.
Gaussian distributions described by these parameters
provide excellent fits to the distributions in $Y$ 
of the neutron-calibration events.

A nominal 90\% acceptance band (chosen before data-taking began) is given by a
region that extends from $Y_{\mathrm{max}}$ 1.28$\sigma_{\mathrm{NR}}$ above 
to $Y_{\mathrm{min}}$ 3$\sigma_{\mathrm{NR}}$ below the fit $Y_{\mathrm{NR}}$.  
For recoil energies below $\sim10$~keV, 
the band is truncated from
below at the ionization yield 
$Y_{\mathrm{min}} = Q_{\mathrm{min}}/E_{\mathrm{R}}$ corresponding to the
ionization-search threshold $Q_{\mathrm{min}}$. 
The nuclear-recoil efficiency $\epsilon_{\mathrm{NR}}$
may therefore be calculated for any 
recoil energy $E_{\mathrm{R}}$:
\begin{equation}
\epsilon_{\mathrm{NR}} = \int_{Y_{\mathrm{min}}(E_{\mathrm{R}})}
                            ^{Y_{\mathrm{max}}(E_{\mathrm{R}})}
                            \frac{1}{\sigma_{\mathrm{NR}}{\sqrt{2\pi}}}
                        \exp{\left[ 
                        \frac{\left(y-Y_{\mathrm{NR}} \right)^{2}}
                             {2\sigma_{\mathrm{NR}}^{2}}
                             \right]}
                             dy \; .
\end{equation}                            

It is also possible to calculate the nuclear-recoil efficiency
empirically.  
A wide ``cleaning cut'' encloses the neutron band
and excludes events that are clearly not neutrons.  
This cut results in a sample dominated by neutrons, 
except at low energies, where it also accepts ionization-noise events.
Not all ionization-noise events are neutrons, so the ``raw''
number of nuclear recoils is overestimated and the efficiency
underestimated
at energies where ionization-noise events may fall in 
the nuclear-recoil acceptance region
($< 10$~keV).  
The data are binned in recoil energy, and the
fraction of events accepted in each recoil-energy bin is calculated.
The empirical efficiency matches the 
nominal efficiency well at high energies where it should; 88\% 
of events passing the cleaning cut fall within the nominal 90\% 
acceptance region.  The small difference between the empirical 
efficiency and the nominal one gives an estimate of the 
systematic error on this efficiency.

In order to calculate the efficiency of the nuclear-recoil cut
for the low-background data, changes in ionization noise with time 
(which dominate changes in phonon noise)
must be taken into consideration.  An increase in ionization noise 
results in a higher ionization-search threshold, effectively reducing 
the nuclear-recoil cut efficiency at low energies where the threshold 
cuts into the nuclear-recoil acceptance region.  More significantly, 
higher ionization noise makes nuclear recoils at all energies more likely 
to spill out of the nuclear-recoil acceptance region.  For the 
beginning of the run, when ionization noise was worst, this latter effect 
reduces the efficiency by $\sim20$\%.  Both effects are included when 
calculating the expected nuclear-recoil cut efficiency on a day-by-day 
basis.  Also taken into account is 
the fact that data for part 
of the run was taken with 4-volt ionization bias, while most of the 
data used a 6-volt bias, for which ionization noise is more 
significant.

\subsubsection{\label{multeff}Combining efficiencies} 

For single-scattering events (such as those caused by WIMPs), 
combining the above efficiencies to determine the overall efficiency 
is straightforward.  The time variation of efficiencies other than 
the nuclear-recoil efficiency is 
generally small and does not appear correlated with the variation of other 
efficiencies.
Therefore, the product of the individual efficiencies yields 
the total efficiency for each detector.  
The systematic error due to making the assumption that efficiencies 
are uncorrelated in time should be $<5$\%.
For multiple-scattering 
events, however, care must be taken because some cut efficiencies for 
different detectors are correlated for individual events.
The \chisq-cut efficiency exhibits no correlations
because its energy dependence is dominated by the individual
detector noise and pulse-shape characteristics.  
The
nuclear-recoil-cut efficiencies are also uncorrelated, aside from
correlations introduced by real physics; \eg, multiple scattering of a
neutron.  
The energy-independent
data-quality-cut efficiencies, however, are correlated.  
An example case of how data-quality cuts introduce
correlations is post-trigger pileup.  When a detector has post-trigger
pileup, its neighbor has a higher-than-random chance of also having
post-trigger pileup because the neighbor may be hit by the same particle
or by particles produced by the same incident muon or high-energy
photon.  Therefore, it is necessary to calculate a matrix of the
joint data-quality-cut efficiencies,
with the probabilities of detectors passing cuts 
depending on the number of detectors that triggered.
These efficiencies are calculated directly from the data.

\subsubsection{\label{effcheck}Checks of cut efficiencies} 

\begin{figure}
\psfig{figure=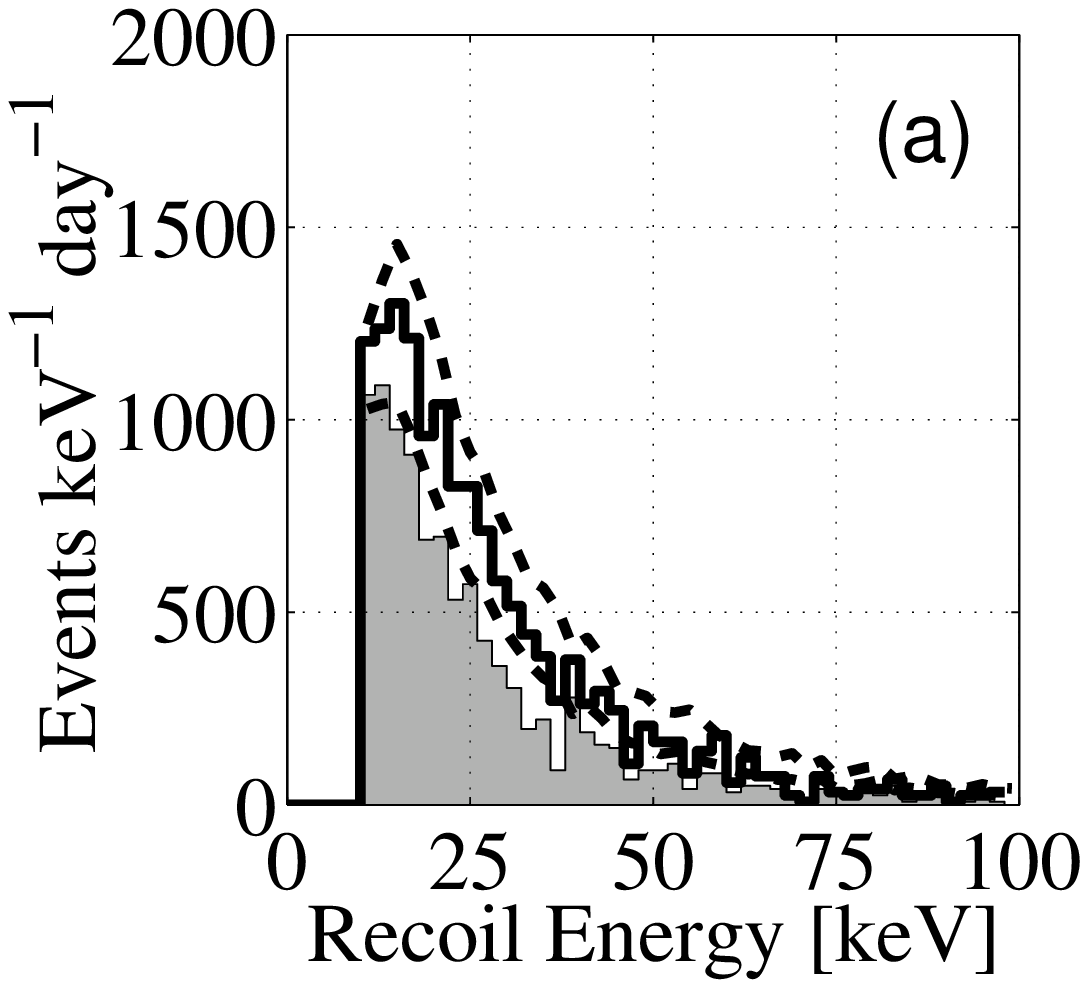,width=1.65in}
\psfig{figure=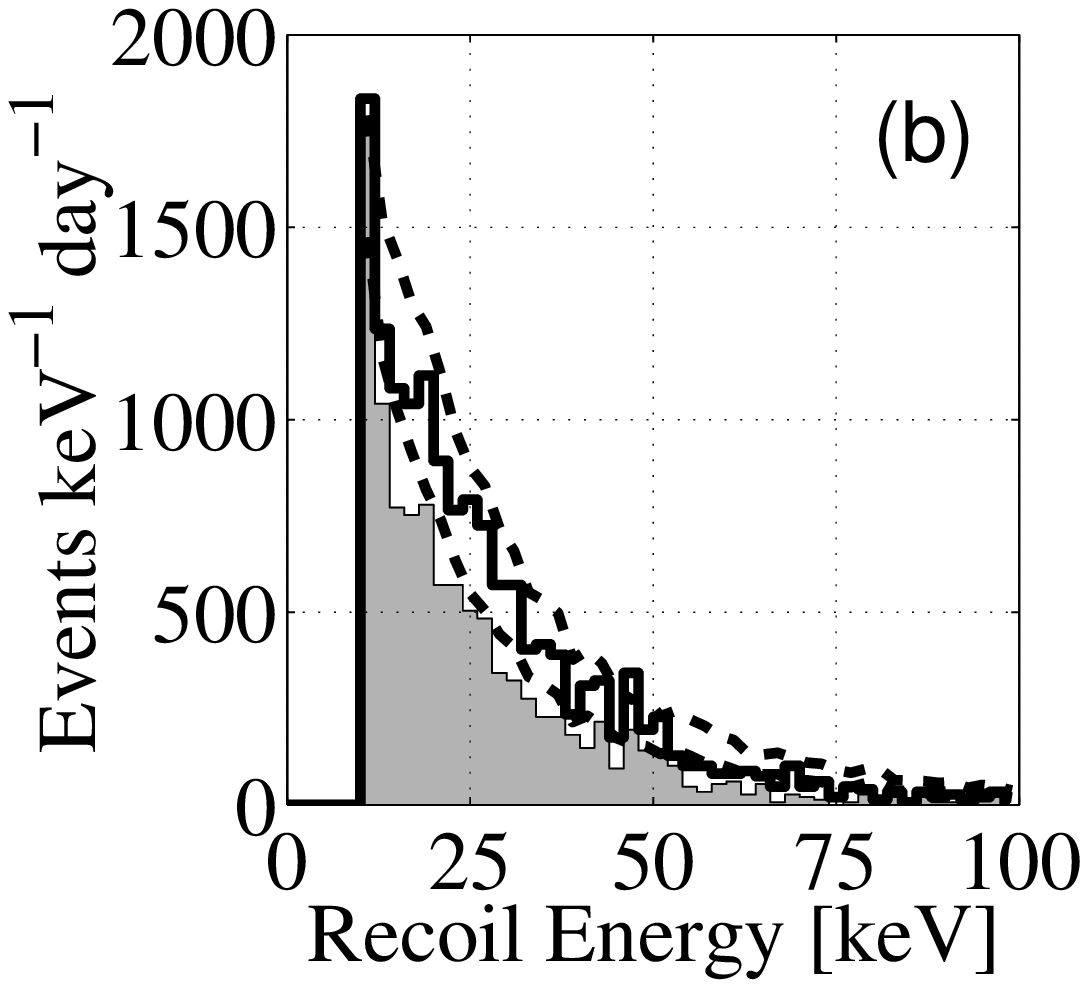,width=1.65in}
\caption{\label{cf252data}
Observed and simulated recoil-energy spectra, coadded over all four detectors, 
with no free parameters, for (a) the first neutron calibration, 
and (b) the second neutron calibration.
Solid lines: observed spectra.  Dashed lines:
simulated, with efficiency corrections applied.  The upper spectra are for
all QIS 
nuclear recoils, while the lower, shaded spectra are for all QI
nuclear recoils.
These same curves, on a logarithmic scale, are shown below in 
Fig.~\ref{cf252multdata}.}
\end{figure}

The absolute accuracy of the efficiency calculation can be checked
using the neutron calibration.
Such a check relies on the accuracy of the neutron Monte 
Carlo simulation; insofar as the simulation may be  
less accurate than the calculated efficiencies, this comparison yields only 
a rough upper limit on the systematic error of the efficiencies.
The observed and simulated spectra for the two neutron-calibration
data sets are shown in Fig.~\ref{cf252data}. 
There are no free parameters in the comparison; the
simulation normalization is set by the source activity and the
efficiencies calculated from the data.  For both calibrations, the
simulated spectra are about 10\% high at low energies, 
and are about 50\% high at high energies. 
Moreover,
although the low-energy cut efficiencies for the two calibrations are 
significantly different, both spectra are
reproduced by the simulation with similar relative errors
after application of the cut efficiencies.  
For both calibrations, the fraction of events classified as QI is 
underestimated at low energy, 
owing to the 
conservative model that describes how ionization is shared between 
the two electrodes.

\begin{figure}
\psfig{figure=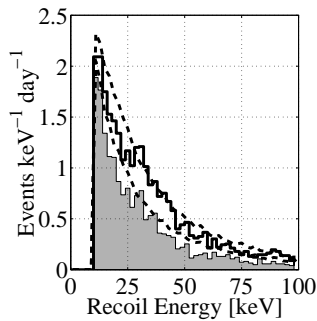,width=1.65in}
\caption{\label{mucoincndata}
Muon-coincident-neutron recoil-energy spectra, 
coadded over BLIPs 4--6, for the entire run, with no free parameters.  
Solid: observed spectra.  Dashed:
simulated.  The upper spectrum is for
QIS nuclear recoils, 
while the lower, shaded spectrum is for QI nuclear recoils.
These same curves, on a logarithmic scale, are shown below in 
Fig.~\ref{cf252multdata}.
}
\end{figure}

The accuracy of the nuclear-recoil efficiency can also be checked by
comparing the simulated and observed spectra for muon-coincident
neutrons.  As discussed in Sec.~\ref{expBackgrounds}, these
neutrons are produced by muons that interact
in the copper cans of the cryostat or in the internal lead shield
after passing through the veto.  This
data set offers the advantage that it is acquired at the same time as
the WIMP-search data set, and thus the efficiencies are exactly the
same, with the exception that no veto-anticoincidence cut is applied.
Figure~\ref{mucoincndata} shows the simulated and observed
muon-coincident-neutron spectra for the same energy cuts and event
categories as shown for the neutron-calibration data.  
Similar to the neutron-calibration data, predicted spectra are 
slightly harder 
than observed spectra, 
with simulated spectra about 10\% high at low energies, 
and about 40\% high at high energies, presumably dominated by 
inaccuracies in the Monte Carlo simulations. 

\begin{figure}
\psfig{figure=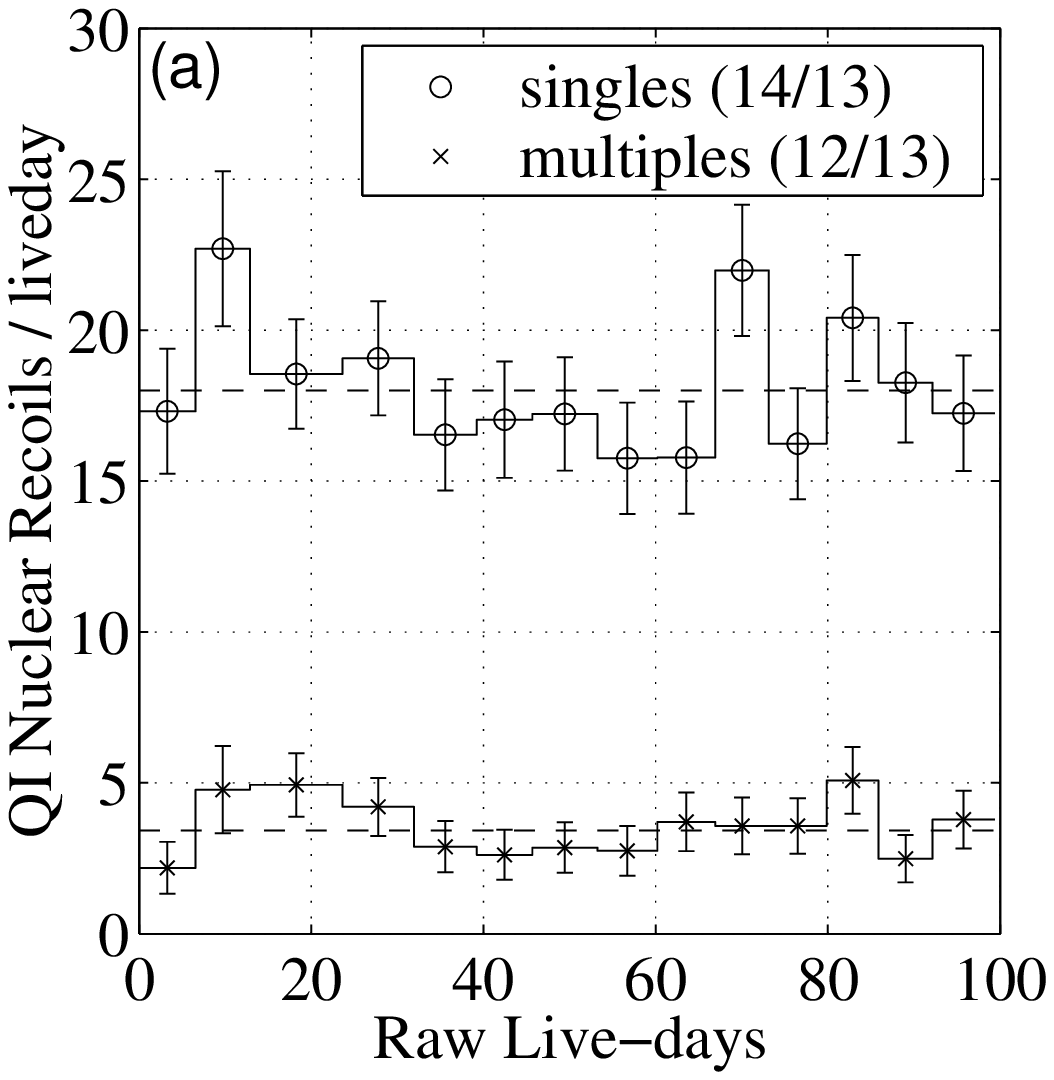,width=1.65in}
\psfig{figure=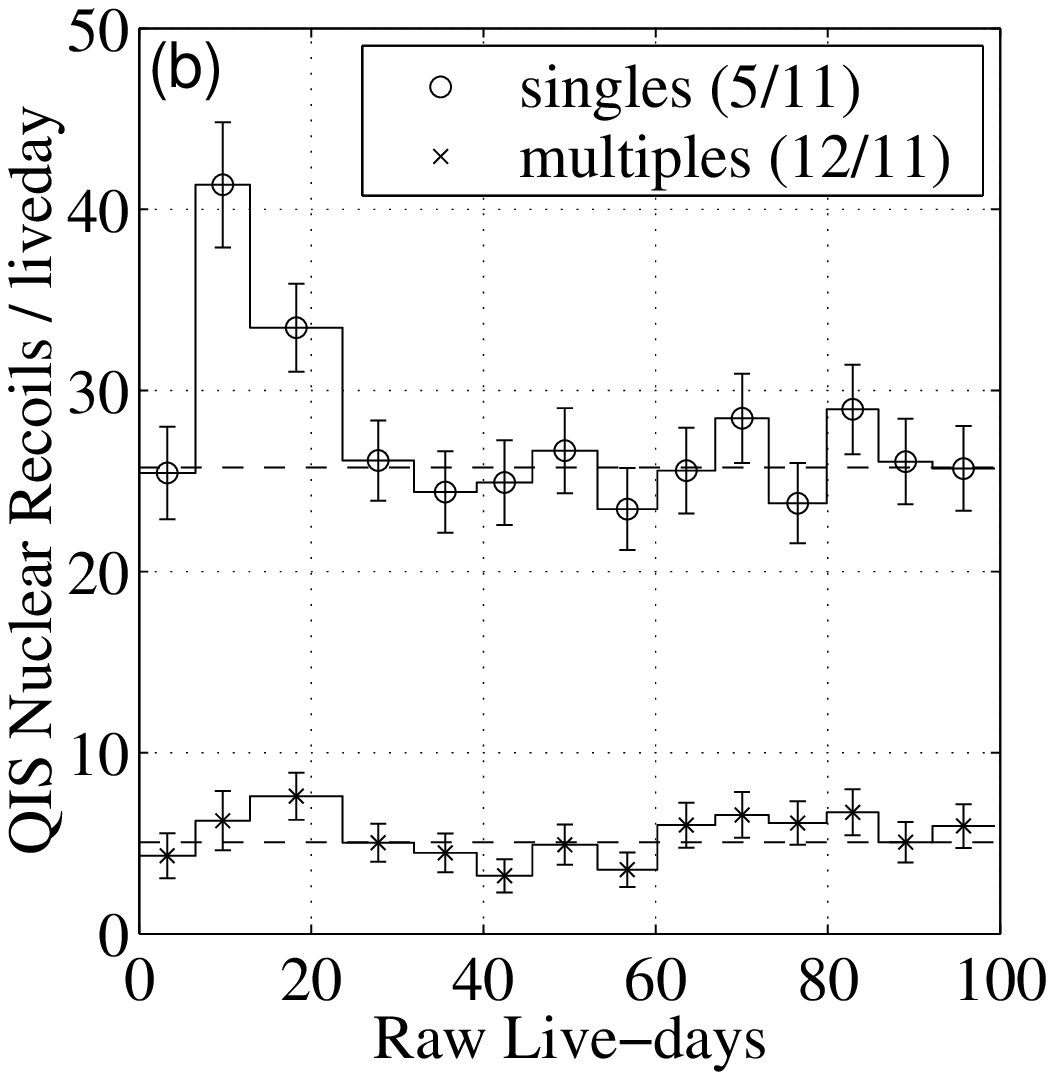,width=1.65in}
\caption{\label{mucoincnratesum}
Rates of muon-coincident single-scatter (upper data) and multiple-scatter 
(lower data) neutron candidates 
vs. time, coadded over BLIPs~4--6, for recoil energies between 
10--100~keV.  
Each bin corresponds to approximately 5~live-days.  
Statistical uncertainties are shown as error bars.  
The \chisq\ and degrees of freedom of the data relative
to the mean (dashes) calculated from the data are shown
as a fraction in the legend.
(a) Events with at least one hit fully contained in the inner electrode.  
(b) Events with at least one hit with any energy in 
the inner electrode (QIS events).  
The increased number of veto-coincident shared-electrode events passing 
the nuclear-recoil 
cut during data taken with 4-volt 
ionization bias (second and third bins) is consistent with other evidence leading 
to the discarding of the 4-volt shared-electrode data set from 
dark-matter analysis.
}
\end{figure}

The stability of the nuclear-recoil acceptance over time is checked by
Fig.~\ref{mucoincnratesum}, which shows the rates of muon-coincident
nuclear-recoil candidates, coadded over the three good detectors, 
as a function of time in blocks of
approximately 5 live-days.  
The rate of shared-electrode candidates is much higher for the data 
at 4-volt ionization bias, which corresponds to the second and third 
bins in the plot.  This evidence of likely contamination for the 
4-volt data, combined with further evidence of worse contamination in 
detector BLIP3 and in the outer-electrode data during this time 
period, leads us to discard the 4-volt shared-electrode data from 
dark-matter analysis.
The rates of the single-scatter (multiple-scatter) candidates
are otherwise stable to 10\% (20\%), consistent
with statistical fluctuations.
In particular, the rates show no statistically significant
change at either the April 3 power outage or the refrigerator
warmup/cooldown cycle in June; these events occurred at roughly 29 and 65 raw
live-days, respectively. 

Overall, the checks of the various cut efficiencies suggest that 
the efficiencies are accurate and stable at about the 10\% level.
Such accuracy is more 
than sufficient because the statistical uncertainties are 
considerably larger.

\subsection{Low-background data}
\label{bgnd}

\begin{figure}
Veto-Coincident Inner-Electrode-Contained\\
\psfig{figure=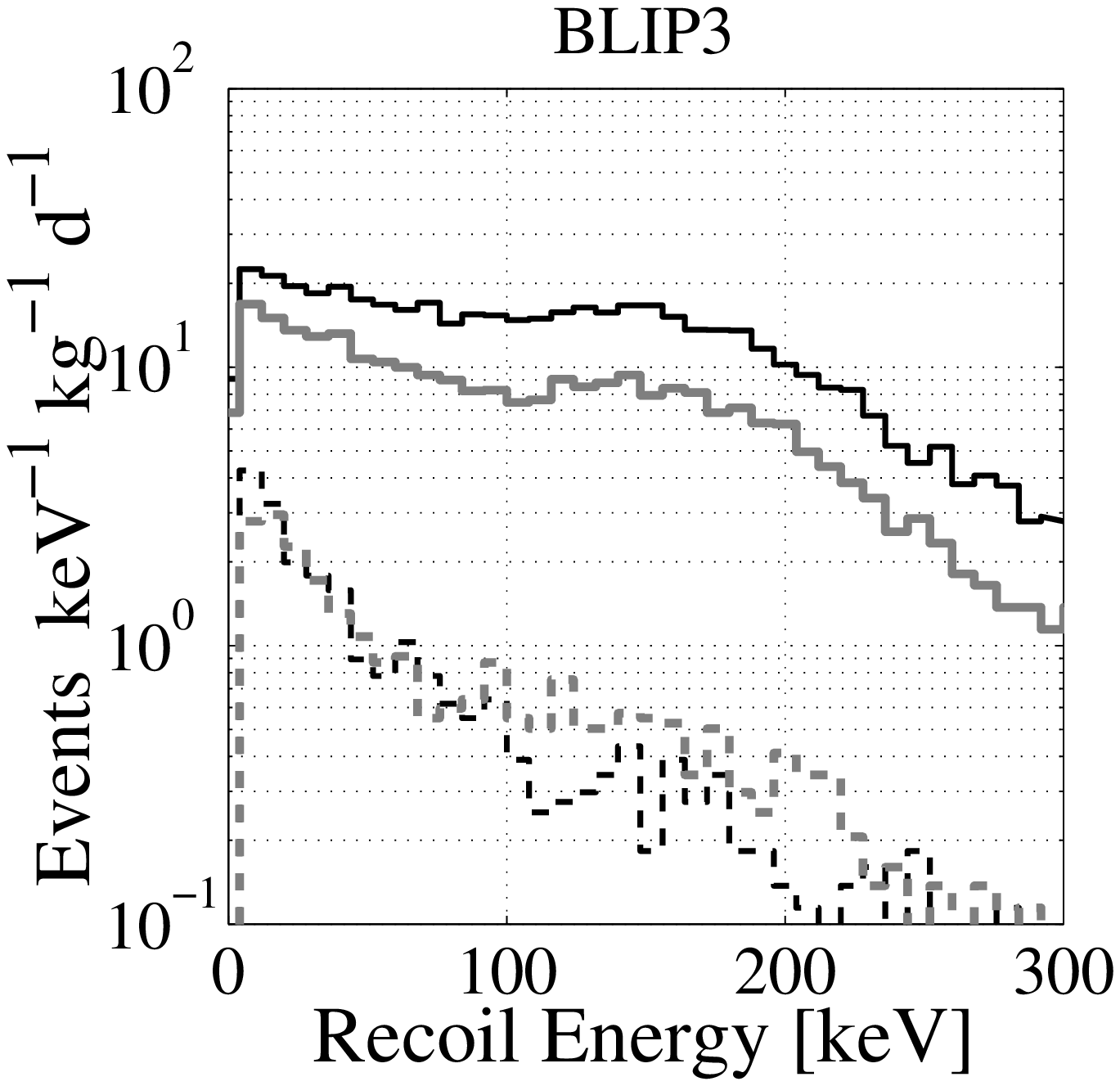,width=1.65in}
\psfig{figure=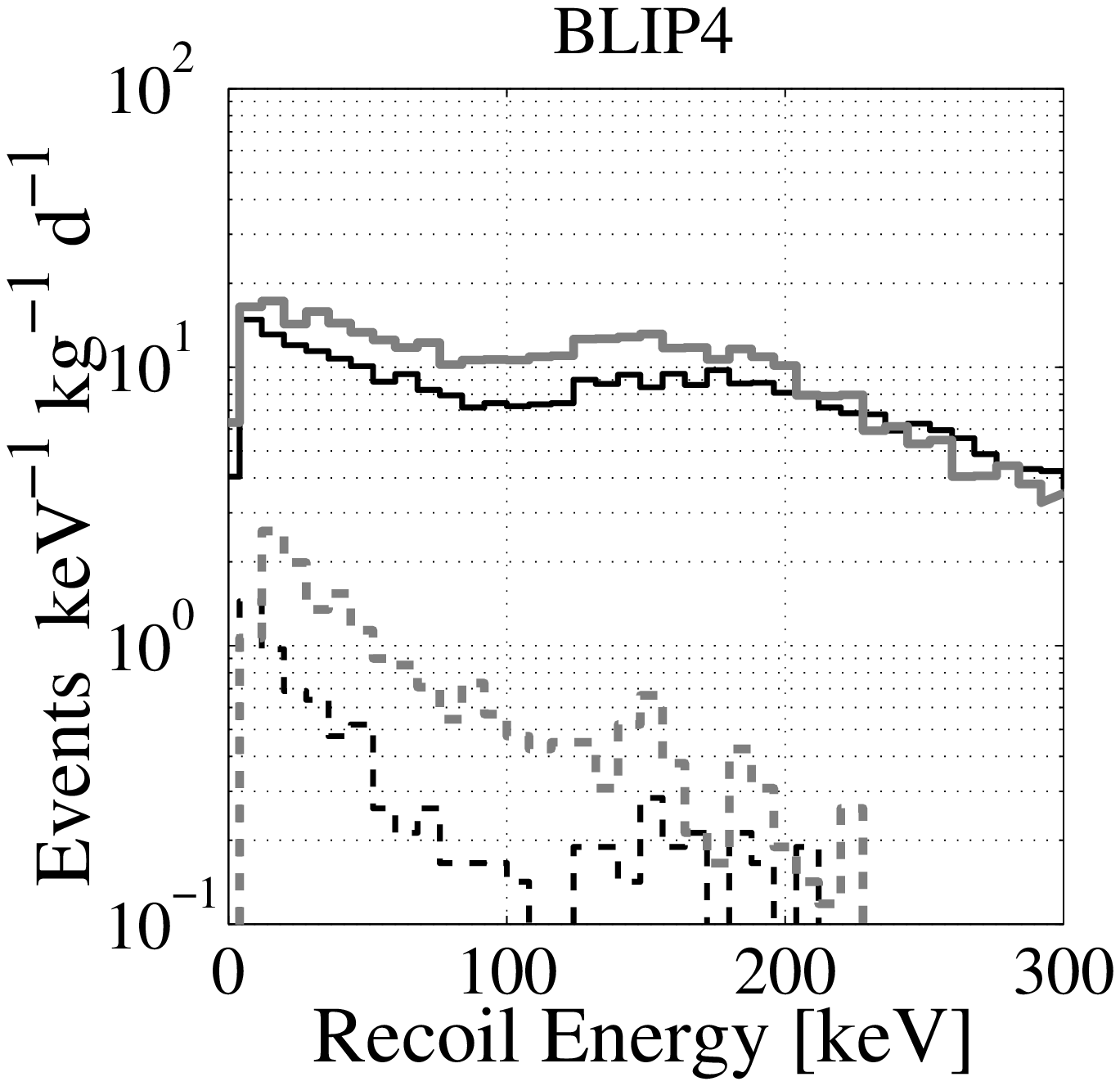,width=1.65in}
\psfig{figure=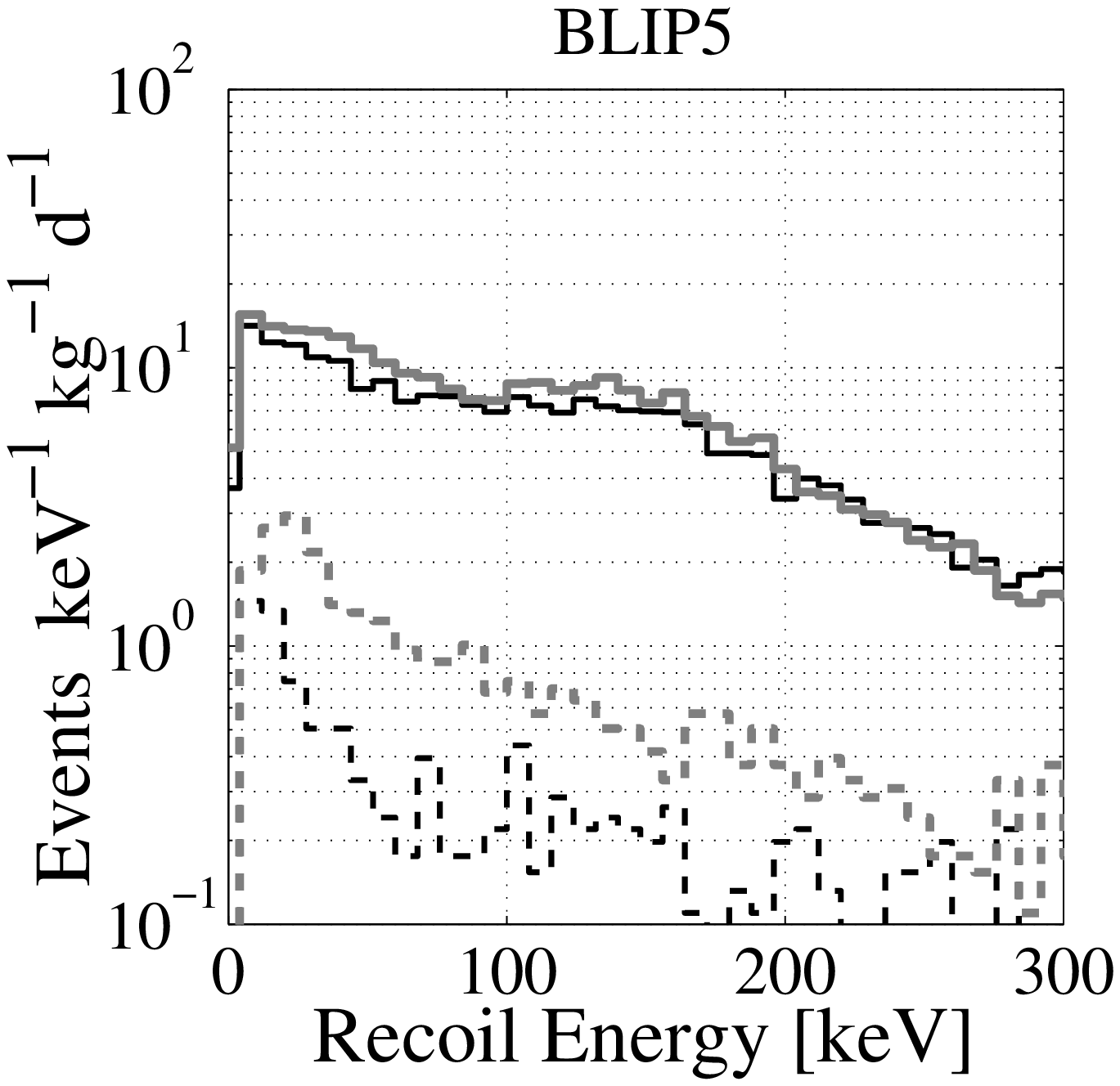,width=1.65in}
\psfig{figure=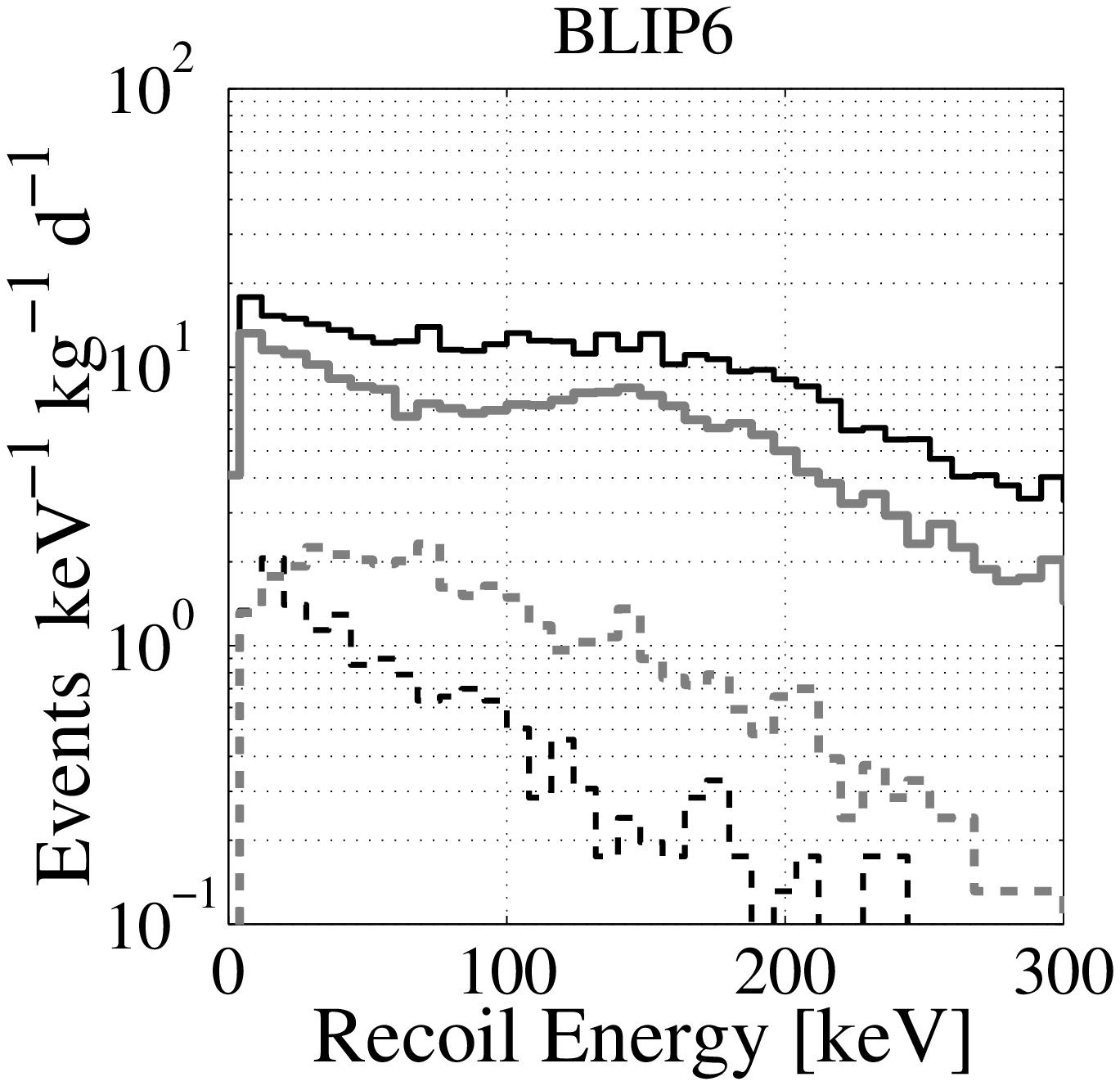,width=1.65in}
\caption{\label{qicoincspec}
Recoil-energy spectra for veto-coincident inner-electrode contained events. 
 Dark solid:
single-scatter photons.  Dark dashed: single-scatter electrons.  Light
solid: photons belonging to double scatters.  Light dashed: electrons
belonging to double scatters.}
\end{figure}

\begin{figure}
Veto-Coincident Shared-Electrode  \\
\psfig{figure=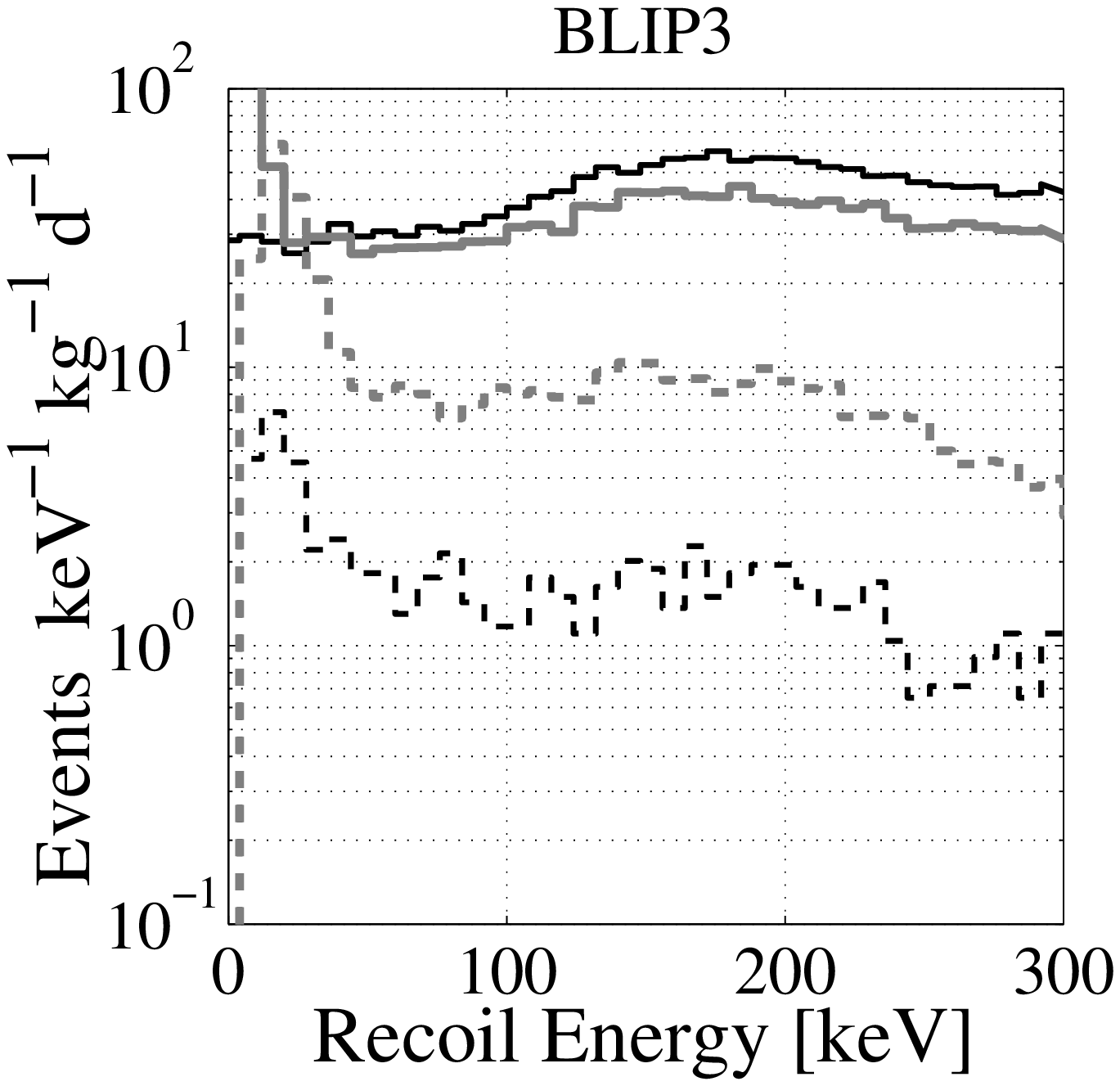,width=1.65in}
\psfig{figure=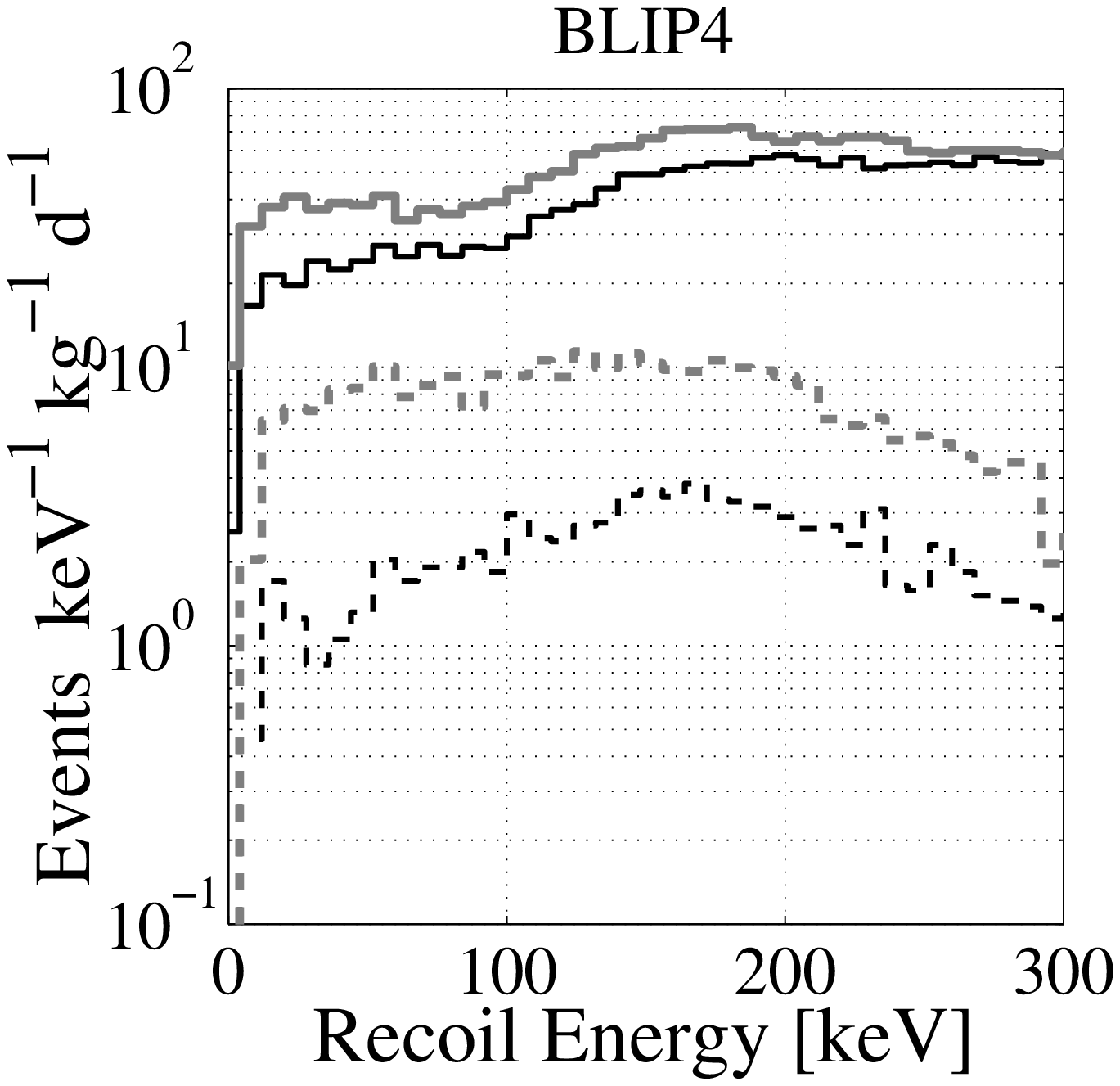,width=1.65in}
\psfig{figure=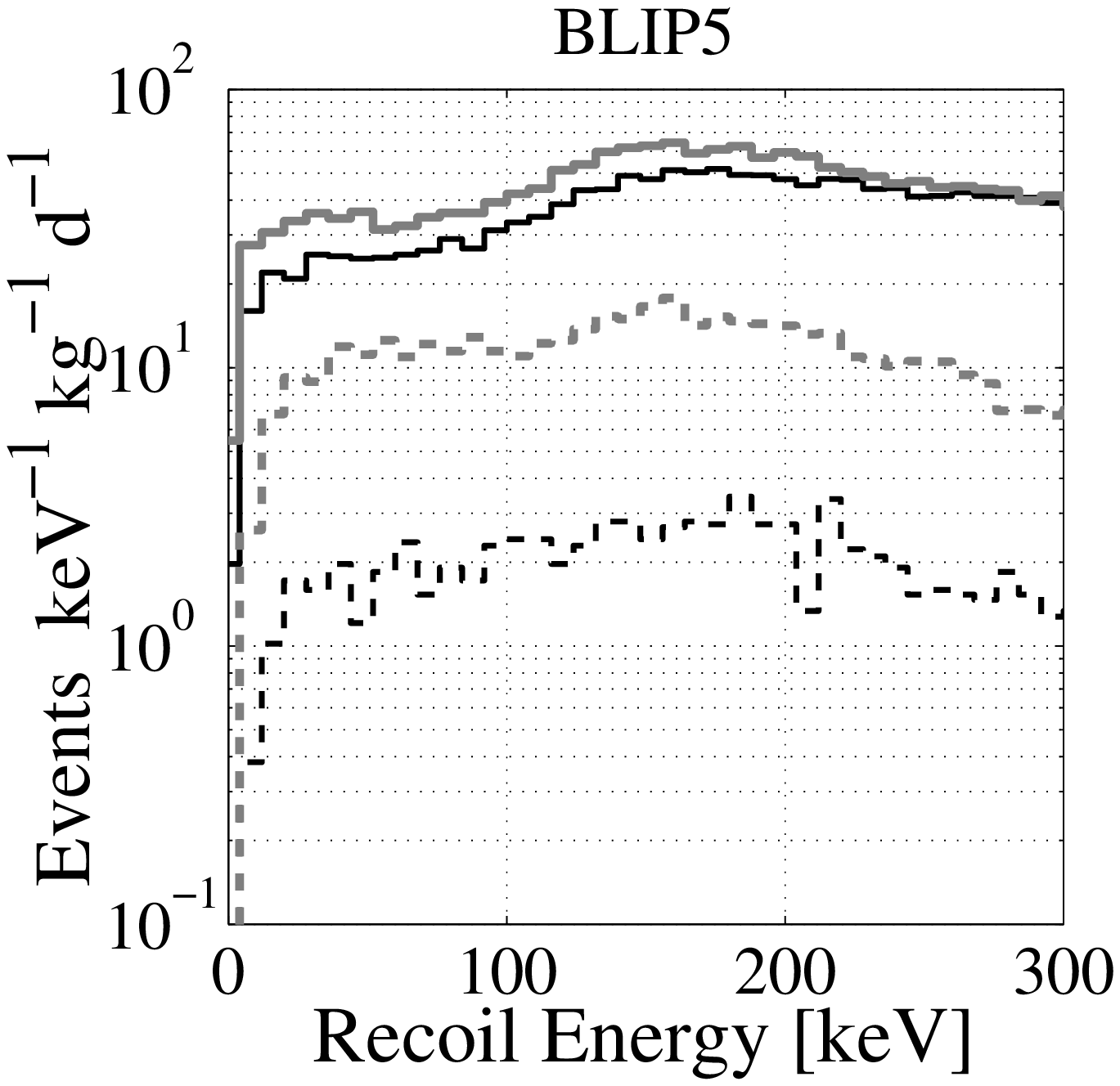,width=1.65in}
\psfig{figure=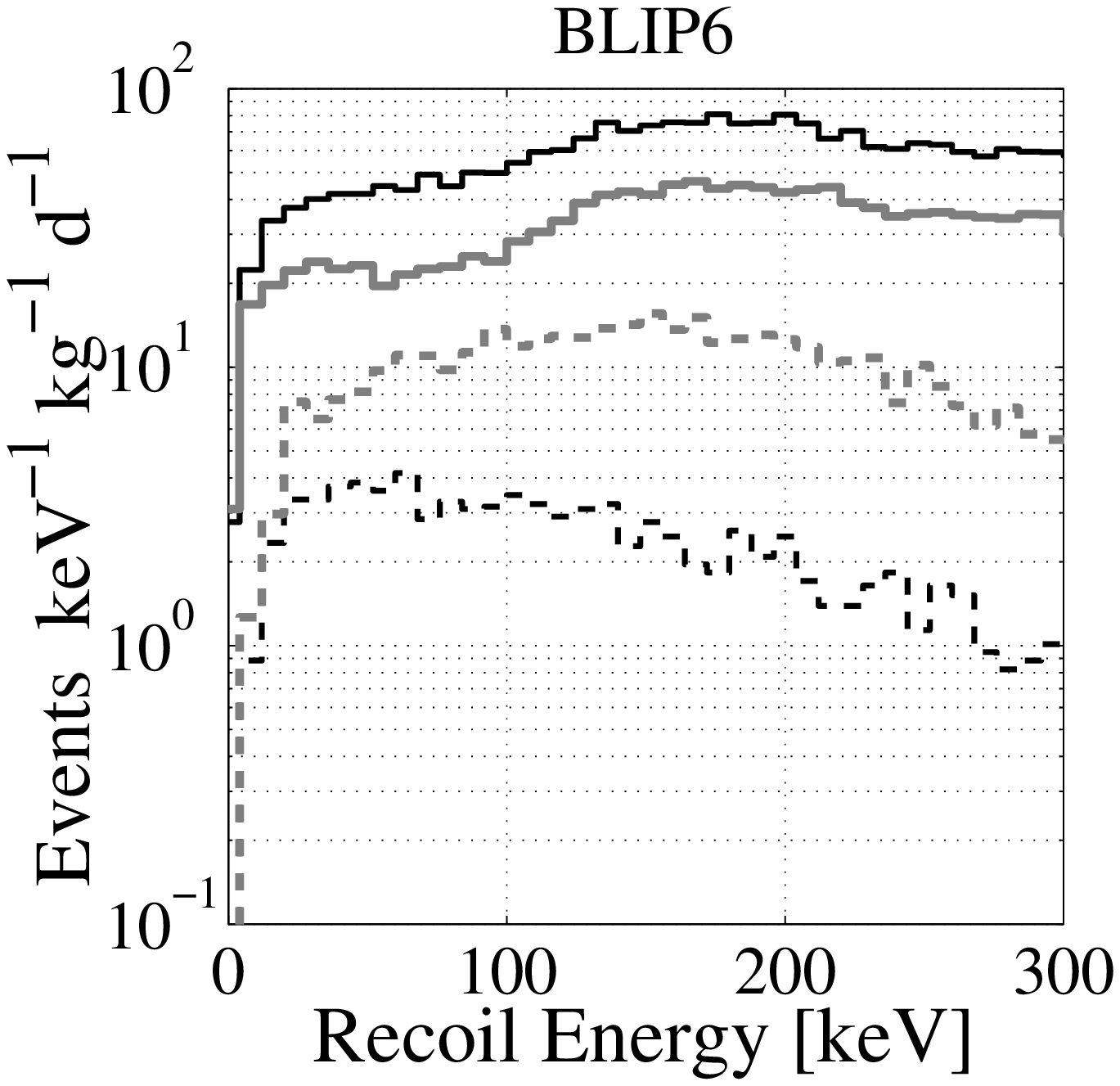,width=1.65in}
\caption{\label{qsharecoincspec}
Recoil-energy spectra for veto-coincident shared-electrode events.  Legend as
in Fig.~\protect\ref{qicoincspec}.}
\end{figure}

At the experiment's current shallow site, most events are induced by 
muons and tagged by the muon veto.
The observed
electromagnetic backgrounds coincident and anticoincident with the veto
are 60 \dru\ and 2 \dru.  
Recoil-energy spectra for the veto-coincident data are shown in
Fig.~\ref{qicoincspec} and \ref{qsharecoincspec}.  
Events with ionization yields consistent with bulk electron recoils 
are histogrammed as photons, while events with ionization yields
inconsistent with bulk electron recoils and nuclear recoils
are histogrammed as electrons.
The relative single- and double-scatter rates reflect the geometry, 
with BLIPs 3 and 6, the detectors on the top and bottom of the stack, 
exhibiting lower double-scatter photon
fractions than BLIPs 4 and 5, the detectors with two nearest neighbors.  
Also, compared to the
veto-anticoincident data, the electron double-scatter fractions are
quite high, indicating most veto-coincident electrons are produced in
showers or are ejected from the detectors and surroundings.  
The photon spectrum incident on the detectors is expected to 
decrease with decreasing energy at
low energy due to the presence of many shielding layers.
The shared-electrode events reflect the incident spectrum
because internal multiple scatters are included in this set, 
increasing the number of events with the full photon energy deposited 
in the detector.  In contrast, the spectrum of
inner-electrode-contained  photons 
increases with decreasing energy
at low energy, as expected from the fact that such events
are dominated by Compton scattering of high-energy photons.

\begin{figure}
Veto-Anti-Coincident Inner-Electrode-Contained\\
\psfig{figure=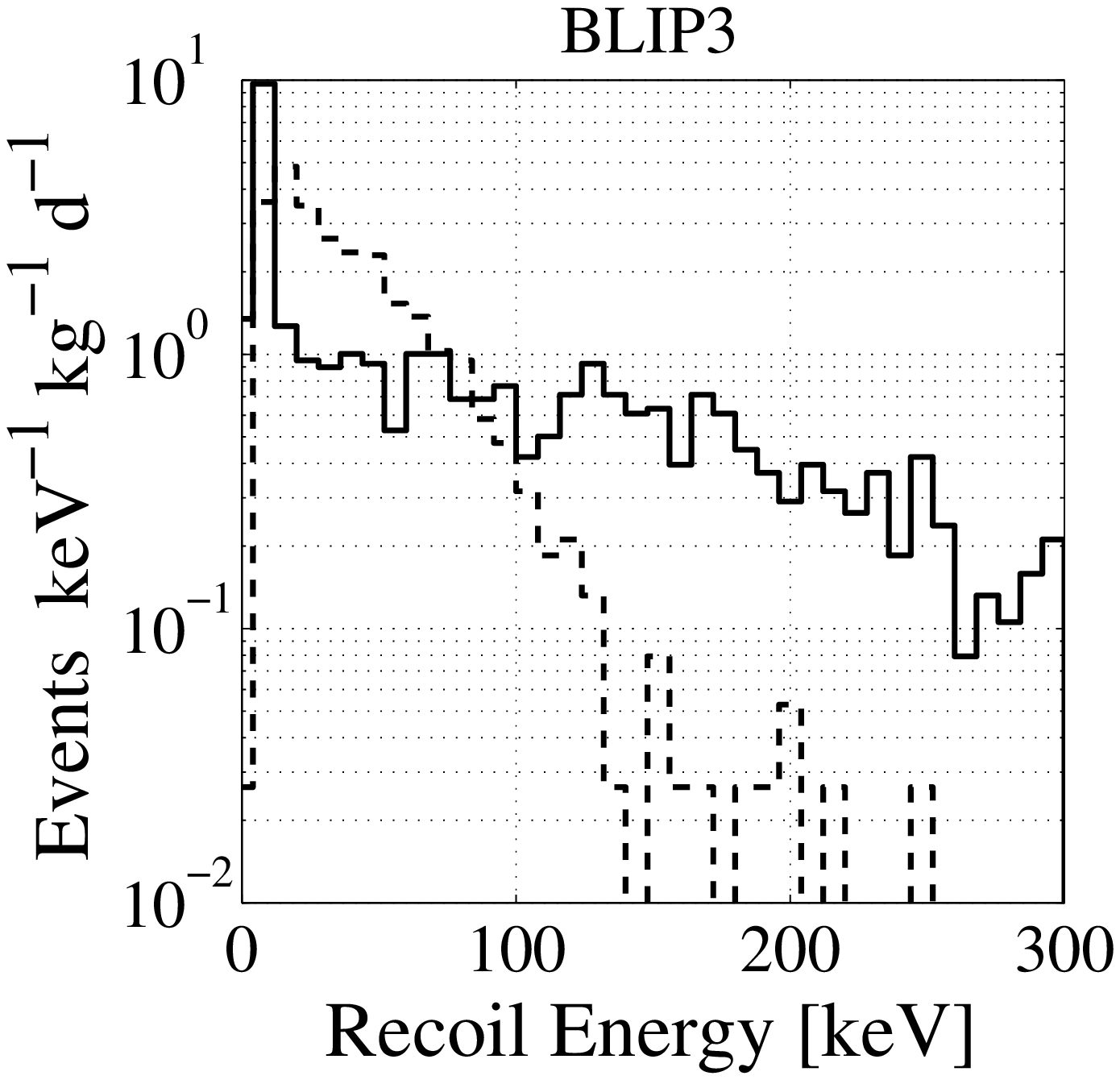,width=1.65in}
\psfig{figure=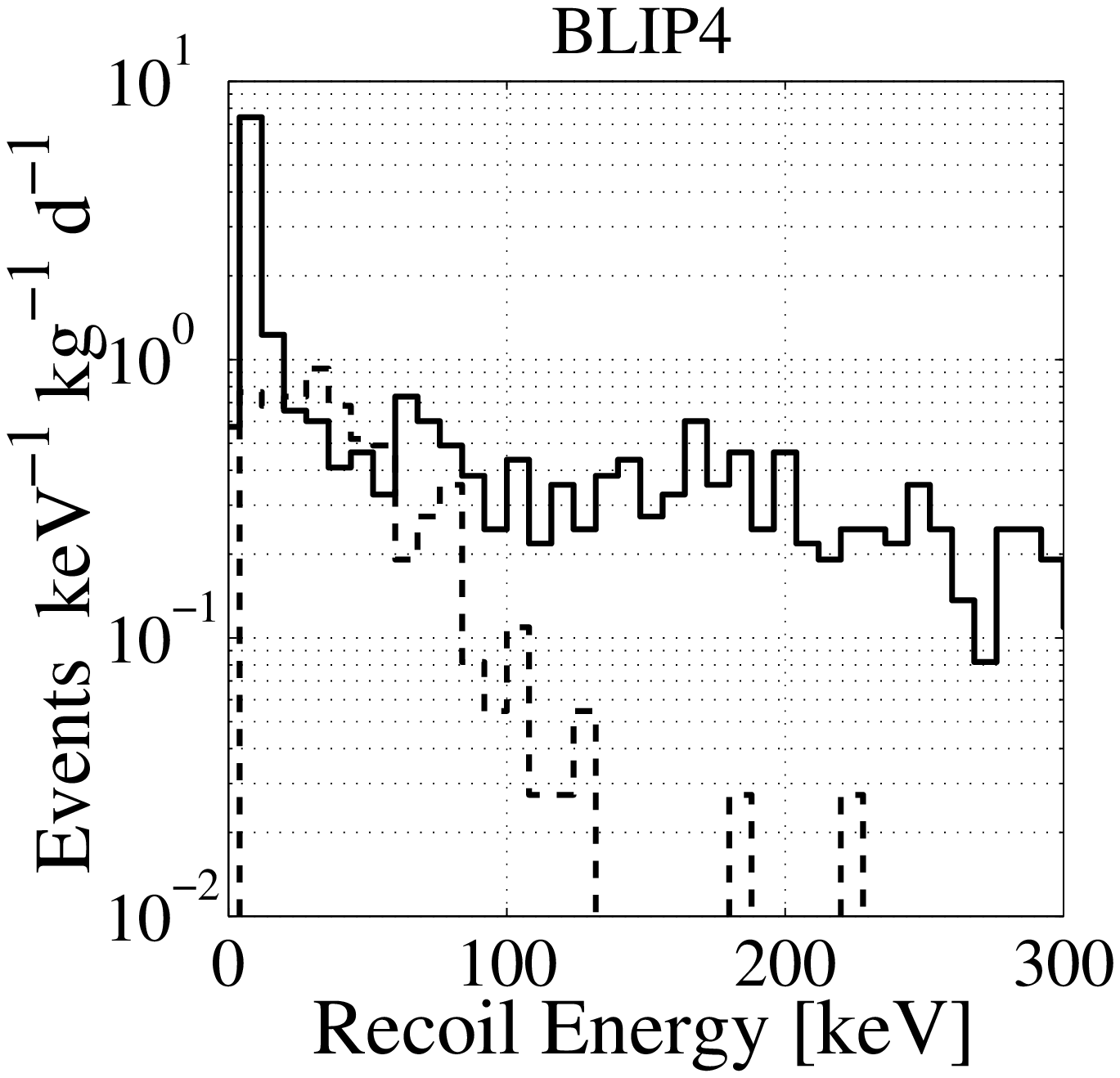,width=1.65in}
\psfig{figure=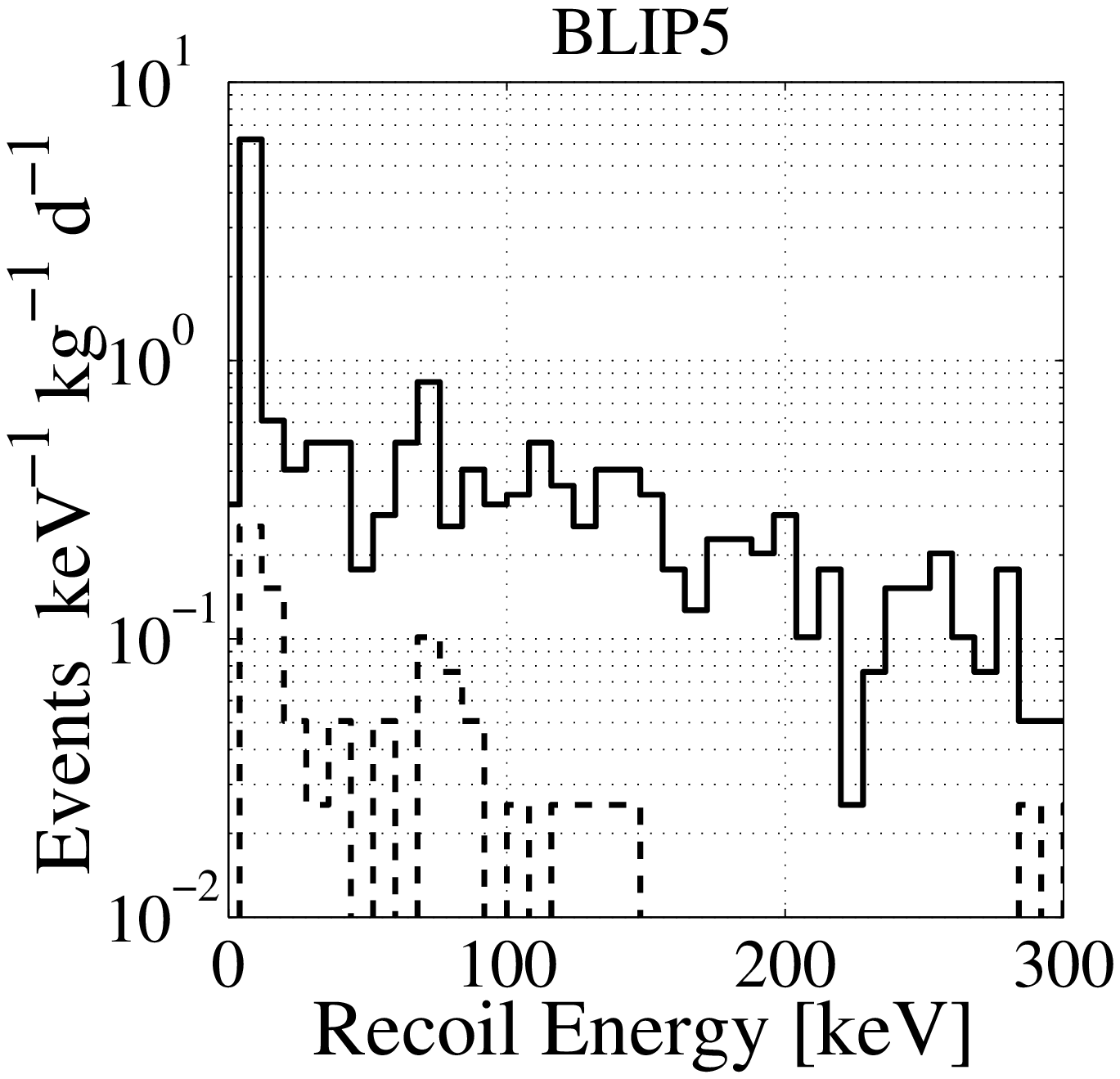,width=1.65in}
\psfig{figure=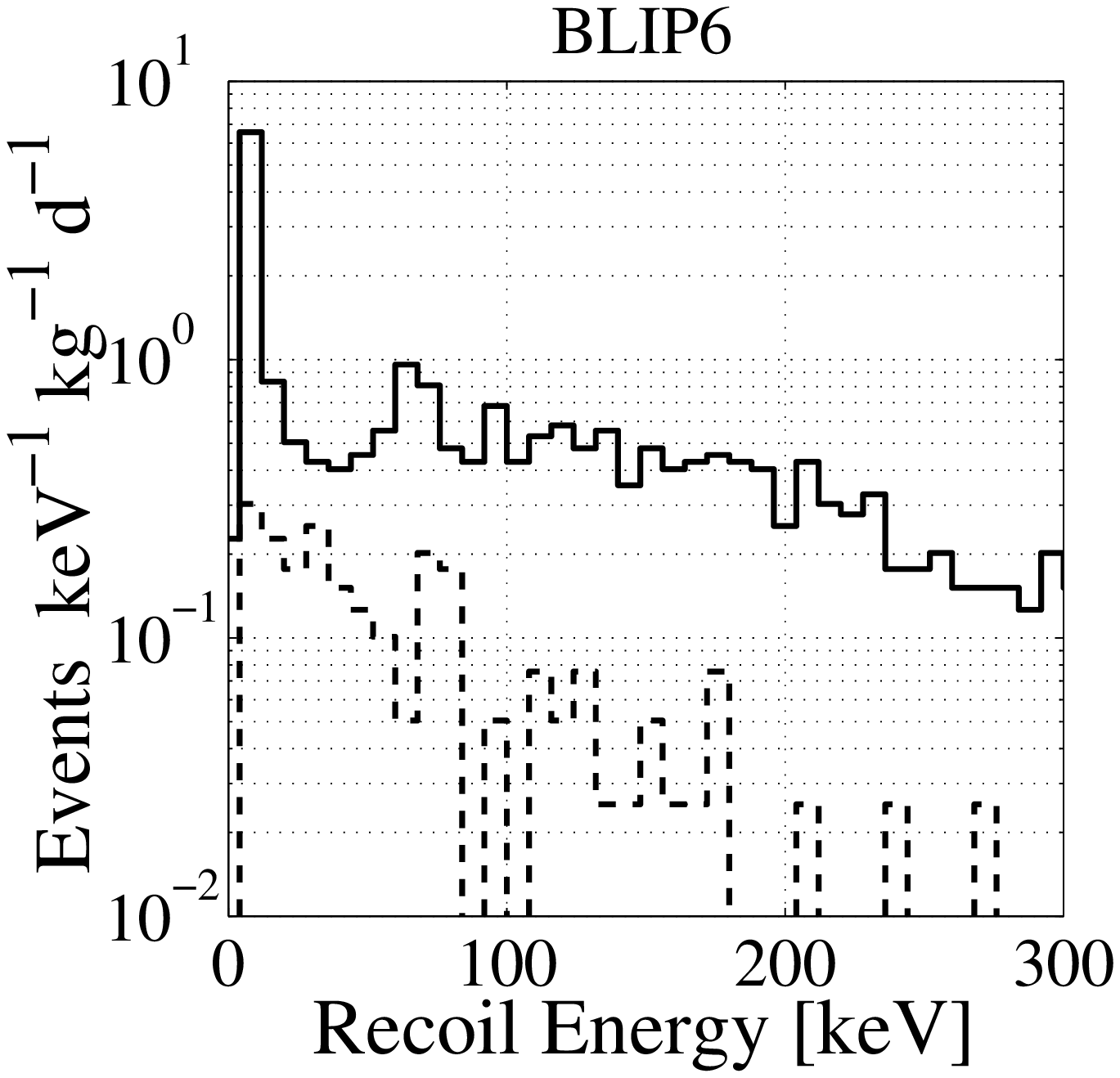,width=1.65in}
\caption{\label{qiantisinglespec}
Single-scatter
photon and electron recoil-energy spectra for veto-anticoincident
inner-electrode-contained events. Solid: photons.  Dashed: electrons.
}
\end{figure}

\begin{figure}
Veto-Anti-Coincident Shared-Electrode  \\
\psfig{figure=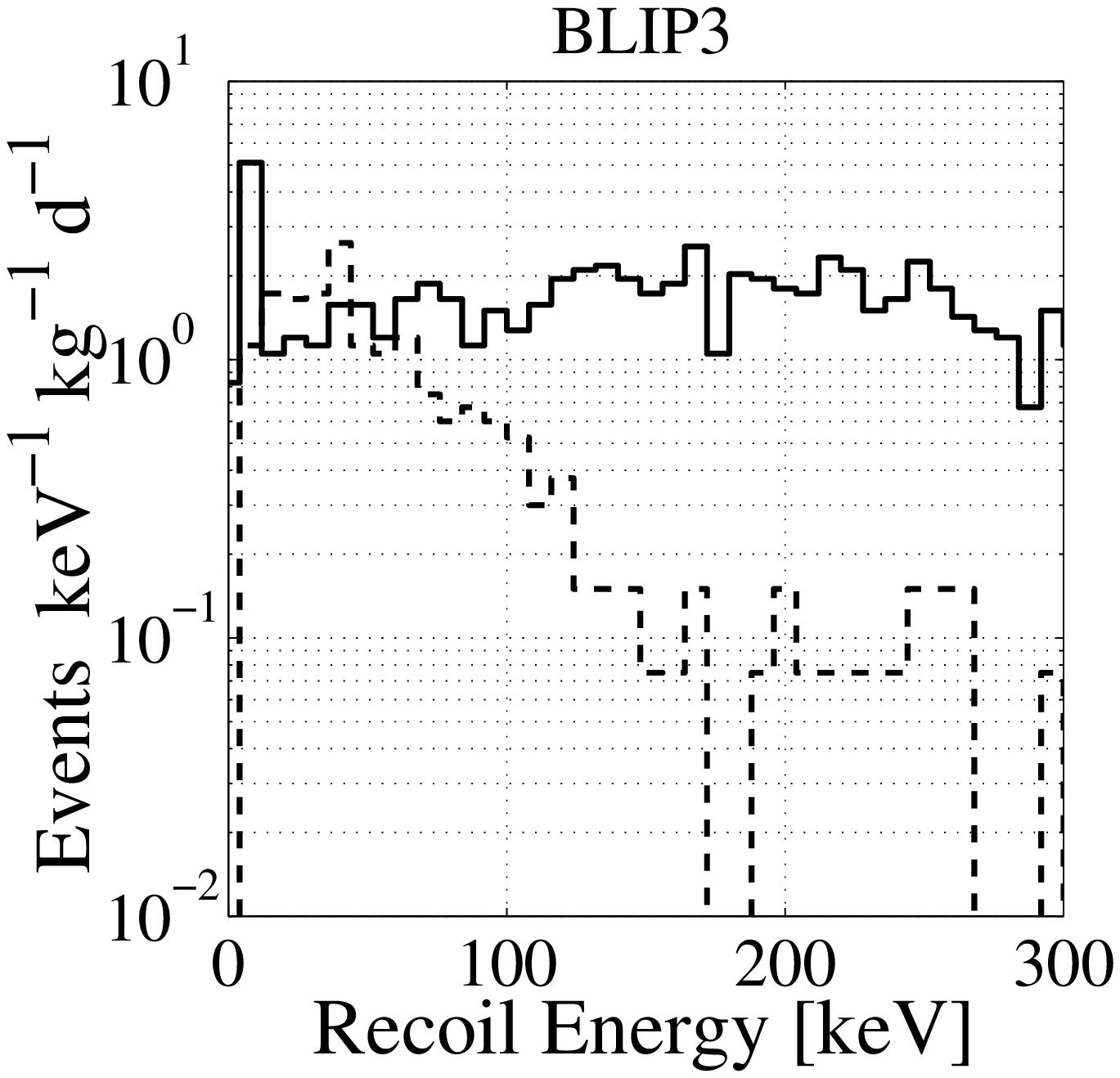,width=1.65in}
\psfig{figure=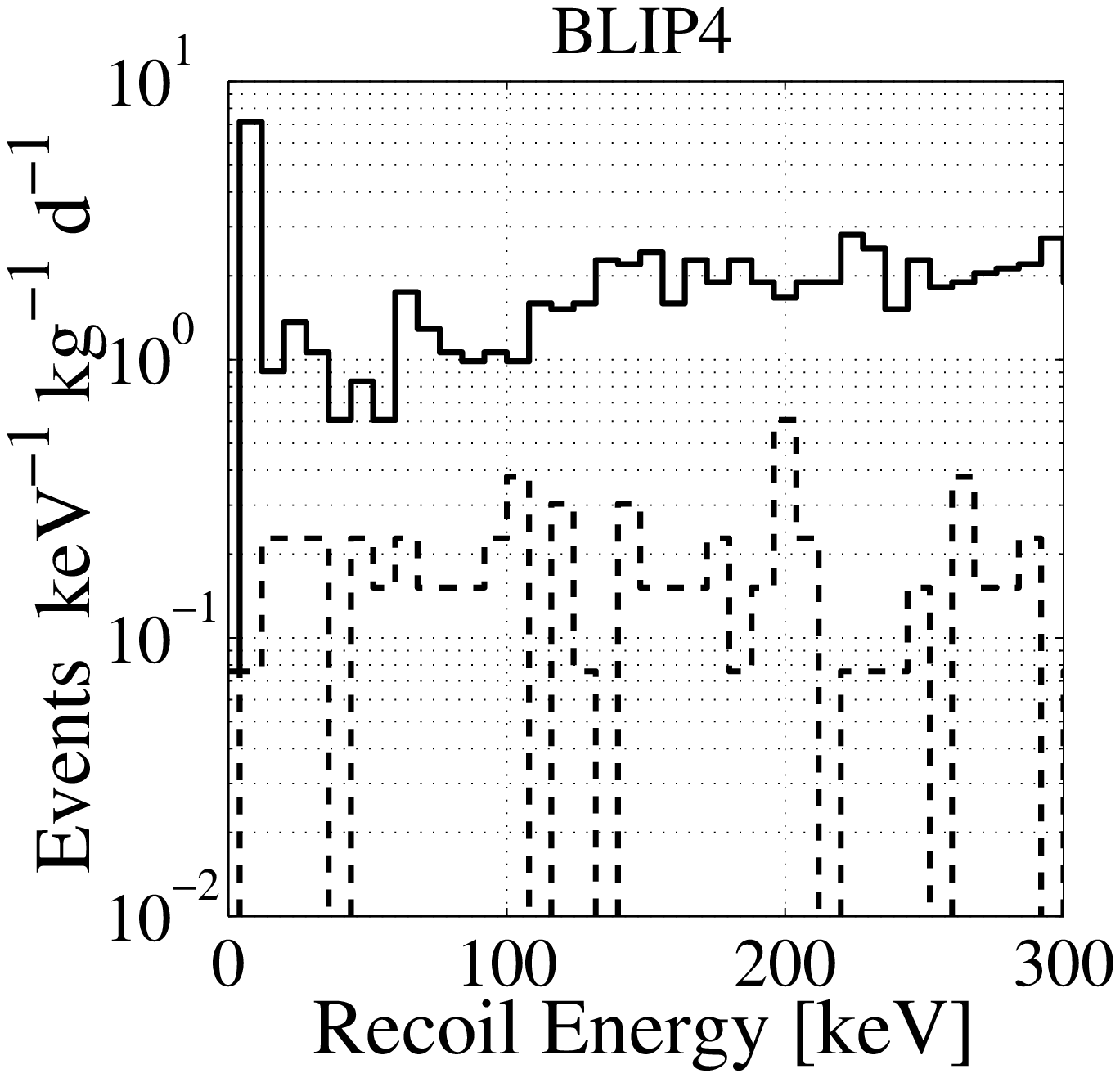,width=1.65in}
\psfig{figure=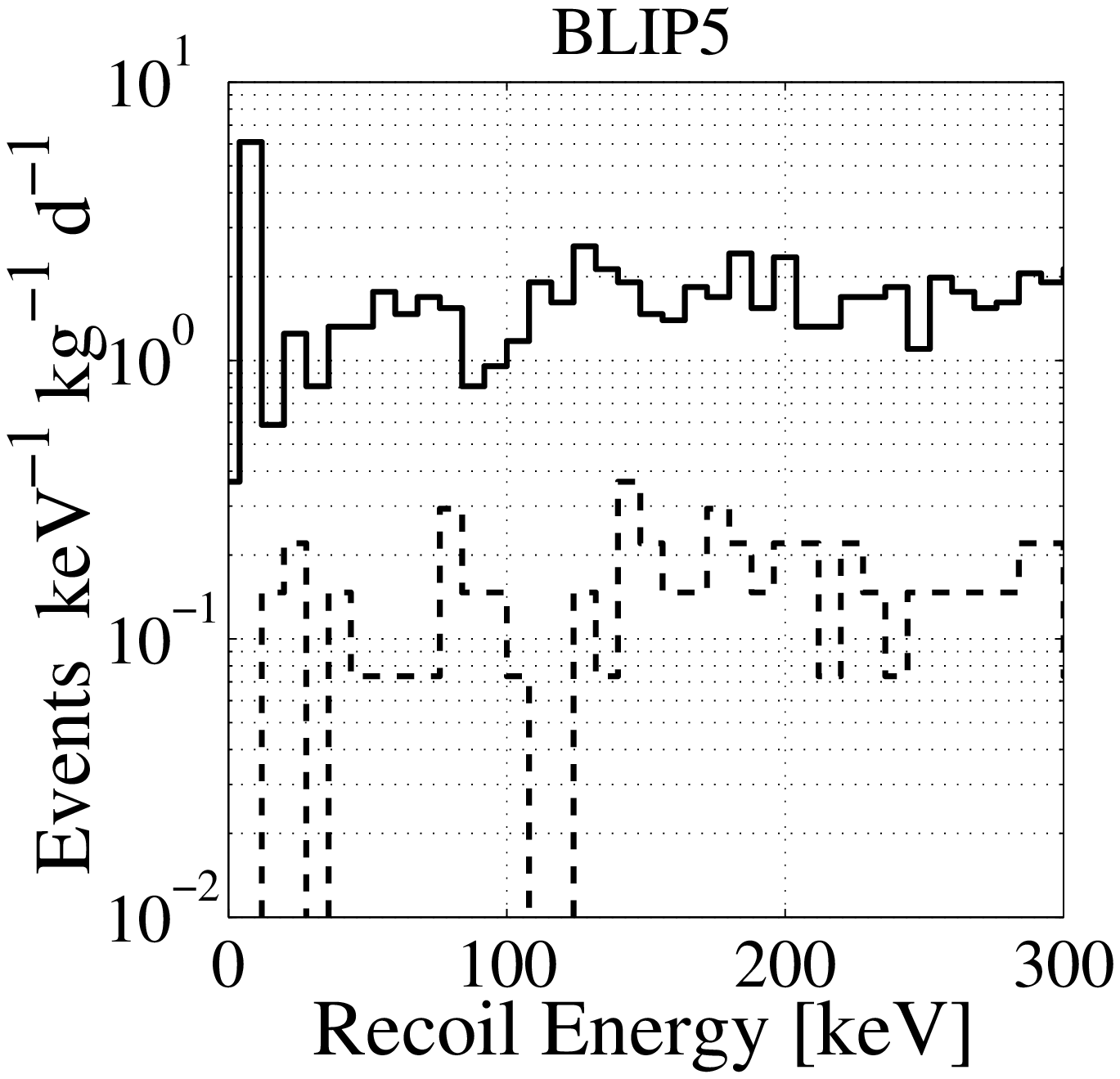,width=1.65in}
\psfig{figure=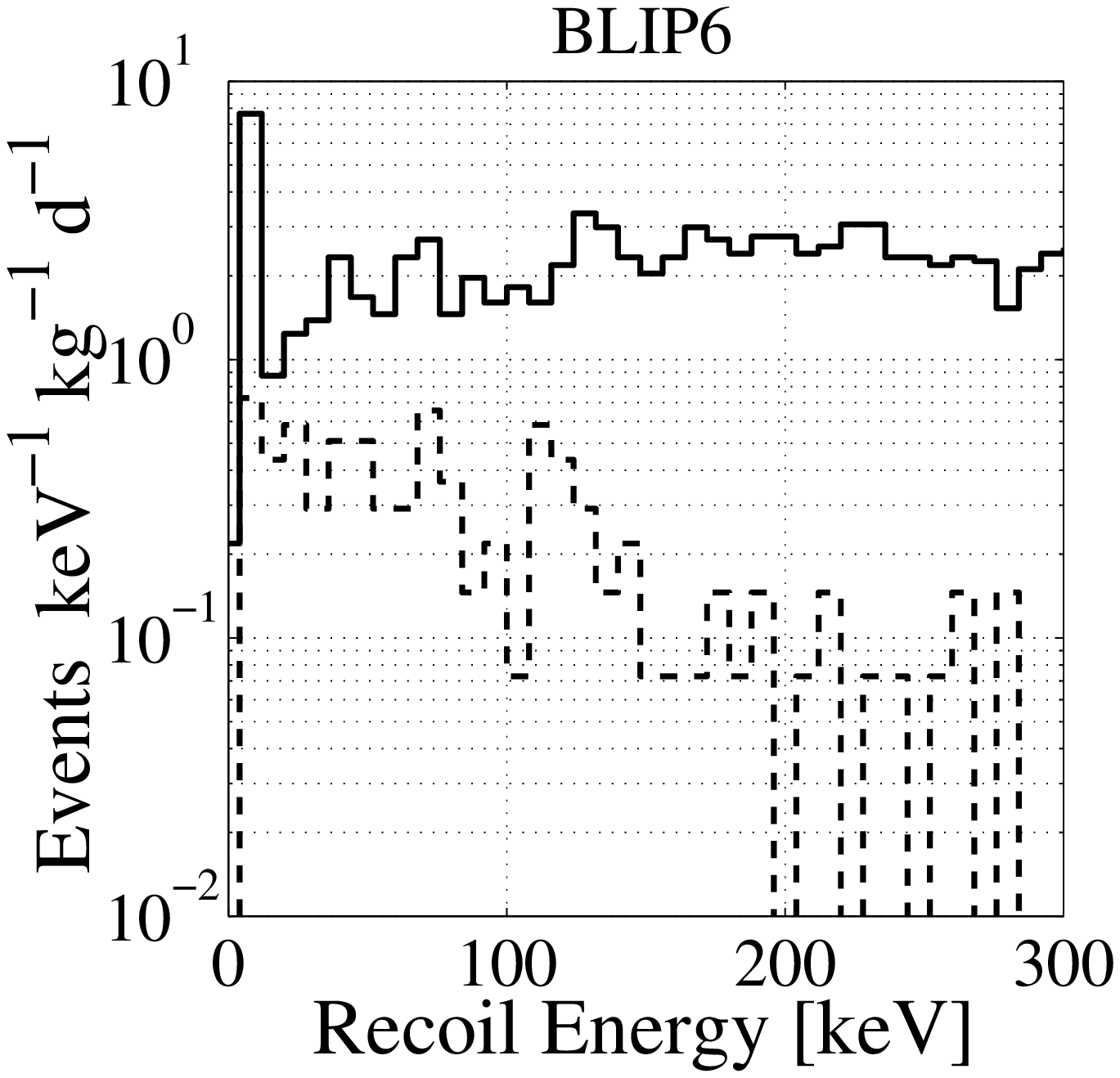,width=1.65in}
\caption{\label{qshareantisinglespec}
Single-scatter
photon and electron recoil-energy spectra for veto-anticoincident
shared-electrode events. Solid: photons.  Dashed: electrons.}
\end{figure}

\begin{figure}
\psfig{figure=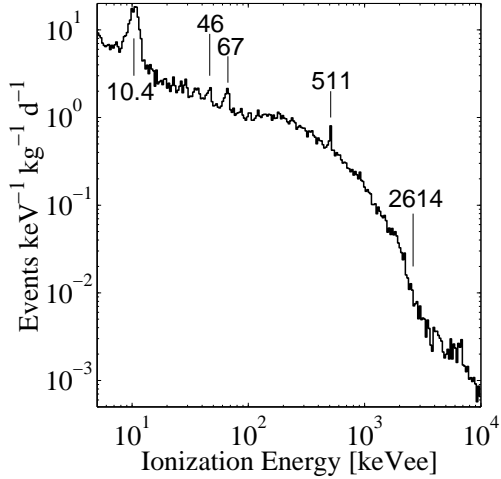,width=2.6in}
\caption{\label{qantiallspec}
Spectra for veto-anticoincident
 events with no other cuts applied, showing the sum of the ionization
 electron-equivalent
 energy in all four detectors.  Bin widths are logarithmic and roughly 
 correspond to the energy resolution at high energies.  Significant 
 spectral lines at 10.4~keV (from internal Ga), at 67~keV (from $^{73m}$Ge), 
 and at 511~keV (from positron annihilation) are indicated.  
 The line at 46~keV (from \pb210) is significant 
 only when a cut selecting events in 
 the outer electrode is applied.  See also Fig.~\ref{Elines}.
 The rate of events above the 
 2.6~MeV endpoint of U/Th is much lower than the rate below this 
 energy, suggesting that a significant fraction of the lower-energy 
 events are due to U/Th contamination.}
\end{figure}

The dominant muon-anticoincident electromagnetic
background is due to natural radioactivity,
long-lived cosmogenic activation, or possibly thermal-neutron activation. 
For the data set
described here, the veto efficiency for muons that pass through the
detectors was $>99.9\%$. 
The muon-induced veto-anticoincident event rate is therefore $<0.1$~\dru, 
far less than the observed total anticoincident rate of $\sim$1~\dru\
(see Fig.~\ref{qiantisinglespec} and \ref{qshareantisinglespec}).
Attempts to simulate this radioactivity-induced background level, assuming
reasonable amounts of radioisotopes in the construction materials, have thus
far failed to yield a rate as high as that observed.
Because the energy of $\sim$ MeV photons is rarely fully contained 
in these low-mass detectors, high-energy spectral lines 
that could otherwise be used 
to determine the abundance of particular radioactive 
contaminants are not visible, as shown in Fig.~\ref{qantiallspec}.

The rate of $\alpha$-particles interacting in the detectors is about 0.8 per
live-day per detector, and about 0.2 per live-day in the fiducial volume of
each detector.  
No evidence of alpha decays in the bulk of the detectors is seen,
consistent with expectations based on the purity of the materials.
Because $\alpha$-particles result in high-energy depositions,
well above the energy region of a potential WIMP signal,
they do not provide a significant background for the WIMP search.
The recoiling nuclei from $\alpha$-decays may result in
low-energy events.  
We have tagged several such events by each one's coincidence
with an $\alpha$-particle in an adjacent detector.  
Because the recoiling nuclei interact in the detector's dead layer,
they result in little
or no ionization and hence yield events outside the nuclear-recoil
acceptance region.

\subsubsection{Muon-anticoincident nuclear recoils}

\begin{figure} 
\psfig{figure=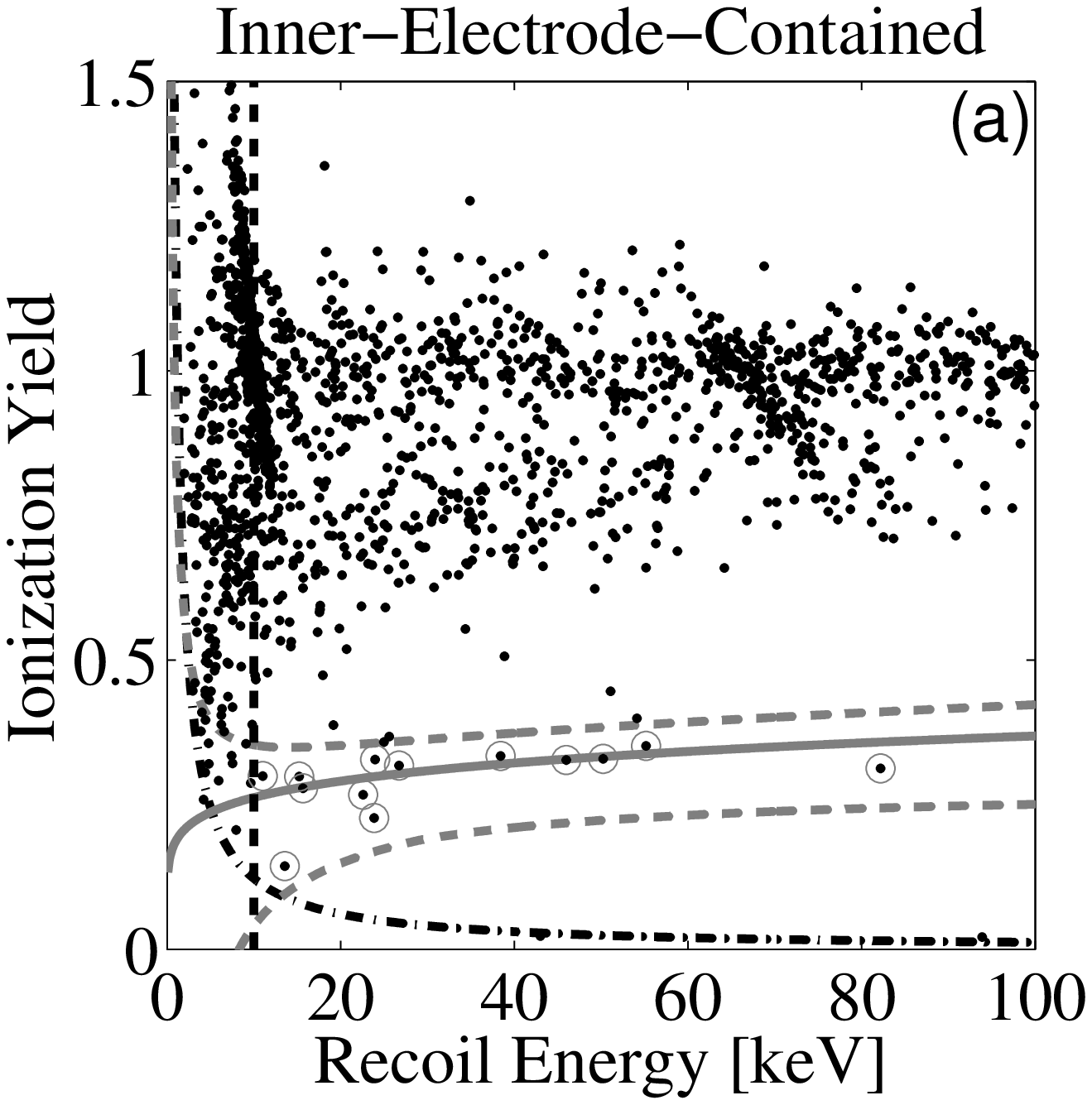, width=6.1cm}
\psfig{figure=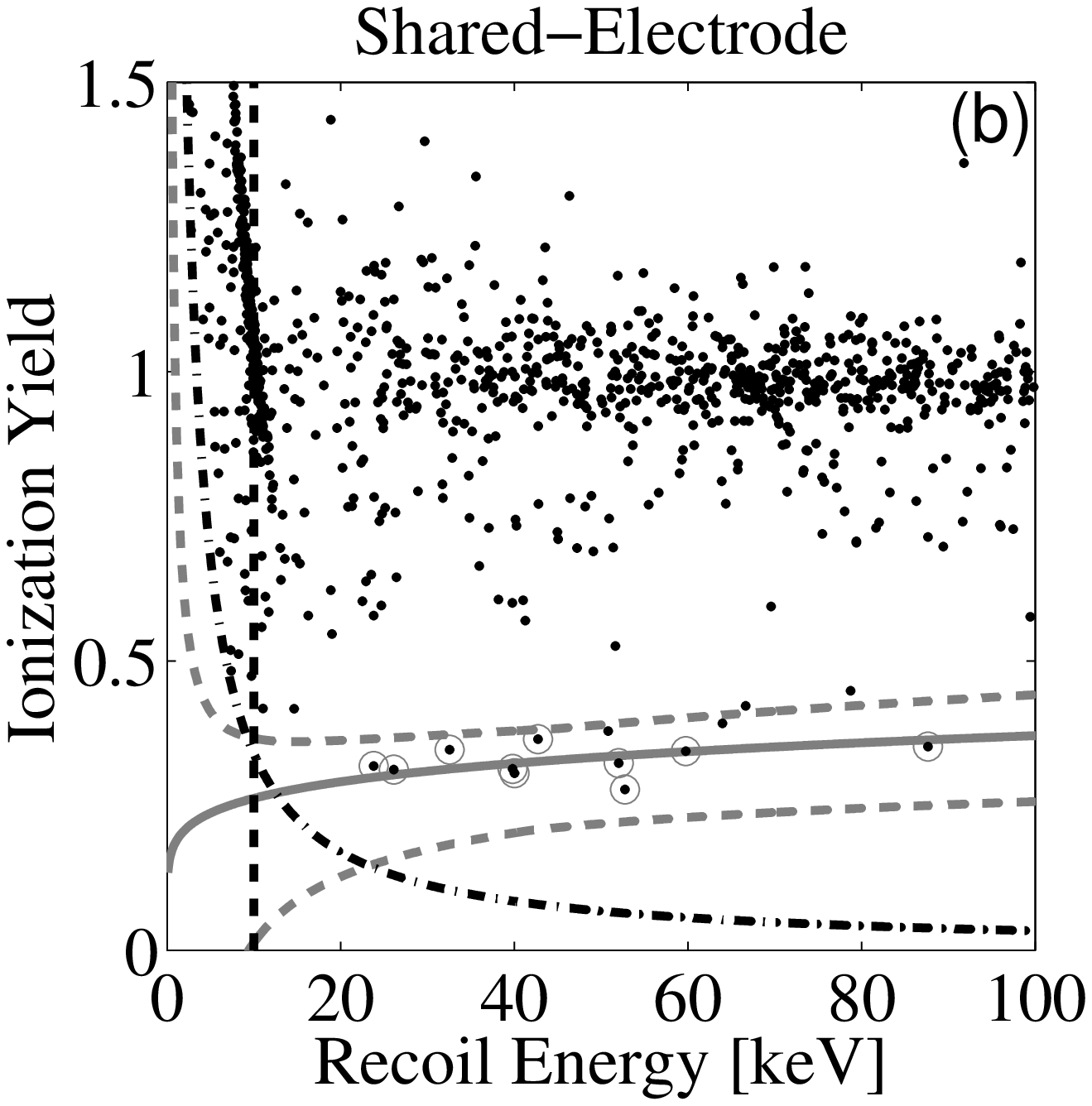, width=6.1cm}
\caption{\label{highbiasyplot}
Ionization yield ($Y$) vs. recoil energy for
veto-anticoincident single scatters in the 3 uncontaminated Ge detectors.  
Solid curve: expected position
of nuclear recoils.  Dashed curves: mean nominal 90\% nuclear-recoil
acceptance region.  Dashed line: 10~keV analysis threshold.
Dot-dashed curve: mean threshold for separation of ionization signal
from amplifier noise.  Circled points: nuclear recoils.  
(a) Events with energy fully contained in the detectors' 
inner electrodes.
(b) Events with energy shared between the detectors' 
inner and outer electrodes.
The presence of 2 uncircled events within the mean nuclear-recoil band 
is due to slight differences in the size of the band for different 
detectors.
About half the 3 QI (4 QS) events just above the acceptance region are 
likely to be nuclear recoils, 
since the top of the nuclear-recoil band is 1.28$\sigma$ 
above its center, yielding 90\% acceptance.
}
\end{figure}

\begin{figure}
\psfig{figure=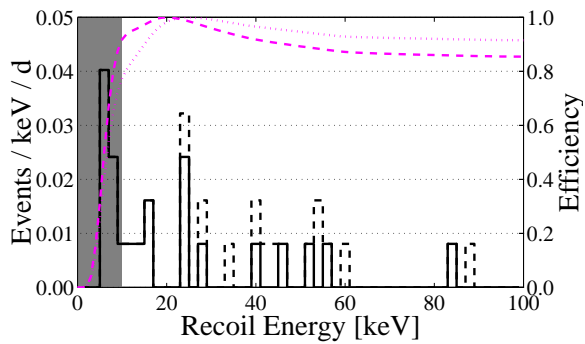, width=7.7cm}
\caption{\label{spec}
Histogram of inner-electrode-contained (solid) 
and shared-electrode (dashed) veto-anticoincident
single-scatter nuclear recoils observed in 
the 3 uncontaminated Ge detectors (left-hand scale). 
The nuclear-recoil efficiencies (right-hand scale) for the
QI (dashed) and QIS (dotted) data are each peak-normalized to 1;
with this normalization, the QIS data corresponds to 0.26~kg effective 
mass, and the QI data corresponds to 0.20~kg effective mass.
Shaded: 10 keV analysis threshold.  
}
\end{figure}

Figure~\ref{highbiasyplot} shows plots of ionization yield vs. recoil
energy for the muon-anticoincident events triggering on any single detector
(the WIMP multiple-scatter rate is negligible).  Bulk electron recoils
(primarily due to photon interactions)
lie at ionization yield $Y \simeq 1$.  Low-energy electron events form a
distinct band at $Y\sim 0.75$, leaking into the nuclear-recoil
acceptance region below 10~keV.  Between 10 and 100~keV, 23 QIS (13 QI)
unvetoed nuclear-recoil candidates are observed,
corresponding to 15.8 (11.9)~kg~d exposure. 
Figure~\ref{spec} displays the recoil-energy spectrum of unvetoed
single-scatter nuclear-recoil candidates for the Ge data set, 
along with the overall efficiency.  

\subsubsection{Expected nuclear-recoil-band contamination}

\begin{figure}
\psfig{figure=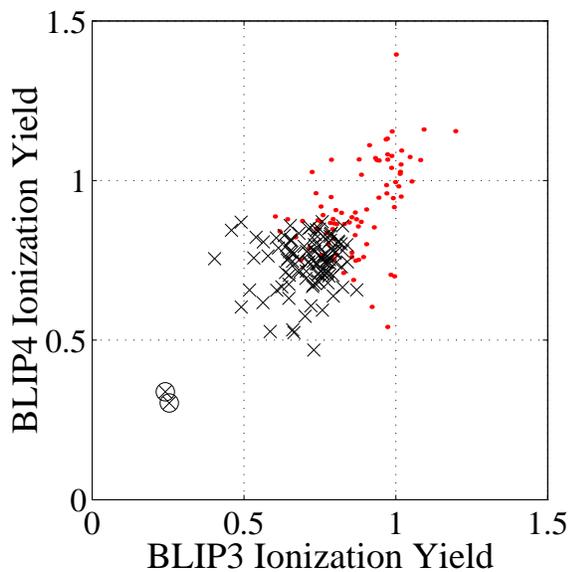,width=3in}
\caption{\label{b3b4y2plot}
BLIP4 ionization yield vs. BLIP3 ionization yield for events ($\times$'s) 
used as the
electron-calibration data set.  This set consists of all veto-anticoincident 
double-scatter events in BLIP3 and BLIP4 with both hits between 
10--100~keV, at least one QIS hit, 
and no hit that appears as a bulk electron recoil ($Y\sim 1$).
Events with one or more apparent bulk electron recoils 
that fulfill all other criteria 
are shown as dots.
Two events (circled) pass
nuclear-recoil cuts for both BLIP3 and BLIP4.
Based on the expected neutron background, about one double-scatter 
neutron should be in this data set. 
The large separation from the main distribution 
of the two events tagged as nuclear recoils in both BLIP3 and BLIP4
suggests they are, in fact, neutrons; 
in the analysis, they are conservatively
assumed to be misidentified electrons.  
}
\end{figure}

The observed photon and
electron event rates can be combined with the photon- and
electron-calibration data to set upper limits on the
expected numbers of misidentified single-scatter photons and electrons
in the low-background set.
As shown in Table~\ref{antisingleleak}, 
photon misidentification should contribute a
negligible number of nuclear-recoil candidates.  The estimate on the 
amount of electron misidentification is not nearly so useful, for two reasons. 
First, the electron calibration is
statistics-limited: even if no nuclear-recoil candidates had been seen in
the electron calibration, the 90\%~CL upper limits would
still be nonnegligible.  Second, the two electron-calibration events 
with both hits in the nuclear-recoil acceptance region 
(see Fig.~\ref{b3b4y2plot}) may well be multiple-scatter neutrons
(about one multiple-scatter neutron is expected in this data set).
However, to be conservative, Table~\ref{antisingleleak} lists these 
events as misidentified electrons.
With this conservative assumption and low statistics, it is possible for
all of the low-background 
nuclear-recoil-candidate events to 
be misidentified electrons.
However, the most likely 
number of misidentified electrons, even with this conservative 
assumption, is only about 6~QIS (3.6~QI) events.
Most of the single-scatter nuclear-recoil candidates are probably 
nuclear-recoil events.

\begin{table}
\begin{ruledtabular}
\begin{tabular}{l|rrrrr} 
Event set     & $N_{\mathrm{c}}$ & $N_{\mathrm{l}}$ & $N_{\mathrm{b}}$
              & $\left<\mu_{\mathrm{l}}\right>$ & $\mu_{\mathrm{l,}90}$  \\ \hline
\multicolumn{6}{l}{Inner-Electrode-Contained Photons}  \\ \hline
10 -- 30~keV  &  4661 & 2  & 490 & 0.2 & 0.6 \\
30 -- 100~keV &  5609 & 0  & 498 & 0.0 & 0.2 \\ \hline
10 -- 100~keV & 10270 & 2  & 988 & 0.2 & 0.5 \\ \hline
\multicolumn{6}{l}{Shared-Electrode Photons}  \\ \hline
10 -- 30~keV  &  2430 & 0  & 172 & 0.0 & 0.2 \\
30 -- 100~keV &  4466 & 1  & 508 & 0.1 & 0.4 \\ \hline
10 -- 100~keV &  6896 & 1  & 680 & 0.1 & 0.4 \\ \hline
\multicolumn{6}{l}{Inner-Electrode-Contained Electrons}  \\ \hline
10 -- 30~keV  &  95   &  2  & 101 & 2.1 & 5.9 \\ 
30 -- 100~keV &  61   &  0  & 180 & 0.0 & 7.0 \\ \hline
10 -- 100~keV & 156   &  2  & 281 & 3.6 & 9.7 \\ \hline
\multicolumn{6}{l}{Shared-Electrode Electrons}  \\ \hline
10 -- 30~keV  &  23  &  1  &  31 & 1.3 &  5.8 \\ 
30 -- 100~keV &  20  &  0  &  78 & 0.0 &  9.7 \\ \hline
10 -- 100~keV &  43  &  1  & 109 & 2.5 & 10.3 \\ 
\end{tabular}
\end{ruledtabular}
\caption{\label{antisingleleak}
Veto-anticoincident inner-electrode and shared-electrode 
single-scatter photon and electron
misidentification estimates.  The first two columns list the numbers
of properly identified calibration events $N_{\mathrm{c}}$ 
and calibration events misidentified as nuclear recoils
$N_{\mathrm{l}}$ in BLIPs~4--6 (BLIPs~3--4) for the 
photon-calibration (electron-calibration)
data sets.
The third column lists the number of single-scatter background events 
$N_{\mathrm{b}}$
in the given data set and energy range. 
The final two columns list the resulting expected number of events 
misidentified as nuclear recoils $\left<\mu_{\mathrm{l}}\right>$ as well as 
the Bayesian 90\%~CL upper limit $\mu_{l,90}$ on this quantity.
The expected misidentification for the full energy range 
need not be equal to the sum of the expected misidentification for 
the two smaller energy ranges. }
\end{table}

\subsubsection{Consistency tests}

The self-consistency of the hypothesis 
that the nuclear-recoil candidates are all 
veto-anticoincident nuclear recoils
is tested by comparing the distributions of 
various event parameters to their expected distributions using the 
Kolmogorov-Smirnov (KS) test (see~\cite{numrecipes} or~\cite{eadie}).

\begin{figure}
\psfig{figure=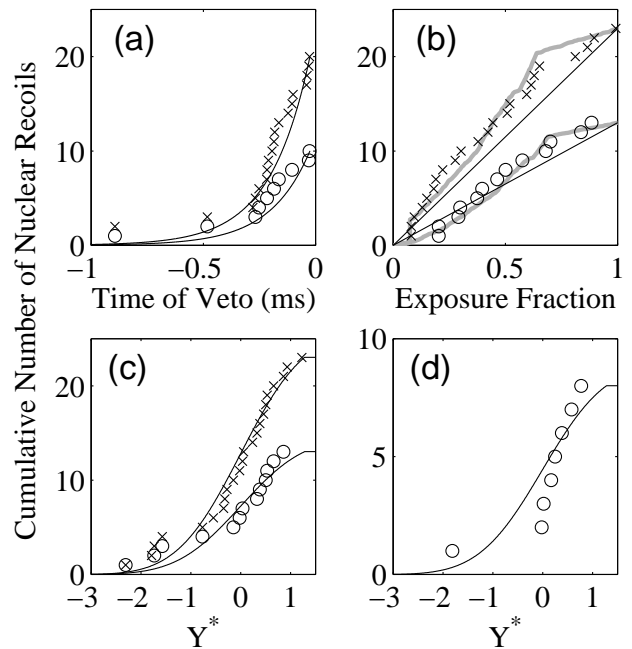,width=3.3in} \\
\caption{\label{kstests}
Comparisons of expected integral distributions (curves) 
to actual integral distributions 
for veto-anticoincident QIS ($\times$'s) and QI (circles) nuclear-recoil
candidates. 
(a) Last veto-trigger-time for ionization triggers.  
(b) Exposure fraction.
The dark lines show the expectations if the rate of events should be 
uncorrelated with changes in veto efficiency with time, while the grey 
curves indicate the expectations if the rate of events should be 
linearly correlated with changes in the veto efficiency.
(c) Single-scatter $Y^{\ast}$ distributions.  
(d) Multiple-scatter $Y^{\ast}$ distributions.  As quantified in the text, 
all distributions are consistent with expectations.}
\end{figure}

Figure~\ref{kstests} shows the cumulative distribution of
the last
veto-trigger times for the 20 QIS (10 QI) ionization-trigger nuclear-recoil 
candidates 
(three of the nuclear-recoil candidates are phonon-trigger events).
These times should follow an exponential distribution if
the veto-trigger times are uncorrelated with the
event times.  The KS test indicates that 42\% (55\%) of experiments 
should
observe distributions that deviate further from the expected 
exponential
distribution for the QIS (QI) events.  

It is also possible to test the time distribution of the events.  
The integrated exposure, the number of kg-days of data taken up to 
the time of an event, takes into account the cut efficiencies and the 
numbers of detectors that were live for each event.
Any unvetoable set of events (such as those due to WIMPs)
should be
uniformly distributed in exposure.
For events caused by cosmic-ray muons that avoid being 
vetoed due to the small residual veto inefficiency, the time dependence of 
the veto efficiency must be included in the calculation of the expected
fraction of events observed as a function of the cumulative exposure.
For events caused by particles much less likely to be vetoed 
(such as neutrons produced outside the veto), the time 
dependence of the veto efficiency is likely negligible.
The KS test indicates 51\% (60\%) of experiments should observe
distributions that deviate further from the distribution expected 
for QIS (QI) events for a constant veto efficiency. 
For QIS (QI) events whose veto probability is directly proportional to 
the veto probability for muons, the KS test indicates that 30\% (82\%) 
of experiments should observe  distributions that deviate further 
from the expected distribution.  The time distribution of the events 
agrees with expectations under each of these hypotheses.

The distribution in ionization yield of the nuclear recoils can be
compared to the expected distribution.  The normalized deviation,
$\ystar$,
is defined by
\begin{equation}
\ystar \equiv \frac{Y - Y_{\mathrm{NR}}(\er)}
                   {\sigma_{\mathrm{NR}}(\er)}  ,
\end{equation}
where $Y_{\mathrm{NR}}(\er)$ is the expected ionization yield of a nuclear 
recoil and $\sigma_{\mathrm{NR}}(\er)$ is the
standard deviation of $Y$ for nuclear recoils, 
both functions of $\er$.  
The usefulness of $\ystar$ is that it puts nuclear recoils at different
$\er$ on the same footing.  In the absence of cuts in $Y$ defining
the acceptance region, the expected distribution is a simple Gaussian
with mean $\mu=0$ and standard deviation $\sigma=1$.  
The ionization-threshold cut that defines the
nuclear-recoil band truncates the distribution in an $\er$-dependent
manner that is calculated for each of the 23 QIS (13 QI) 
single-scatter nuclear recoils.
Figure~\ref{kstests}c shows the expected and actual distributions.  
The KS test
indicates that 76\% (77\%) of experiments should
observe distributions that deviate further from the expected
distribution.  This level of agreement is important because
misidentified electron events would be expected to have a distribution
either flat in $Y$ or weighted toward high $Y$.

The single-scatter nuclear-recoil candidate events are consistent in every way 
with being nuclear recoils, and the expected contamination from 
misidentification is only a few events, even under the conservative 
assumption that there are no neutrons in the electron-calibration data 
set.
It therefore appears that the nuclear-recoil candidates are mostly, if 
not entirely, actual nuclear-recoil events.
In order to set a conservative upper limit on the number of WIMPs in 
the data set, we will assume that all these nuclear-recoil candidates 
are nuclear-recoil events.

\section{\label{sect:n}Estimate of Neutron Background}

\begin{figure}
\psfig{figure=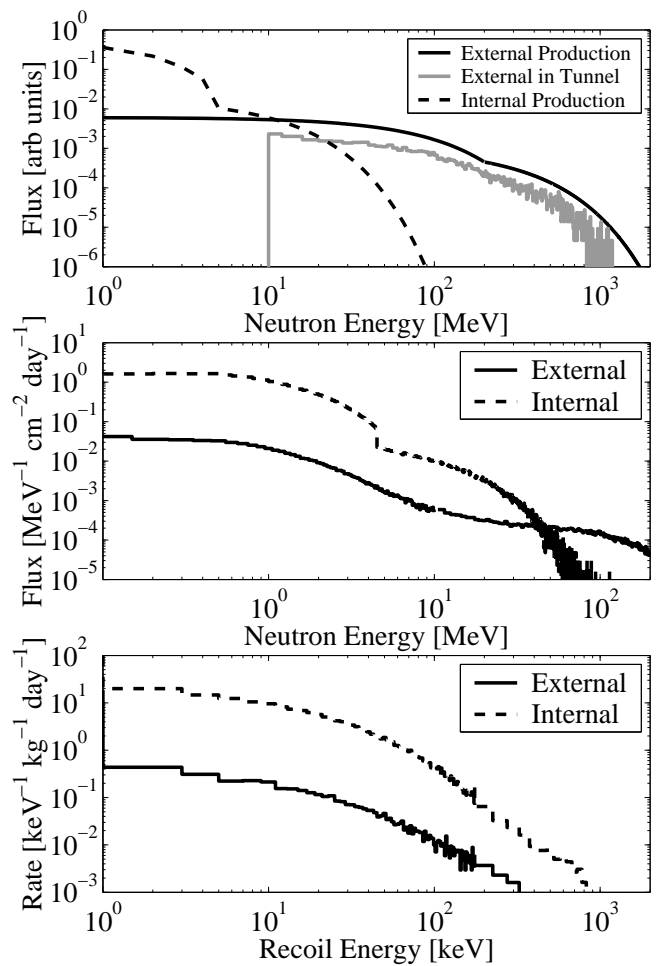,width=3.4in}
\caption{\label{prodspec}
Top: Arbitrarily normalized expected production spectra 
of internal (dashed curve) and external (solid curve) neutrons.
The resulting simulated spectrum of external neutrons 
after propagating through the tunnel rock (grey solid 
curve) is cut-off artificially at 10~MeV.
Neutrons below this energy are unimportant because 
a negligible number of lower-energy neutrons 
penetrate the experiment's shielding.
Middle: Expected spectra of internal and external neutrons incident on the 
detectors.  Below about 4~MeV, the two spectral shapes match closely.
Bottom: Resulting simulated recoil-energy spectra in Ge for both internal and 
external neutrons.  Note that an incident neutron can impart at most 
$1/18$ of its energy to Ge in a single elastic scatter.
Despite the extremely different production spectra of the 
primary neutrons, the recoil-energy spectra below 100~keV are nearly identical,
as explained in the text.
}
\end{figure}

As described in Sec.~\ref{expBackgrounds}, 
a significant unvetoed neutron background is expected due to neutrons 
produced outside the muon veto by high-energy photonuclear and 
hadronic shower processes induced by cosmic-ray muons.
The expected production spectrum 
\begin{equation}
dN(E) \propto \left\{ \begin{array}{lcc}
                      6.05\ \exp(-E/77\ {\mathrm{MeV}})\ dE &  &
                      E < 200\ {\mathrm{MeV}} \\
                      \exp(-E/250\ {\mathrm{MeV}})\ dE &  &
                      E > 200\ {\mathrm{MeV}}
                     \end{array} \right.   
\label{extneutspec}
\end{equation}
is shown in the top graph of Fig.~\ref{prodspec}. 
The spectrum is based on a compilation of measurements
shown in Fig.~4 of~\cite{khalchukov1983},
whose authors note that ``the spectra do not depend on the
projectile ($\pi$, p, n, $\gamma$) and its energy provided the latter
is greater than 2~GeV.'' Hence, this single two-component
spectrum is used for the 
high-energy photonuclear and hadronic shower processes.  
The production rate of 4~\iru, 
which would yield an integral flux of these neutrons into the
tunnel of \scinot{2}{-6}~\percmsqpersec, is quite uncertain; 
the true production rate and flux could be as much as two times larger or
smaller.
Monte Carlo simulations of the CDMS experiment 
indicate that $\sim$40\% of these externally 
produced neutrons are tagged as muon-coincident due to their 
interactions in the veto scintillators.  However,
additional uncertainty arises because an unknown fraction of the hadronic
showers associated with neutron production may also trigger the veto.
Furthermore, the energy spectrum may differ somewhat from that given 
in Eq.~(\ref{extneutspec})
due to contributions from projectiles with energies $<2$~\gev.
Due to these uncertainties in both the rate and the energy spectrum,
no quantities that depend significantly on the neutron production 
spectrum should be considered reliable for neutron background 
estimation.

Fortunately, the low-energy spectrum of neutrons incident on the 
detectors due to these high-energy external neutrons 
does not depend significantly on the details of the 
production spectrum.
The low-energy part of the incident spectrum, 
made up of secondary and tertiary neutrons, is evaporative,
just like the spectrum of low-energy neutrons resulting from negative muon
capture~\cite{khalchukov95}.  
For this reason, the incident spectrum due to external neutrons
(shown in Fig.~\ref{prodspec})
is essentially the same at low energies ($<5$~MeV)
as that due to the veto-coincident, ``internal'' neutrons which, as
explained in Sec.~\ref{expBackgrounds}, arise from negative muon capture and
low-energy photonuclear interactions of muons within the shield. 
While the internal neutron spectrum is taken from the 
literature~\cite{Singer,dasilvathesis}, the
incident spectrum due to high-energy external neutrons is obtained by
simulating the propagation and showering of these neutrons within the
shield.
Good agreement at low energy between the two spectra indicates that
secondary production is well simulated.  
Studies of simulations confirm that 
the spectrum of secondaries at the detectors is largely
insensitive to features in the primary spectrum~\cite{TAPthesis}.  
The spectral shape of
primaries affects only the absolute rate and the high-energy tail 
($\agt 5$~MeV) 
of the incident energy spectrum of the secondary neutrons.

The detector recoil-energy spectra in the range of interest ($< 100$~keV) 
are dominated by interactions with low-energy neutrons ($\alt 5$~MeV)
due to simple kinematics
and the suppression of neutron cross-sections at high energy.  Therefore,
the expected recoil-energy spectra below 100~keV due to external and internal
neutrons are almost identical in shape, as shown in Fig.~\ref{prodspec}. 
The predicted spectral shape of all neutron interactions is therefore 
insensitive to the relative numbers of interactions arising from 
neutrons that originate internally versus externally.
Other normalization-independent predictions include
the fraction of neutrons that scatter in multiple detectors,
and the relative rates of neutron interactions in Ge and Si.
These results are also nearly independent of the primary neutron spectrum 
and are almost the same for internal and external neutrons.  
Only these 
normalization-independent quantities are used to estimate the neutron
background in the low-background data.  

Comparison of Monte Carlo results with the calibration and internally-produced 
neutron data sets provides
checks of the accuracy of the neutron simulations,
particularly for these normalization-independent quantities, 
as well 
as checks of the efficiency calculations described in Sec.~\ref{effcheck}.
As discussed in Sec.~\ref{multeff},
calculation of the efficiency for multiple-scatter events is nontrivial due to
correlations in the cuts for detector combinations. 
Estimates of the systematic uncertainty of these efficiency 
calculations combine to give an overall systematic uncertainty of 8\% on the 
expected measured fraction of neutron interactions that are identified as 
multiple scatters.
These uncertainties are
due primarily to the 10\% uncertainty on the fiducial-volume 
efficiency at low energies
(which results in a 5\% uncertainty 
on the expected fraction of neutrons identified as multiple scatters), 
and a possible 5\% uncertainty on the correlated efficiencies 
discussed in Sec.~\ref{multeff}.

Studies of the Monte Carlo simulation, including comparisons to 
standard cross sections and to results from GEANT4 simulations,
indicate that inaccuracies 
in the Monte Carlo simulation should not cause an error on the 
predicted neutron multiple-scatter fraction larger than 10\%.
In particular, a negligible error should result from
the fact that the simulation ignores the possibility 
that an external neutron may be accompanied by other external 
neutrons from the same shower.
Using an approximate muon energy spectrum~\cite{Cassiday} and
muon ionization loss~\cite{Lohmann}, along with results of a calculation
of neutron yield and multiplicity distribution per muon~\cite{YFWang},
we find that a neutron generated at SUF depth 
by a muon with energy $>10$~GeV 
is accompanied on average by only 10 other neutrons in the same shower.  
This average is not very sensitive to the low-energy cutoff in
muon energy.  
Because our Monte Carlo simulation shows that external neutrons 
reaching the experimental shielding
have only a $10^{-4}$ probability of hitting
a detector,  
the neutron production multiplicity has a negligible effect on
the probability of detecting multiple scatters.
Furthermore, a simple calculation assuming an
isotropic neutron flux, isotropic elastic scattering, and an appropriate
interaction cross section, 
verifies the multiple-scatter fractions predicted by the Monte Carlo
simulation
for the simple case of the neutron calibration.
Combining the uncertainty on the efficiencies with the possible 
systematic error of the Monte Carlo simulation
results in an overall systematic uncertainty on this fraction of 13\%.

\begin{figure}
\begin{center}
First Neutron Calibration\\    
\psfig{figure=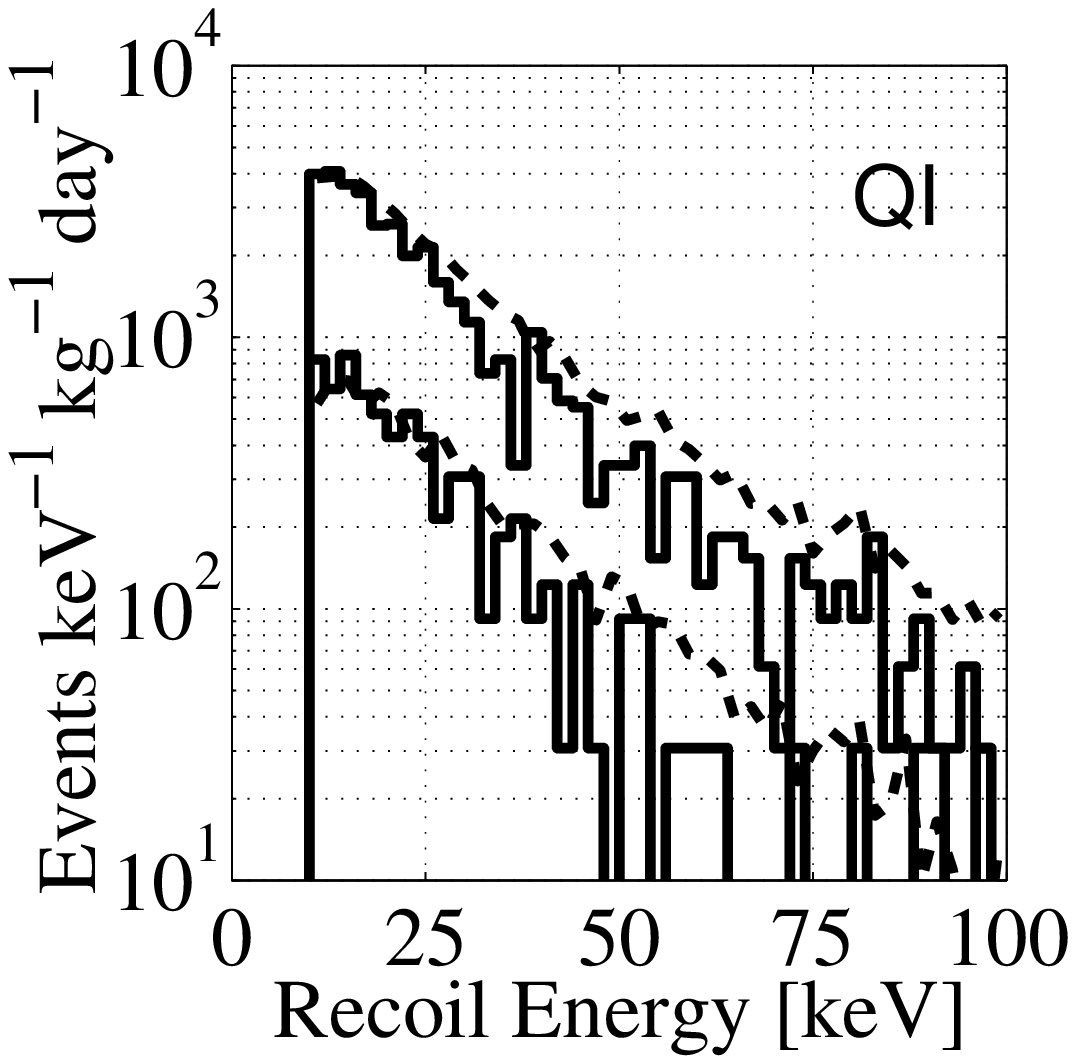,width=1.65in}
\psfig{figure=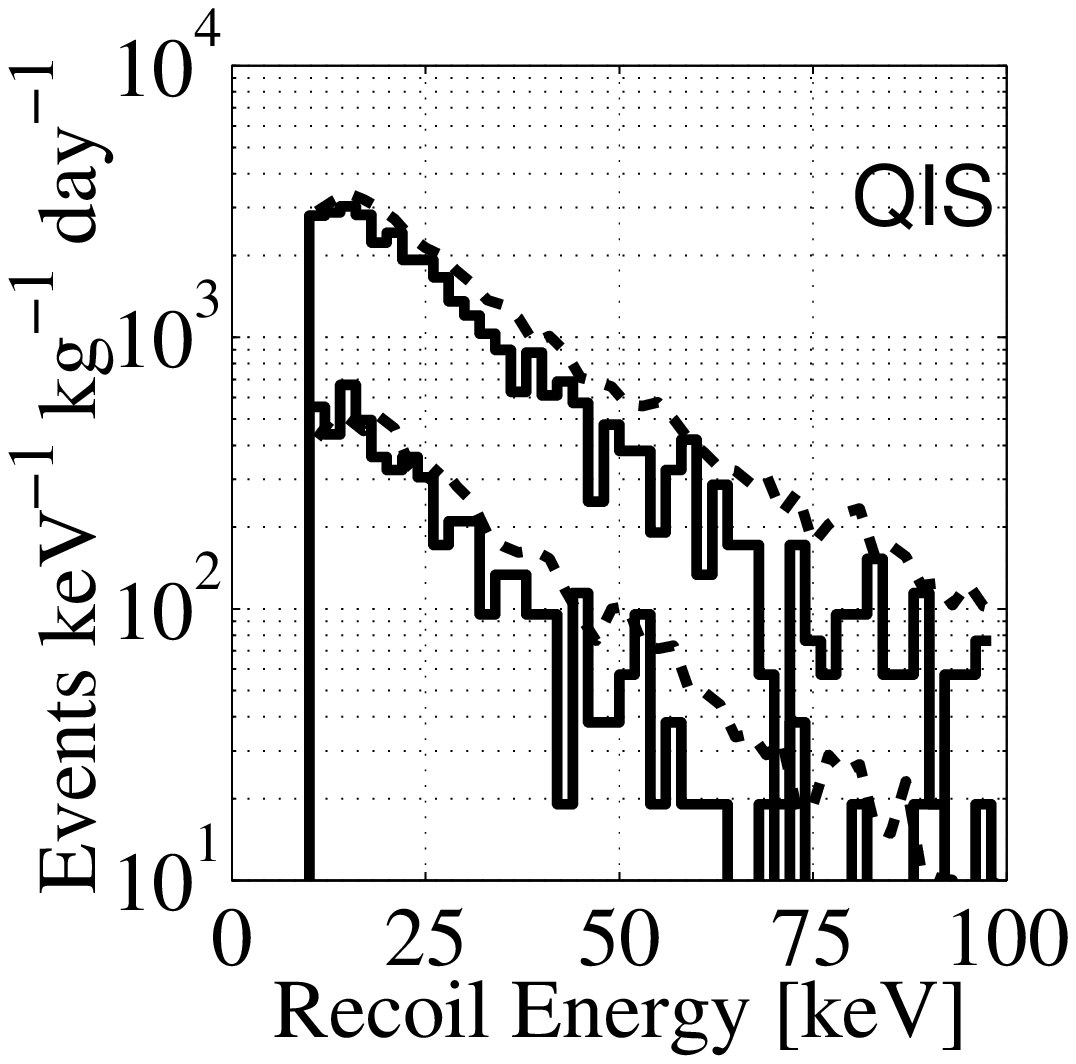,width=1.65in}\\
Second Neutron Calibration\\
\psfig{figure=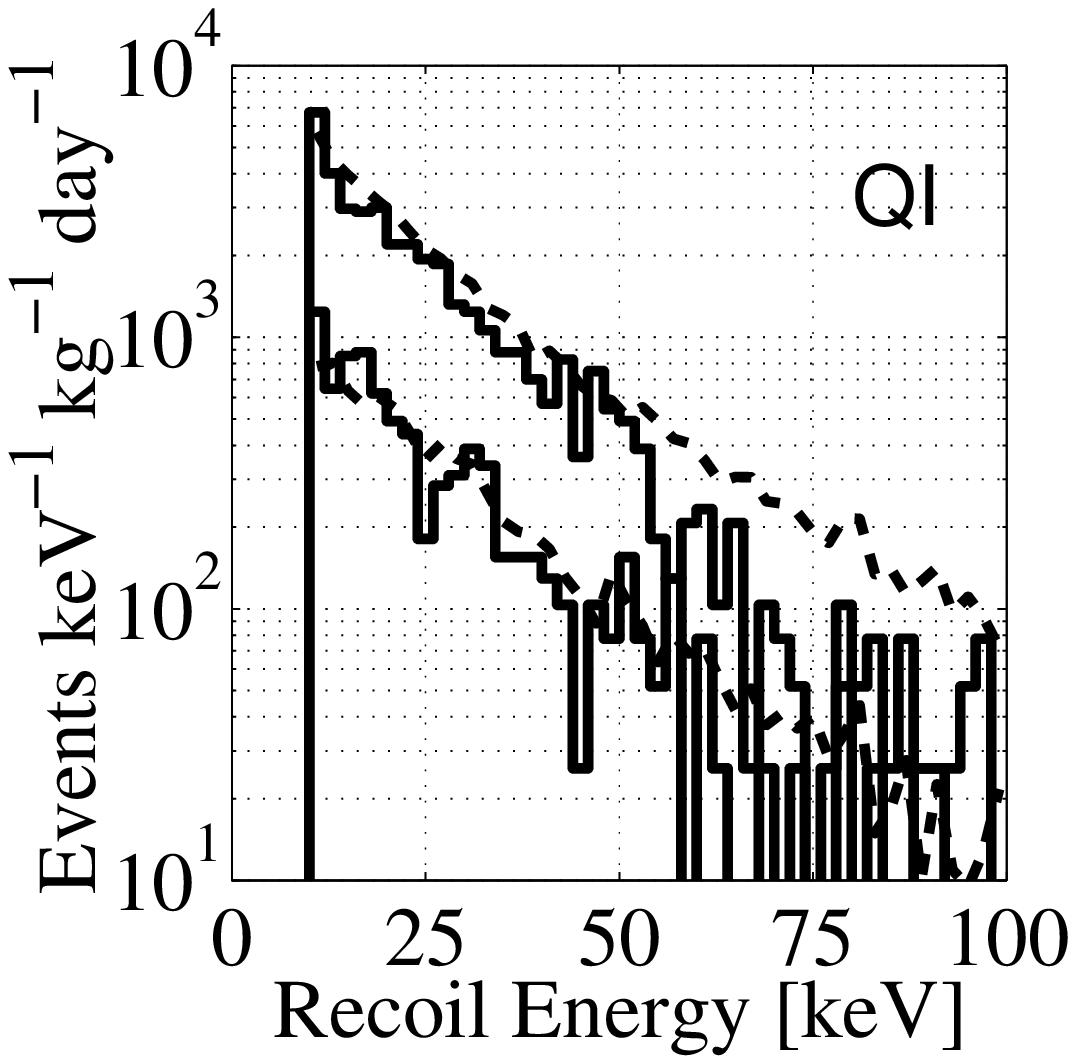,width=1.65in} 
\psfig{figure=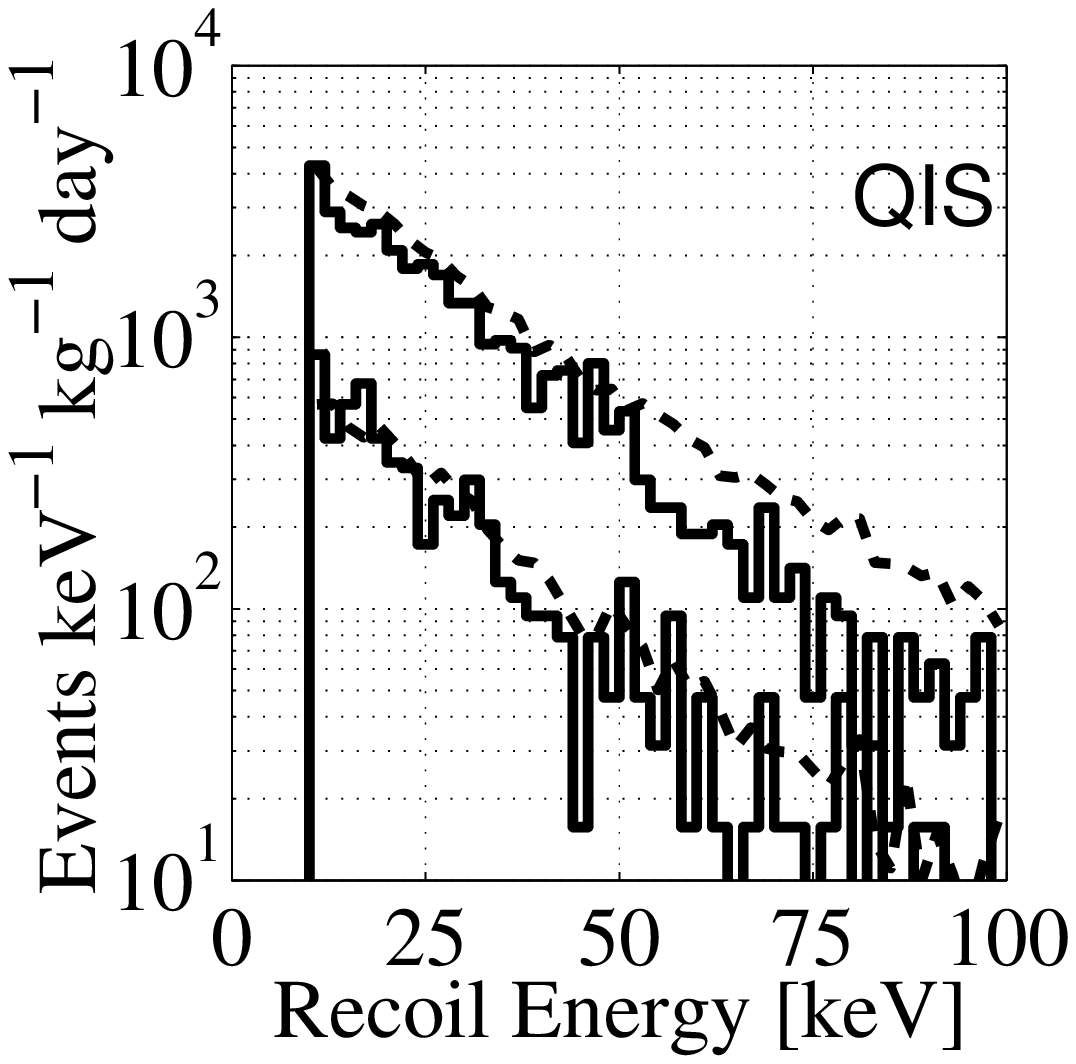,width=1.65in} \\
Veto-Coincident Neutrons\\
\psfig{figure=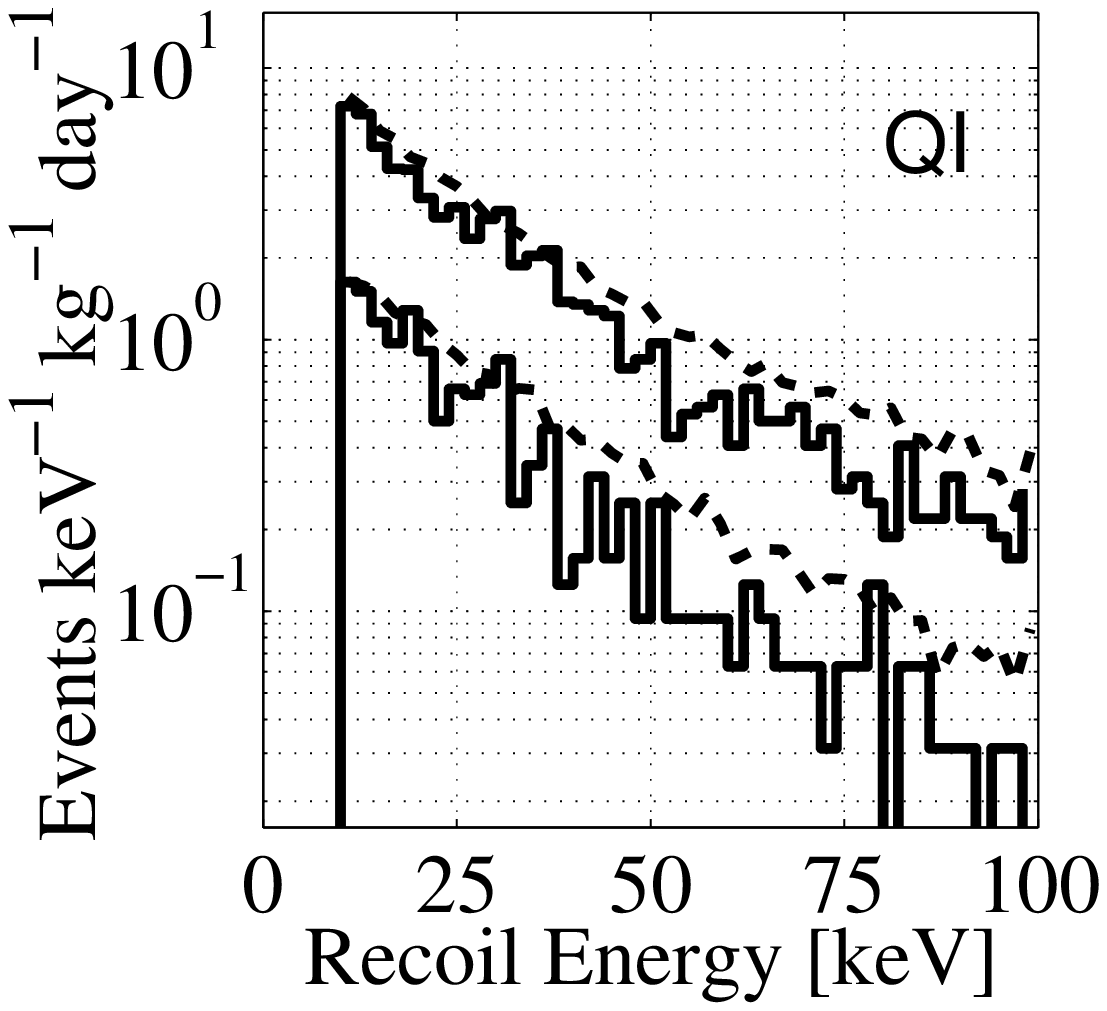,width=1.65in}
\psfig{figure=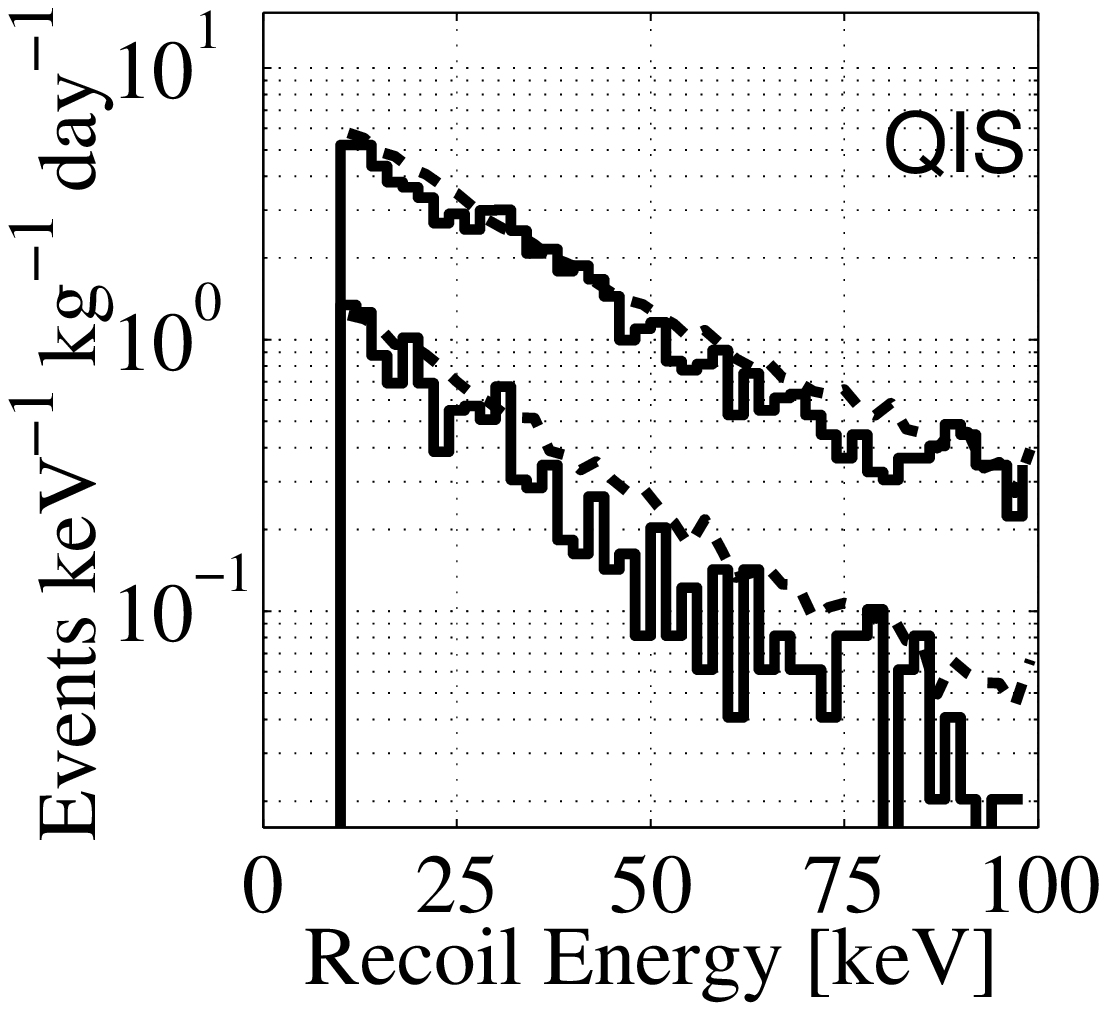,width=1.65in} 
\caption{\label{cf252multdata}
Observed and simulated 
neutron-calibration and veto-coincident spectra, 
coadded over detectors, with no free parameters.  
In each plot, spectra both for all scatters (top) and for multiple-scatters 
(bottom) are shown, for both data (solid) and simulations 
(dashes).  
Figures in the left column show events with at least one QI scatter; 
figures in the right column show events with at least one QIS scatter.
Top: first neutron calibration.
Middle: second neutron calibration.
Bottom: veto-coincident (internal) neutrons.
The calibration data is coadded over all four detectors; the 
veto-coincident data is coadded over BLIPs 4--6.
}
\end{center}
\end{figure}

\begin{table}
\begin{center}
\begin{ruledtabular}
\begin{tabular}{l|rrr} 
Source  & Ge Singles & Ge Multiples & Si Singles \\ 
\hline 
\multicolumn{4}{l}{Internal} \\ \hline 
copper          &  72 (76) &  8 (6) &  142 (177) \\ \hline 
1998 inner lead &          &        &  125 (155) \\ 
1999 inner lead &  75 (79) &  8 (6) &            \\ \hline 
outer lead & {$\sim 6$} (6) &
             {$\sim 0.8$} (0.6) &  
             {$\sim 11$}   (14)   \\ \hline 
total           & 153 (161) & 17 (13) & 278 (346)\\ \hline
\multicolumn{4}{l}{External} \\
\hline 
rock       & 3.0 (3.2) & 0.3 (0.2) & 5.0 (6.3) \\ 
\end{tabular}
\end{ruledtabular}
\caption[Simulated neutron-interaction rates]
{Expected rates of neutron interactions per kg-day between 10-100~keV 
(20-100~keV) for Ge (Si) detectors at SUF.  Numbers in parentheses
indicate the rates expected for ideal detectors with 
energy-independent efficiency,
no dead periods, and both hits of a multiple-scatter required to be in
the fiducial volume 
(the last requirement causes the rate of multiple-scatters to 
be smaller for these ``ideal'' detectors than for the actual detectors).  
As discussed in the text, the expected rate of external neutrons is quite 
uncertain.  The rate of internal neutrons is much better determined, 
with systematic uncertainties $\sim10$\%.  
Only the prediction for neutrons from the outer lead has a 
significant statistical uncertainty ($\sim25$\%). 
Because the mass of the inner lead shield 
was increased between the 1998 Si data run and the 1999 Ge data run, 
the fraction of interactions due to neutrons produced in
the inner lead is slightly greater 
for the Ge detectors than for the Si detector.
}
\label{neutbgndtable}
\end{center}
\end{table}

Based on the neutron simulations, Table~\ref{neutbgndtable} shows the 
expected neutron-background rates. 
The simulated and observed multiple-scatter-neutron spectra are shown in
Fig.~\ref{cf252multdata}.  
All recoils of a
multiple-scatter event are required to be between 10 and 100~keV 
for the event to pass cuts.
Each histogram is filled for each recoil of a
multiple-scatter event; \eg, a double scatter adds two entries to the
histogram.  
For the neutron calibrations, 
the simulation predicts a 20\% higher overall rate than is observed,
along with a slightly harder energy spectrum than is observed.
For the veto-coincident neutrons, comparisons are hampered by the 
fact that 
the fraction of neutrons coincident with other muon-induced particles 
is unknown.
Accurate measurement of the rate of these coincidences is complicated 
by the fact that interactions of several~MeV
in one detector produce crosstalk of $\sim$10~keV in neighboring detectors,
potentially making electron-recoil events indistinguishable from 
neutron-induced events.
These problems, combined with the fact that the production of the 
muon-induced particles other than neutrons is not as yet simulated,
results in a 20\% systematic uncertainty on the measured rate of 
veto-coincident neutrons, and a 20\% systematic uncertainty on the 
measured fraction of neutrons that multiply scatter. 

\begin{table}
\begin{ruledtabular}
\begin{tabular}{lccc} 
              &   First            &  Second           & Veto-         \\
              &   Neutron          &  Neutron          & Coincident \\
Event Set     &   Calibration      &  Calibration      & Neutrons \\
\hline
all QI NRs       &    0.82            &     0.80          &  0.81  \\
multiple QI NRs  &    0.86            &     0.93          &  0.73  \\
all QIS NRs      &    0.79            &     0.77          &  0.88  \\
multiple QIS NRs &    0.86            &     0.91          &  0.77   \\
\end{tabular}
\end{ruledtabular}
\caption{\label{nspecacc}
Scaling factors that must be applied to the results of the 
simulation to match the total rates observed in BLIPs~4--6.  
Data sets include both QIS and QI nuclear recoils (NRs),
and multiple scatters with at least one QI scatter 
(``multiple QI NRs'') 
and those with at least one QIS scatter (``multiple QIS NRs'').
Statistical uncertainties are 6--7\% for multiple-scatters and
2--3\% for all events.
As can be seen, the overall rates predicted are accurate to $\sim20$\%,
and the predicted fractions of events that are multiple scatters
are accurate to $\sim10$\%.
}
\end{table}

Table~\ref{nspecacc} lists the overall scale factors by which 
the simulated spectra must be scaled to match the data.  
Comparisons of the ratios of single-scatter events to
multiple-scatter events for the calibration and internally produced 
neutrons provide checks of the accuracy of the prediction of the same 
ratio for 
veto-anticoincident neutrons.  
For each data set, the ratios agree 
with those predicted to within the combined systematic and statistical
uncertainties.
The good agreement between data and the results of the Monte Carlo 
simulations builds
confidence in the predictive power of 
using normalization-independent results of the Monte Carlo simulation
for estimating the
external neutron background.  
The predicted ratios of the different classes of neutron events,
together with the observed number of Ge multiple-scatter neutrons and the number
of neutron events in the Si detector, 
should provide a dependable estimate of the expected number of neutron 
single-scatters in the Ge data set.

\subsection{\label{antidouble}Ge multiple-scatter data set} 

\begin{figure}
\psfig{figure=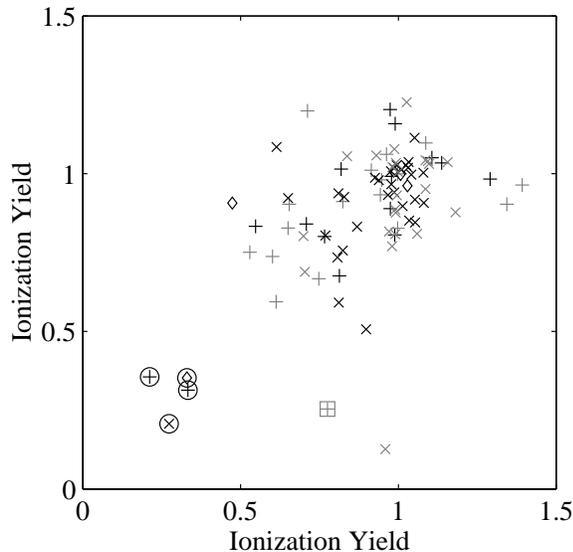, width=7.7cm}
\caption{\label{multiy2}
Scatter plot of ionization yields for veto-anticoincident 
multiple scatters in
the 3 uncontaminated Ge detectors with at least one QI
(black) or QS (grey) 
scatter and with both scatters between 10 and 100~keV.  
Events are double-scatters in BLIP4 and BLIP5 (the top and middle 
uncontaminated detectors, $+$), 
in BLIP4 and BLIP6 (the top and bottom uncontaminated detectors, $\diamond$),
or in BLIP5 and BLIP6 (the middle and bottom uncontaminated detectors, $\times$).
The ionization yield of the higher-numbered detector is plotted on 
the $x$ axis.
Circled events are tagged as nuclear recoils in both detectors.  
The boxed event is tagged as a nuclear recoil in only BLIP4.
Bulk recoils and surface events lie at $Y \simeq 1$ and $Y
\sim 0.75$, respectively.
Both events with ionization yield $Y<0.45$ in only one of the two 
detectors hit have the low-yield hit in the outer electrode, 
consistent with expectations for misidentification of electron recoils 
in the outer electrode.}
\end{figure}

Figure~\ref{multiy2} displays a scatter plot of ionization yields 
in one detector versus those in another for 
low-background multiple scatters.  
The four Ge multiple-scatter nuclear-recoil candidates 
should all be multiple-scatter neutrons.  
WIMPs interact too weakly to multiply scatter.
It is also
highly unlikely that these events are misidentified low-energy
electron events.  Figures~\ref{highbiasyplot} and \ref{multiy2}
demonstrate excellent separation of low-energy electron events from
nuclear recoils. 
As shown in Fig.~\ref{kstests}d, the multiple-scatter nuclear-recoil 
candidates have $\ystar$ values consistent 
with those expected for nuclear recoils
(a KS test indicates 9\% of experiments should result in a 
distribution less similar to expectations).
Finally, three of the events have both hits with energy in the inner 
electrode, consistent with expectations for neutrons.
If these events were due to misidentification of electron-induced 
events, more hits would likely be in the outer electrode
since misidentification 
occurs much more often for hits in the outer electrode, as shown
in Fig.~\ref{ygammacalhist}.

The expected number of misidentified multiple-scatter electron recoils
may be estimated quantitatively.
As described above, BLIP3/BLIP4 multiple scatters with too little 
ionization in both BLIP3 and BLIP4 to be photons  
may be used as a
low-statistics electron calibration. 
Of the 216 hits tagged as electrons (or neutrons) in BLIP3 or BLIP4, only 4 
pass the nuclear-recoil cut,
so the expected fraction of electron misidentification $\beta_{\beta}= 
4/216$ under the conservative assumption that none of the hits are 
neutrons.
In using the electron calibration to
estimate the number of double-scatter nuclear-recoil candidates
arising from misidentified electrons, it is important to make use of
the fact that, while the double-scatter electrons do cluster around
$Y\sim 0.75$, there is no correlation between the two 
detectors'
deviations from this central value of the ionization yields, as seen 
in
Fig.~\ref{b3b4y2plot} --- the
electron events do not form a line with slope~1.  In order to be
misidentified as a double-scatter neutron, a double-scatter electron
must therefore be misidentified in {\em both} detectors; such
misidentification is suppressed by a factor $\beta_\beta^2$ rather
than only $\beta_\beta$.

The lack of correlation between the ionization yields in the two detectors 
is expected because
energy deposited in the first detector is not a strong function of the electron
energy --- it depends on the track length in the crystal, which may be
short for a high-energy electron if it is backscattered.  The
ionization yield is, however, 
well correlated with the
track length: shorter tracks are also likely to be more shallow.
Thus, for double-scatter electrons, the ionization yield for one
scatter, while correlated with the deposited energy, may not be
a good predictor of the actual electron energy, and thus may not be a
good predictor of the ionization yield observed in the second recoil.

As shown in Fig.~\ref{multiy2}, most veto-anticoincident double 
scatters between BLIPs 4, 5 and 6 appear to be photons, with 
ionization yield $Y\sim 1$ for both hits.  
Note that most multiple-scatter photon events do not appear on this plot,
either because energy is deposited in three or more detectors, 
or because at least one energy deposition is outside 
the 10--100~keV energy range.
Monte Carlo simulations of generic sources of radioactive 
contamination, such as U/Th in the detector housing, 
suggest that for every single scatter resulting in a 
recoil between 10--100~keV, there are $\sim$0.07
double scatters with both recoils between 10--100 keV, 
and there are an additional $\sim$0.6 multiple-scatter events. 
The fraction of photon events that appear as double scatters appears
consistent with expectations from these simulations if one takes into 
account the large number of 10.4~keV photons unlikely to multiple-scatter.

There are also 16 events with both hits having ionization yield $Y$ lower than 
typical photons, and an additional 21 events with one of the two hits 
having lower $Y$ than typical photons.
To be conservative, we count the total number of 
$16\times2+21=53$~low-$Y$ hits as 
yielding an effective $N_{\beta}=26.5$ double-scatter surface-electron events.
The expected number of misidentified 
surface-electron-recoil double-scatter events is therefore only 
$N_{\beta}\beta_\beta^2 = 26.5 \times (4/216)^{2}=0.009$.
The upper limit at the 90\% confidence level on the number of
double-scatter electrons expected to be misidentified as 
double-scatter neutrons is $b_{\mathrm{d}}= 0.05$ events.  
Even if the misidentification were somehow correlated between the two 
detectors, the expected number of misidentified 
electron-recoil hits would be only $N_{\beta}\beta_\beta = 26.5  \times 
(4/216)=0.5$,
again under the conservative assumption that neither of the 
calibration-set nuclear-recoil candidates are neutrons.  
Misidentified electrons provide truly negligible contamination
of the four neutron multiple-scatter 
events.
The Ge multiple-scatter data therefore provides a reliable estimate of 
the neutron background.

\subsection{\label{silicon}Si data set} 

\begin{figure}
\psfig{figure=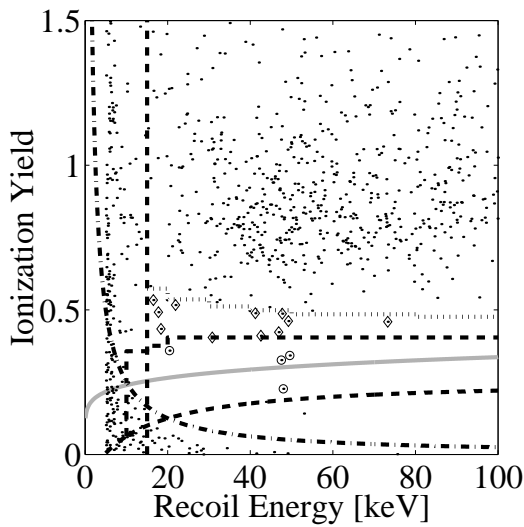,width=2.75in}
\caption{\label{alexyplot}
1998 Si ZIP detector veto-anticoincident data after cuts.
Four nuclear-recoil candidate events (circled) lie near the
center of the nuclear-recoil band (light solid curve), 
within the nuclear-recoil-acceptance region (bordered by dashed curves),
and above both the ionization threshold (dot-dashed curve) 
and nuclear-recoil analysis threshold (vertical dashed line).
Eleven additional events (diamonds),
of which $\sim$1 should be a nuclear recoil,
lie in the band (bordered by the dotted curve) 
just above the nuclear-recoil band.
These 11 events are consistent with the expected 
distribution of 
surface events based on \insitu\ calibrations with photon sources.
Events below the ionization threshold are likely dominated by events 
with poor charge collection in the outer ionization electrode.
Events with recoil energies $E_{\mathrm{R}}<5$~keV are not shown.
}
\end{figure}

An earlier run consisting of 33~live
days taken with a 100~g Si ZIP detector between April and July, 1998,
also measured the neutron background.  The Si run yields a
1.5~kg~d exposure after cuts.  The total low-energy electron
surface-event rate is 60~kg$^{-1}$~d$^{-1}$ between 20 and 100 keV.
As shown in Fig.~\ref{alexyplot}, 
four nuclear-recoil candidates are observed in the Si data set.  
Detailed analysis of this data is
described elsewhere~\cite{alexsheffield,clarkethesis}.

\begin{figure}
\psfig{figure=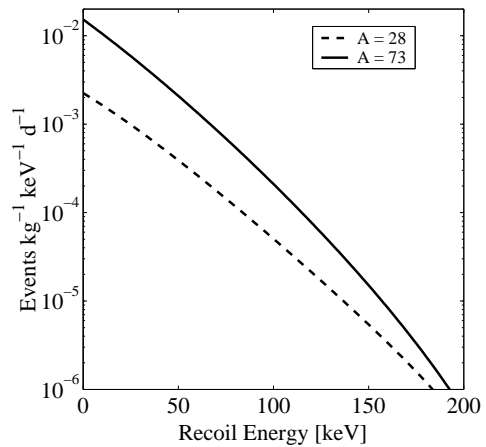,width=2.5in}
\caption{\label{Avariation}
Expected differential recoil-energy
spectra for Si
($A = 28$) and for Ge ($A = 73$), for a 100-\gev\ WIMP 
with WIMP-nucleon cross-section $\sigma = 10^{-42}$~cm$^{2}$
under standard assumptions listed in Sec.~\ref{sectlikelihood}.}
\end{figure}

The four nuclear-recoil candidates observed in the 1998 Si ZIP data
cannot be WIMPs: whether their interactions with target nuclei are
dominated by spin-independent or spin-dependent couplings, WIMPs
yielding the observed Si nuclear-recoil rate would cause far more
nuclear recoils in the Ge data set than were observed.  
The WIMP-nucleus cross-section
scales as $A^2$ for WIMPs with spin-independent interactions.
Expected recoil-energy spectra in Ge and Si for a WIMP
with spin-independent interactions are shown in
Fig.~\ref{Avariation}.
Ge and Si differ by a factor of 5 to 7 in
differential rate between 0 and 100~keV.  
After including the effects of energy thresholds and efficiencies,
one expects of order 
90 (70) times the number of WIMPs in the 15.8~kg~d QIS (11.9~kg~d QI) 
Ge data set as in the 1.5~kg~d Si data.
The argument is more
complicated for spin-dependent interactions, but it also holds that
there should be many more nuclear recoils in the 1999
Ge data set than are observed.  Furthermore, the
spin-dependent cross section corresponding to the observed Si event
rate is significantly larger than expected from the MSSM.

It is possible, however, that not all of the Si nuclear-recoil candidates 
are neutrons. 
As shown in Fig.~\ref{alexyplot},
the separation between the nuclear-recoil band and the electron-recoil 
band is not as large for the Si data as it is for the Ge data.
A calibration of the Si detector with a \c14\ electron 
source at a test facility provides a high-statistics estimate of the possible 
electron contamination.
Based on the statistical uncertainties of this calibration, 
the upper limit on the expected number
of unrejected surface events is 0.26~events (90\%~CL).  
However, the systematic uncertainties are larger, since this 
calibration was made with a collimated source 
and was taken under different conditions 
than the low-background data.
A simple and conservative estimate of the contamination is made using 
data taken with a $^{60}$Co photon source at SUF
under essentially the same conditions as the low-background data.
Assuming that all events passing nuclear-recoil cuts are due to the 
small number of electrons present in the calibration sample leads to an 
expectation of 2.2~low-background contamination events and an
upper 
limit of 7.3~expected low-background contamination
events at the 90\% confidence level.
For comparison, this assumption results in 13~(an upper 
limit of 17)~events expected in the 
band just above the nuclear-recoil band below 30~keV,
and 4.9~(an upper limit of 8.8)~events 
expected in this band above 30~keV.
As shown in Fig.~\ref{alexyplot}, these predictions are
in agreement with the 11~events in that band.

The measurement of the unvetoed neutron background from the 1998 Si data set 
is consistent with the measurement from the Ge multiple-scatter data set. 
However, the large systematic uncertainty on the Si data 
means the Ge data set dominates our combined measurement.  
We note that new Si and Ge ZIP detectors~\cite{tarekltd9} 
perform significantly better than the Si ZIP of the earlier design 
used in 1998.

\subsection{\label{sect:Nconsist}Neutron consistency tests}

The fact that the observed 
number of single-scatter nuclear-recoil events in Ge is about as large 
as
the expected background
suggests that all 
such events may be due to neutrons.  Although this possibility is of 
course not 
assumed in calculating limits on the 
WIMP-nucleon cross section,
it is important to test the consistency of this possibility.

\begin{figure}
\psfig{figure=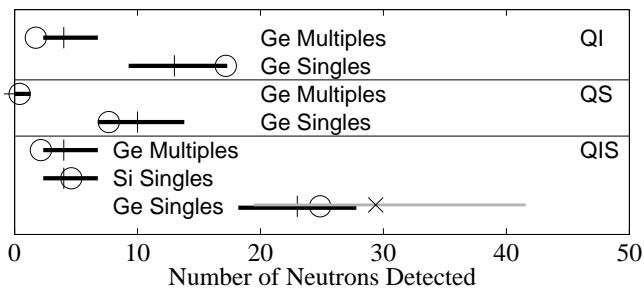,width=3.4in}
\caption{\label{mcvsdatarates}
Schematic comparison of predicted numbers of neutrons to observed 
numbers (crosses), with Feldman-Cousins 68\% CL confidence 
intervals~\protect\cite{feldmancousins} (dark lines).
Predictions are made by normalizing the simulation 
by the neutron background that best fits 
$N_{\mathrm{s}}$, $N_{\mathrm{d}}$,
and $N_{\mathrm{Si}}$ jointly (circles).
An additional prediction for QIS Ge singles 
($\times$, with light line indicating the 68\% CL confidence interval) 
is based on the
neutron background that best fits 
$N_{\mathrm{d}}$ and $N_{\mathrm{Si}}$ jointly.
Top: inner-electrode-contained (``QI'') events.  
Middle: shared-electrode (``QS'') events.
Bottom: events that are either contained in the inner electrode or
shared between the electrodes (``QIS events''),
together with Si events.}
\end{figure}

In fact, there
is good agreement between predictions from the Monte Carlo
simulation and the relative observed numbers of $N_{\mathrm{d}}= 4$ QIS 
(4 QI) Ge double
scatters, $N_{\mathrm{Si}}= 4$ Si single scatters, 
and $N_{\mathrm{s}}=23$ QIS (13 QI) Ge single scatters.
Schematically,
the data and simulation can be compared in two ways: by normalizing the
simulation by the neutron-background rate that best fits $N_{\mathrm{s}}$,
$N_{\mathrm{d}}$,
and $N_{\mathrm{Si}}$ jointly; 
or by normalizing by the neutron-background rate
that best fits $N_{\mathrm{d}}$ and $N_{\mathrm{Si}}$ 
and predicting $N_{\mathrm{s}}$.  
The latter is
the intuitive interpretation of using the Ge doubles and Si events to
predict the neutron background in the Ge singles set.  These
comparisons are shown in Fig.~\ref{mcvsdatarates}.

More rigorously, a likelihood-ratio test can be used to compare
the default hypothesis, that the 
$N_{\mathrm{s}}$, $N_{\mathrm{d}}$, and $N_{\mathrm{Si}}$
events are due to
a neutron background with relative rates given by the simulation,
to an alternate hypothesis, that the three event sets arise from three different
background sources.  Effectively, the latter hypothesis
corresponds to three arbitrary background sources for the three event
types, the most general possible hypothesis.  
This test indicates that a neutron
background should result in a less likely combination of Ge QIS (QI) single
scatters, Ge QIS (QI) multiple scatters, and Si single scatters $\gtrsim 
48\%$ (21\%)
of the time, with only weak dependence on the assumed true neutron
background~\cite{golwalathesis}.  
The self-consistency of the division of the neutrons into their five 
categories can also be tested. 
A neutron
background should result in a less likely combination of Ge QS single
scatters, Ge QI single scatters, Ge QS multiple scatters, 
Ge QI multiple scatters,
and Si single scatters $\gtrsim 30\%$ of the time.

\begin{figure}
\psfig{figure=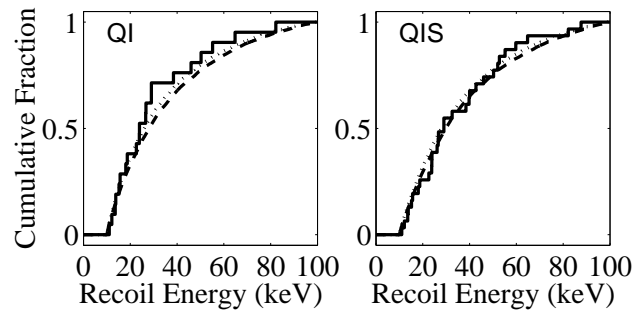,width=3.3in}
\caption{\label{neutspecdata}
Observed Ge nuclear-recoil integral
recoil-energy spectra (solid), 
including single-scatter and multiple-scatter hits,
for QI events (left) and QIS events
(right).  Observed spectra agree well with 
expectations from either
the external-neutron (dashed curves) or the internal-neutron (dotted 
curves) simulations.}
\end{figure}

Finally, as shown in Fig.~\ref{neutspecdata},
the observed nuclear-recoil spectral shape is consistent with 
expectations for neutrons whether the neutrons are 
produced internally or externally to the veto;
recall that the expected internal and external neutron recoil-energy spectra
should be similar because the recoil-energy spectrum is fairly independent
of the high-energy tail of the external-neutron spectrum.
Kolmogorov-Smirnov tests indicate that
the deviation between the observed and simulated nuclear-recoil spectral
shapes 
using the QIS (QI) events
should be larger in 86\% (39\%) of experiments for external neutrons,
and the deviation should be larger in 61\% (67\%) of experiments for 
internal neutrons.
These results should be taken only as support
for the consistency of the data with the neutron simulation; 
they do not alone disfavor an interpretation that 
some (or even all) events may be due to WIMPs.
The spectra are also
consistent with a combination of WIMPs and neutrons,
or with WIMPs alone if the WIMP mass $M \gtrsim 100$~\gev.

\section{\label{ConRegion}Calculating the Confidence Region}

The 90\%~CL excluded region for the WIMP mass $M$ and 
WIMP-nucleon cross section $\sigma$
is derived using an extension of the approach of Feldman and
Cousins~\cite{feldmancousins}.  The above arguments require accounting
for the component of the $N_{\mathrm{s}}$ observed Ge single scatters 
(with energies $E_{i}, i=1, \ldots, N_{\mathrm{s}}$) 
that is due to the unvetoed neutron flux $n$.  This
flux is constrained by the number $N_{\mathrm d}$ of double scatters
in Ge and the number $N_{\mathrm{Si}}$ of nuclear recoils in Si.  To
determine the 90\%~CL excluded region in the plane of $M$ and $\sigma$
alone, the parameter $n$ is projected out.  For a grid of physically
allowed values of $M$, $\sigma$, and $n$, the expected distribution of
the likelihood ratio 
\begin{equation}
R = \frac{ {\mathcal L}( E_{i}, N_{\mathrm d}, N_{\mathrm
{Si}} | \,\sigma, M, \tilde{n}) }
{{\mathcal {L}}( E_{i}, N_{\mathrm d}, N_{\mathrm
{Si}} | \,\widehat{\sigma}, \widehat{M}, \widehat{n})}
\label{ratios}
\end{equation}
is calculated by
Monte Carlo simulation in order to determine the critical parameter
$R_{90}$ such that 90\% of the simulated experiments have $R>R_{90}$.
Here $(\widehat{\sigma}, \widehat{M}, \widehat{n})$ is the
set of physically allowed parameters that maximizes the likelihood
${\mathcal L}$ for the given observations, while $\tilde{n}$ is the physically
allowed value of $n$ that maximizes the likelihood
${\mathcal L}$ for the given parameters $M$ and $\sigma$ and the
observations.
The 90\%~CL region excluded
by the observed data set consists of
all parameter space for which the observed likelihood ratio
$R_{\mathrm{data}}\leq R_{90}$.  
The 90\%~CL excluded region is projected into two dimensions 
conservatively by excluding only those
points excluded for all possible values of $n$.

\subsection{\label{sectlikelihood}Likelihood function}

The likelihood function
consists of functions $g$ describing the Poisson 
probabilities of obtaining the numbers of events 
actually detected,
combined with a function $f$ describing the probabilities of the events' 
energies:
\bea
{\mathcal L} & = & g_{\mathrm{s}}(N_{\mathrm{s}} | n,\sigma,M)
               g_{\mathrm{d}}(N_{\mathrm{d}} | n)
               g_{\mathrm{Si}}(N_{\mathrm{Si}} | n,\sigma,M) 
               \nonumber \\
         &   & \times \prod_{i}
    f_{\mathrm{s}}(E_{i} | n,\sigma,M) 
    .
\label{likefunction}
\eea
The energy spectrum of the multiple-scatter events
is ignored because it cancels in the likelihood ratio.
The energy spectrum of the Si events is also ignored,
as it would influence the likelihood ratio very weakly.

The expected energy spectrum of detected WIMPs, $w_{\mathrm{s}}(E)$,
and their total number, $w$, are calculated 
by making standard (but probably over-simplifying) assumptions
following~\cite{lewin}: WIMPs reside in an isothermal halo with 
WIMP characteristic velocity 
$v_0 = 220$~km~s$^{-1}$, 
Galactic escape velocity $v_{\mathrm{esc}}= 650$~km~s$^{-1}$,
mean Earth velocity $v_{\mathrm{E}} = 232$~km~s$^{-1}$, 
and local WIMP density $\rho = 0.3$~GeV~c$^{-2}$~cm$^{-3}$.
The energy spectrum of detected WIMP events also depends on the 
detection efficiency $\epsilon(E)$ and the nuclear form factor $F^{2}$.
We use the Woods-Saxon (Helm) form factor $F^{2}$, 
with thickness parameters $a=0.52$~fm, $s=0.9$~fm, and 
$c=1.23A^{1/3}-0.6$~fm,
as recommended by Lewin and Smith~\cite{lewin}.  

The resulting WIMP energy spectrum is well approximated by an exponential 
with a cut-off energy:
\bea
w_{\mathrm{s}}(E) = {\cal N} e^{-E/\langle E\rangle} \epsilon(E) 
                   F^{2}(E) H(Q_{\mathrm{max}} - E)  ,
\eea
where $H(x)$ is the Heaviside step function (0 for $x<0$ and 1 for $x>0$),
$Q_{\mathrm{max}}$ is the maximum
possible recoil energy from a WIMP of velocity $v_{\mathrm{esc}}$,
${\cal N}$ is a normalization constant, and $\langle E\rangle = E_0 r/c_2$
in the notation of Lewin and Smith~\cite{lewin}.  At low energies near the
spectrum peak, this form differs $< 5$\% from Eq.~3.13 of
Lewin and Smith.  We use this approximation in order to speed up
the calculation of the confidence region.

The neutron contribution to the energy spectrum,
$n_{{\mathrm{s}}}(E)$, is given by a best-fit function to the results of 
the external neutron Monte Carlo simulation including detection 
inefficiencies.

The Monte Carlo simulations, 
including the possible 13\% systematic error on the fraction of 
neutrons that multiple scatter, set the expected fraction of 
single scatters 
$\beta_{\mathrm{QIS}}= 0.91$ ($\beta_{\mathrm{QI}}= 0.90$) 
amongst the Ge neutron events with at least one QIS (QI) scatter.
Simulations also set the ratio
$\gamma_{\mathrm{QIS}}= 0.17$ ($\gamma_{\mathrm{QI}}= 0.24$) 
of the number of
neutrons expected in Si to the number expected in Ge
with at least one QIS (QI) scatter.
The expected ratio $\alpha$ of 
WIMPs detected in Si to those detected in Ge, given the 
relative exposures in  each, depends weakly on the WIMP mass.  
For WIMPs with masses $M \agt 30$~\gev\, 
$\alpha_{\mathrm{QIS}}\approx 0.011$ ($\alpha_{\mathrm{QI}} \approx 0.015$).
The expected electron background in Si $b_{\mathrm{Si}}$
is conservatively set to 7.3 events 
(corresponding to the 90\% CL upper 
limit on the  background expected in the 20-100~keV region
under the most conservative possible assumption).
This treatment of the Si data is not correct (it is overly conservative).
Ignoring the Si data, or using a better (and more complicated) treatment 
would result in a lower limit.
We conservatively neglect possible electron contamination in the 
Ge single data.  
We also neglect the possibility of electron 
contamination in the multiple-scatter Ge data, since the analysis 
presented in Sec.~\ref{antidouble}
indicates that the expected double-scatter contamination
$b_{\mathrm{d}}<0.05$ at the 90\% confidence level.

With these constants set, the expectation values for the 
observables are
\bea
\langle N_{\mathrm{s}} \rangle  & = & n \beta + w , \label{eqnNs} \\
\langle N_{\mathrm{d}} \rangle  & = & n (1-\beta) , \\
\langle N_{\mathrm{Si}} \rangle & = & n \gamma + w \alpha + b_{\mathrm{Si}} 
\label{eqnNsi} .
\eea
The pertinent contributions to the likelihood function are
\bea
g_{\mathrm{k}} = \frac{ e^{-\langle N_{\mathrm{k}} \rangle}
     \langle N_{\mathrm{k}} \rangle^{N_{\mathrm{k}}} }
     { N_{\mathrm{k}} ! }
\eea
for k $\equiv$ s, d, and Si, and
\bea
f_{{\mathrm{s}}}(E|n,\sigma,M) & = & \eta n_{{\mathrm{s}}}(E)  + 
                    (1-\eta) w_{{\mathrm{s}}}(E) ,
\eea
where $\eta = n \beta / (n \beta + w)$ is the fraction of single-scatter Ge
events expected to be neutrons.
Dropping factors that cancel in ratios yields
\bea
{\mathcal L} & \propto & 
         e^{[ -n(1+\gamma) - w(1+\alpha) - b_{\mathrm{Si}} ]} n^{N_{\mathrm{d}}} 
	 (n \gamma + w \alpha + b_{\mathrm{Si}})^{N_{\mathrm{Si}}}  
	 \nonumber \\
         &   & \times \prod_{i=1}^{N_{\mathrm{s}}}  \left[ 
               n \beta n_{\mathrm{s}}(E_{i})  + 
	       w       w_{\mathrm{s}}(E_{i}) 
	       \right]   .
\label{eqnlikelihood}
\eea

\subsection{Calculating an upper limit assuming arbitrary background}

Despite the evidence given above that the Ge single-scatter 
background is dominated by events due to neutrons, it is informative 
to calculate exclusion limits without using any information about 
the expected background.
A near-optimal classical method, practical when there are relatively small 
numbers of events detected, is Yellin's ``Optimum Interval'' 
method~\cite{yellin}.
Effectively, the method excludes the worst of the background by basing 
the limit on the interval in allowed energy that yields the lowest 
upper limit, while assessing the proper statistical penalty for 
the freedom to choose this optimum interval.
The limit is essentially set by a region of the energy spectrum with 
few events compared to the number expected from the WIMP energy 
spectrum.

Every possible interval is considered, with intervals characterized 
by the numbers $m$ of events in them, 
and $C_{m}(x,\mu)$ is defined as the probability that all 
intervals with $\le m$ events have
a computed expectation value of the number of events that is less than $x$,
where $\mu$ is the expected number
of events in the entire range of the measurement.
For each value of $m$, the interval with the largest expected number of 
events $x$ is determined.  For intervals with no events, 
the probability of this maximum expected number
being less than $x$ is
\begin{equation}
C_0(x,\mu) = \sum_{k=0}^m \frac{(kx-\mu)^ke^{-kx}}{ k!}
\left(1 + \frac{k}{\mu-kx}\right), 
\end{equation}
where $m$ is the greatest integer $\le\mu/x$.
For an interval with $m>0$ events, $C_m(x,\mu)$ is determined from 
Monte Carlo simulation.

$C_{\mathrm{max}}$ is defined as the maximum value
of $C_m(x,\mu)$ for any $m$.
High assumed cross section leads to high $C_{\mathrm{max}}$ 
for this experiment's data;
so if $C_{\mathrm{max}}$ is ``unreasonably" high, the assumed cross-section can be
rejected as being too high.  The expected probability distribution of
$C_{\mathrm{max}}$, as determined with a Monte Carlo simulation, is used to
compute a 90\% confidence region.

\section{\label{sect:results}Results}

\begin{figure}
\psfig{figure=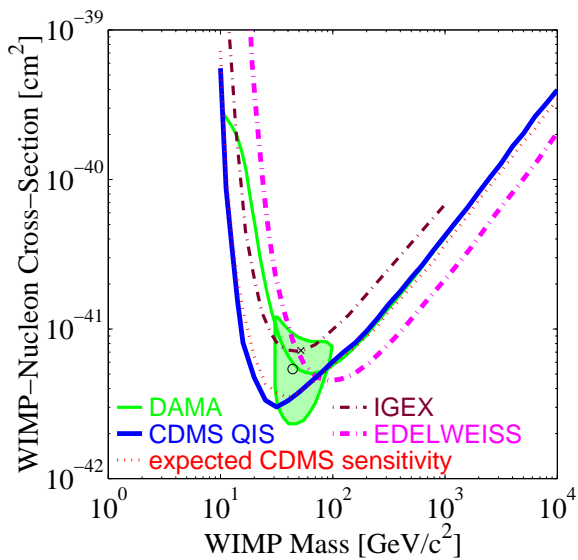, width=7.7cm}
\caption{\label{limitplot}(Color).
Spin-independent $\sigma$ vs. $M$.  The regions above the
curves are excluded at 90\%~CL.  The limits resulting from analysis of
the QIS data (solid dark blue curve) 
are shown.
The (red) dotted curve indicates the CDMS expected sensitivity 
given an expected neutron background of 27~events in Ge, 
and an expected background in Si of 7.2~electrons and 4.6~neutrons.
Solid light (green)
curve: DAMA limit using pulse-shape analysis~\protect\cite{DAMApsa}.
The most likely 
value for the WIMP signal from the annual-modulation measurement 
reported by the DAMA collaboration~\protect\cite{DAMA2000},
calculated including (not including) the DAMA limit using pulse-shape
analysis, is shown as a circle (as an $x$).
The DAMA 3$\sigma$ allowed region 
not including the DAMA limit~\protect\cite{DAMA2000} is shown as a shaded region.  
CDMS limits are the most sensitive upper limits for WIMPs with masses 
in the range 10-70~\gev.  
Above 70~\gev, the EDELWEISS experiment~\protect\cite{edel2000}
provides more sensitive limits (dot-dashed maroon curve).
Also shown are limits from IGEX~\cite{igex2002} (dot-dashed brown 
curve).
These and other results are available via an interactive web 
plotter~\protect\cite{dmplotter}.
All curves are normalized
following~\protect\cite{lewin}, using the Helm spin-independent
form-factor, $A^2$~scaling, WIMP characteristic velocity 
$v_0 = 220$~km~s$^{-1}$, mean Earth velocity $v_E = 232$~km~s$^{-1}$, and 
$\rho =0.3$~GeV~c$^{-2}$~cm$^{-3}$.}
\end{figure}

As shown in Sec.~\ref{sect:Nconsist} above,
the data are fully consistent with the possibility that all detected 
nuclear-recoil events are due to background neutron scatters and not 
WIMPs.  For this reason, the data provide no lower limit on the 
WIMP-nucleon cross section.
Figure~\ref{limitplot} displays the upper limits on the WIMP-nucleon 
cross section calculated under the assumptions on the WIMP halo 
described in Sec.~\ref{sectlikelihood};
these values are the lower envelope of points excluded 
at the 90\% confidence level for all
values of the neutron background $n$.  

\begin{figure}
\psfig{figure=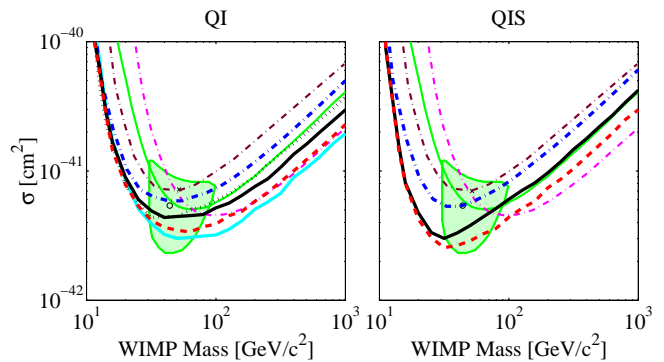, width=3.4in}
\caption{\label{limits}(Color).
Additional upper limits on the spin-independent WIMP-nucleon 
cross-section $\sigma$, based on different treatments of the data,
for both the QI (left) and QIS (right) data.
The regions above the curves are excluded at 90\%~CL.  
In each plot, CDMS limits including estimates of the neutron 
background,
as described in Sec.~\ref{ConRegion},  
are shown as 
black solid curves.
Limits calculated ignoring the 1998 Si data entirely (red dashed curves) 
would be better than these limits.
Limits calculated ignoring all knowledge 
about the neutron background (thick dark blue dot-dashed curves) 
would still be the 
most sensitive upper limits of any experiment for WIMPs with masses 
between 10--45~\gev.
The QI limit is worse than the CDMS QI limit previously 
reported~\cite{r19prl} 
(light blue solid curve) primarily due to the more conservative treatment of the 
1998 Si data. 
The QI limit is better than the expected sensitivity (black dotted curve)
for high WIMP masses because more multiple-scatter neutrons were 
detected than expected.
As in Fig.~\ref{limitplot}, 
the light green solid curve is the DAMA limit using pulse-shape 
analysis~\protect\cite{DAMApsa},
the shaded region is the DAMA 3$\sigma$ allowed region~\protect\cite{DAMA2000},
the circle ($x$) indicates the DAMA best-fit point including (not 
including) the DAMA limit using pulse-shape 
analysis, the thin, dark (brown) dot-dashed curve is the upper limit of the IGEX 
experiment~\cite{igex2002},
and the thin, light (maroon) dot-dashed curve is the upper limit of the 
EDELWEISS experiment~\protect\cite{edel2000}.
}
\end{figure}

Figure~\ref{limitplot} also shows the expected sensitivity of the data 
set, \ie, the expected 90\% CL exclusion limit given no expected WIMP 
signal,
an expected background in the QIS Ge data set of 27 neutron events, 
and an expected background in Si of 7.2 electrons and 4.6 neutrons.
To calculate these expected sensitivities, 
an ensemble of experiments are simulated, and the median 
resulting limit is taken (statistical fluctuations are large, 
so only 50\% of the limits fall within $\pm$ 50\% 
of these median expected sensitivities).  As indicated in the figure,
the upper limit for the QIS data is slightly better than expected at 
low masses and slightly worse than expected at high masses; 
Fig.~\ref{limits} shows that the upper 
limit of the QI data is slightly worse than expected at 
low masses and slightly better than expected at high masses.  
These results are consistent with statistical fluctuations.

For WIMP masses $M\agt 100$~\gev, 
the expected WIMP energy spectrum matches that predicted for 
neutrons, so the estimate of the neutron background 
(based on the number of detected multiple-scatter neutrons and Si 
neutrons)
has a dominant 
effect on the limits.  
Because the QIS data set represents a larger data set yet has no more 
multiple-scatter neutrons than the QI data set, its estimate of the 
neutron background is lower, and the QIS upper limits are slightly worse
than the QI limits.  
For these WIMP masses, 
the upper limits correspond to expectations of $\sim 23$ ($\sim 13$)
WIMP interactions in the Ge single-scatter QIS (QI) data set,
about the same as the actual number of observed events.  
As described above, these data are also consistent with no  
WIMP interactions.

For a low-mass WIMP, estimates of the neutron background have no 
effect.  A low-mass WIMP would result in a sharply falling energy spectrum; 
only the events just above the energy threshold could be WIMPs.  
For this reason, at the lowest masses (10--15~\gev), 
the upper limits for the QI and QIS 
data sets are very similar.  
The smaller statistical uncertainty associated with 
the larger QIS data set makes its limits slightly better than the QI 
upper limits at low mass. 

\begin{figure}
\psfig{figure=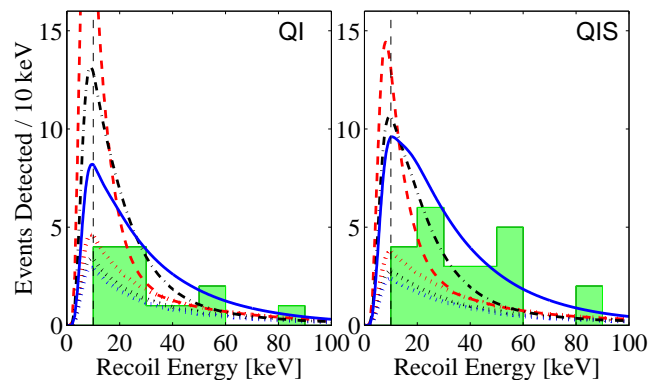, width=3.375in}
\caption{\label{SpecUL}(Color).
Histograms of energies of WIMP-candidate events (green shaded) for both the 
QI (left) and QIS (right) data sets, compared with the spectra expected 
to be detected by CDMS
for WIMPs excluded at exactly the 90\% confidence level.
Spectra for WIMPs with masses of 20~\gev\ (red dashes), 40~\gev\ 
(black dot-dashes), 
and 125~\gev\ (blue solid) are 
shown, including the expected contribution for the neutron 
background $\tilde{n}$ that maximizes the likelihood function
for the given WIMP mass and WIMP-nucleon cross-section (see 
Sec.~\ref{ConRegion}).
These most likely neutron 
backgrounds (shown separately as dotted curves)
correspond to 1.0, 0.7, and 0.6 (1.1, 0.8, and 0.7) multiple-scatter QIS 
(QI) neutrons expected, given the WIMP masses of 
33~\gev\ (top curve), 67~\gev\ (middle curve), 
and 216~\gev\ (bottom curve).
These low expected neutron backgrounds contribute to the unlikelihood 
of the WIMP models considered.
}
\end{figure}

For intermediate WIMP masses, the energy spectrum of the Ge single-scatter events
contributes to the estimate of the neutron background, 
with the number of high-energy events helping to set the neutron 
background.  Because the QIS data set has a slightly harder energy 
spectrum than the QI data set, the QIS data set results in a larger 
neutron estimate and a lower upper limit on the WIMP signal for these moderate 
masses.
Figure~\ref{SpecUL} shows the barely-excluded spectra for a 
sampling of WIMP masses.

\begin{figure}
\psfig{figure=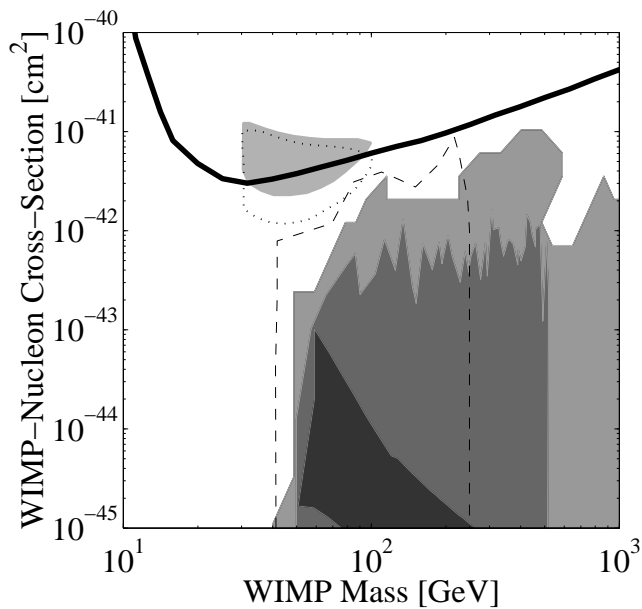, width=3.375in}
\caption{\label{SUSY}
CDMS upper limits on the spin-independent WIMP-nucleon 
cross-section $\sigma$ (dark curve),
shown with 
the DAMA 3$\sigma$ allowed regions including (dotted)
and not including (light shaded region) the DAMA limit~\protect\cite{DAMA2000}, 
as well as with regions of parameter space consistent 
with various frameworks of the MSSM and the standard WIMP 
interactions and galactic halo described above.
The region outlined in dashes~\cite{bottino}
and the lightest theoretical region~\cite{baltz} each shows the 
results from
calculations under an effective scheme, with parameters defined at 
the electroweak scale.
The medium-gray region~\cite{ellis01} arises from constraining the parameter space to 
small values of  $\tan\beta$, the ratio of vacuum expectation 
values of the two Higgs bosons.
The darkest region represents the models allowed in a more constrained
framework (called minimal supergravity or constrained MSSM), 
in which all soft scalar masses are unified at the
unification scale~\cite{ellis01}.
}
\end{figure}

These limits are lower than those of any other experiment for WIMPs with 
10~\gev $< M < 70$~\gev.
According to the calculations presented in~\cite{baltz,bottino,ellis01},
these limits do not appear to exclude any parameter space
consistent with the minimal supersymmetric standard model (MSSM) and allowed by
accelerator constraints.
Figure~\ref{SUSY} compares these limits 
to the regions of parameter space consistent 
with various frameworks of the MSSM.

As shown in Fig.~\ref{limits}, both the QIS and QI  
limits would be lower if the 1998 Si data were ignored.
The conservative estimate of the amount of electron contamination in 
the nuclear-recoil band of the Si data reduces the estimate of the 
neutron background.  
This more conservative estimate of the Si contamination is the main 
reason that the QI limit is worse than that 
previously reported~\cite{r19prl}.

\begin{figure}
\psfig{figure=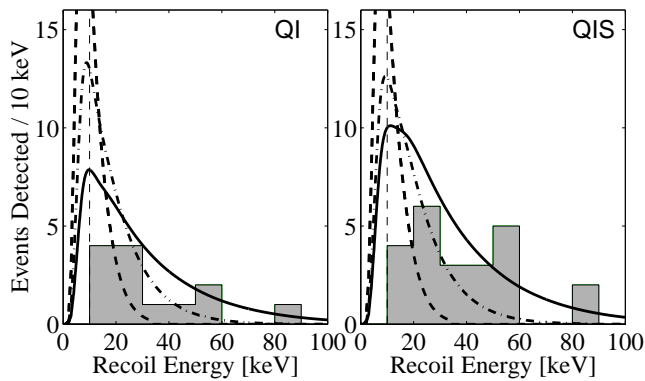, width=3.375in}
\caption{\label{SpecULnosub}
Histograms of energies of WIMP-candidate events (shaded) for both the 
QI (left) and QIS (right) data sets, indicating the spectra expected 
to be detected by CDMS
for WIMPs excluded at exactly the 90\% confidence level if all knowledge 
about the background is ignored.
Spectra for WIMPs with masses of 20~\gev\ (dashes), 40~\gev\ 
(dot-dashes), 
and 125~\gev\ (solid) are 
shown.
}
\end{figure}

Figure~\ref{limits} also shows the upper limits if all knowledge 
about the neutron background is ignored.
The figure shows that even without any background estimation, CDMS 
limits are more sensitive for WIMPs with masses between 10--45~\gev\
than those of any other experiment.  
Figure~\ref{SpecULnosub} shows the barely-excluded spectra for a 
sampling of WIMP masses.

Under the assumptions of standard WIMP interactions and halo,
the QIS (QI) data with estimation of the neutron background
exclude, at $> 99.9$\% ($> 99$\%)~CL,
the most likely value ($M= 52$~\gev, $\sigma=7.2\times10^{-6}$~pb)
for the spin-independent WIMP signal from the annual-modulation measurement 
reported by the DAMA collaboration~\cite{DAMA2000}.
The QIS (QI) data exclude, at $> 99\%$ ($> 95$\%)~CL,
the most likely value 
($M= 44$~\gev, $\sigma=5.4\times10^{-6}$~pb~\cite{DAMA2000})
obtained by combining DAMA's annual-modulation measurement with their
exclusion limit based on pulse-shape analysis~\cite{DAMApsa}.
The CDMS limits without any background estimation exclude, at $90\%$~CL
(at $> 90\%$~CL),
the most likely value 
for the WIMP signal from the DAMA annual-modulation measurement 
with (without) their
exclusion limit based on pulse-shape analysis.

At 90\% CL, these data do not exclude the complete parameter space
reported as allowed at $3\sigma$ by the annual-modulation measurement 
of the DAMA collaboration.  However, compatibility between the annual 
modulation signal of DAMA and the absence of a significant signal in CDMS 
(or in another experiment) is best determined by a goodness-of-fit test, 
not by comparing overlap regions of allowed parameter space.
A likelihood-ratio test can determine the probability 
of obtaining a given combination of experimental results 
for the same parameters.  
The test involves calculating 
$\lambda \equiv {\cal L}_{0} / {\cal L}_{1}$, 
where ${\cal L}_{0}$ is the likelihood of the data assuming compatibility 
and   ${\cal L}_{1}$ is the likelihood without assuming compatibility.
If the data are compatible, 
$-2\ln \lambda$
should follow the $\chi^{2}$ distribution with two degrees of freedom
in the asymptotic limit of large statistics and away from 
physical boundaries.
Under this approximation and
the assumptions of standard WIMP interactions and halo, 
this test indicates the model-independent 
annual-modulation signal of DAMA (as shown in Fig.~2 of~\cite{DAMA2000})
and CDMS data are incompatible at 99.99\%~CL. 
Furthermore, even under the assumption that none of the CDMS events 
are due to neutrons,
a likelihood-ratio test indicates the CDMS data and the DAMA signal are 
incompatible at 99.8\%~CL.
Simply put, a spin-independent WIMP-nucleon cross-section that would 
give rise to the annual-modulation amplitude 
$A = 0.022$~events~kg$^{-1}$~keV$^{-1}$ observed by DAMA 
averaged over 2--6~keV electron-equivalent energy 
should yield $>3$~events~kg$^{-1}$~day$^{-1}$ in Ge, incompatible with 
the 23 CDMS events in 15.8~kg~d even if none of the events are 
due to neutrons.
If the amplitude of the annual modulation observed by DAMA is a large
statistical fluctuation, 
or if part of the modulation is due to something other than 
WIMPs, the CDMS and DAMA results may be 
compatible.  
Furthermore, if the distribution of WIMPs locally is much 
different than assumed (see \eg~\cite{DAMAgalaxy,copikrauss}), 
or if WIMPs interact other than by spin-independent elastic scattering
(see \eg~\cite{smit-wein-01,DAMAinelastic,DAMAmixed}), 
or if WIMP interactions are otherwise different than assumed,
the two results may be compatible.

\section*{Acknowledgments}

We thank Paul Luke of LBNL for his advice regarding surface-event
rejection.  We thank R.~Abusaidi, J.~Emes, D.~Hale, G.W.~Smith, 
J.~Taylor, S.~White, D.N.~Seitz, J.~Perales, M.~Hennessy, 
M.~Haldeman,
and the rest of the engineering and technical staffs at our
respective institutions for invaluable support.  This work is
supported by the Center for Particle Astrophysics, an NSF Science and
Technology Center operated by the University of California, Berkeley,
under Cooperative Agreement No. AST-91-20005, by the National Science
Foundation under Grant No. PHY-9722414, by the Department of Energy
under contracts DE-AC03-76SF00098, DE-FG03-90ER40569,
DE-FG03-91ER40618, and by Fermilab, operated by the Universities
Research Association, Inc., under Contract No. DE-AC02-76CH03000 with
the Department of Energy.

\bibliography{r19prd}

\begin{thebibliography}{76}
\expandafter\ifx\csname natexlab\endcsname\relax\def\natexlab#1{#1}\fi
\expandafter\ifx\csname bibnamefont\endcsname\relax
  \def\bibnamefont#1{#1}\fi
\expandafter\ifx\csname bibfnamefont\endcsname\relax
  \def\bibfnamefont#1{#1}\fi
\expandafter\ifx\csname citenamefont\endcsname\relax
  \def\citenamefont#1{#1}\fi
\expandafter\ifx\csname url\endcsname\relax
  \def\url#1{\texttt{#1}}\fi
\expandafter\ifx\csname urlprefix\endcsname\relax\def\urlprefix{URL }\fi
\providecommand{\bibinfo}[2]{#2}
\providecommand{\eprint}[2][]{\url{#2}}

\bibitem[{\citenamefont{{Bergstrom}}(2000)}]{bergstrom}
\bibinfo{author}{\bibfnamefont{L.}~\bibnamefont{{Bergstrom}}},
  \bibinfo{journal}{Rep. Prog. Phys.} \textbf{\bibinfo{volume}{63}},
  \bibinfo{pages}{793} (\bibinfo{year}{2000}).

\bibitem[{\citenamefont{Kolb and Turner}(1990)}]{kolbturner}
\bibinfo{author}{\bibfnamefont{E.~W.} \bibnamefont{Kolb}} \bibnamefont{and}
  \bibinfo{author}{\bibfnamefont{M.~S.} \bibnamefont{Turner}},
  \emph{\bibinfo{title}{{The Early Universe}}}
  (\bibinfo{publisher}{Addison-Wesley}, \bibinfo{address}{Reading, MA},
  \bibinfo{year}{1990}).

\bibitem[{\citenamefont{Peebles}(1993)}]{peebles}
\bibinfo{author}{\bibfnamefont{P.~J.~E.} \bibnamefont{Peebles}},
  \emph{\bibinfo{title}{{Principles of Physical Cosmology}}}
  (\bibinfo{publisher}{Princeton University Press},
  \bibinfo{address}{Princeton, NJ}, \bibinfo{year}{1993}).

\bibitem[{\citenamefont{de~Bernardis et~al.}(2002)}]{boomerang}
\bibinfo{author}{\bibfnamefont{P.}~\bibnamefont{de~Bernardis}}
  \bibnamefont{et~al.}, \bibinfo{journal}{Astrophys. J.}
  \textbf{\bibinfo{volume}{564}}, \bibinfo{pages}{559} (\bibinfo{year}{2002}),
  \eprint{astro-ph/0105296}.

\bibitem[{\citenamefont{Pryke et~al.}(2002)\citenamefont{Pryke, Halverson,
  Leitch, Kovac, Carlstrom, Holzapfel, and Dragovan}}]{dasi}
\bibinfo{author}{\bibfnamefont{C.}~\bibnamefont{Pryke}},
  \bibinfo{author}{\bibfnamefont{N.~W.} \bibnamefont{Halverson}},
  \bibinfo{author}{\bibfnamefont{E.~M.} \bibnamefont{Leitch}},
  \bibinfo{author}{\bibfnamefont{J.}~\bibnamefont{Kovac}},
  \bibinfo{author}{\bibfnamefont{J.~E.} \bibnamefont{Carlstrom}},
  \bibinfo{author}{\bibfnamefont{W.~L.} \bibnamefont{Holzapfel}},
  \bibnamefont{and} \bibinfo{author}{\bibfnamefont{M.}~\bibnamefont{Dragovan}},
  \bibinfo{journal}{Astrophys. J.} \textbf{\bibinfo{volume}{568}},
  \bibinfo{pages}{46} (\bibinfo{year}{2002}), \eprint{astro-ph/0104490}.

\bibitem[{\citenamefont{Stompor et~al.}(2001)}]{maxima}
\bibinfo{author}{\bibfnamefont{R.}~\bibnamefont{Stompor}} \bibnamefont{et~al.},
  \bibinfo{journal}{Astrophys. J. Lett.} \textbf{\bibinfo{volume}{561}},
  \bibinfo{pages}{7} (\bibinfo{year}{2001}).

\bibitem[{\citenamefont{{Srednicki}}(2000)}]{srednicki}
\bibinfo{author}{\bibfnamefont{M.}~\bibnamefont{{Srednicki}}},
  \bibinfo{journal}{Eur. J. Phys. C} \textbf{\bibinfo{volume}{15}},
  \bibinfo{pages}{143} (\bibinfo{year}{2000}).

\bibitem[{\citenamefont{Jungman et~al.}(1996)\citenamefont{Jungman,
  Kamionkowski, and Griest}}]{jkg}
\bibinfo{author}{\bibfnamefont{G.}~\bibnamefont{Jungman}},
  \bibinfo{author}{\bibfnamefont{M.}~\bibnamefont{Kamionkowski}},
  \bibnamefont{and} \bibinfo{author}{\bibfnamefont{K.}~\bibnamefont{Griest}},
  \bibinfo{journal}{{Phys. Rep.}} \textbf{\bibinfo{volume}{267}},
  \bibinfo{pages}{195} (\bibinfo{year}{1996}).

\bibitem[{\citenamefont{{Ellis} et~al.}(1997)\citenamefont{{Ellis}, {Falk},
  {Olive}, and {Schmitt}}}]{ellis}
\bibinfo{author}{\bibfnamefont{J.}~\bibnamefont{{Ellis}}},
  \bibinfo{author}{\bibfnamefont{T.}~\bibnamefont{{Falk}}},
  \bibinfo{author}{\bibfnamefont{K.~A.} \bibnamefont{{Olive}}},
  \bibnamefont{and}
  \bibinfo{author}{\bibfnamefont{M.}~\bibnamefont{{Schmitt}}},
  \bibinfo{journal}{Phys. Lett.} \textbf{\bibinfo{volume}{B413}},
  \bibinfo{pages}{355} (\bibinfo{year}{1997}).

\bibitem[{\citenamefont{{Edsjo} and {Gondolo}}(1997)}]{gondolo}
\bibinfo{author}{\bibfnamefont{J.}~\bibnamefont{{Edsjo}}} \bibnamefont{and}
  \bibinfo{author}{\bibfnamefont{P.}~\bibnamefont{{Gondolo}}},
  \bibinfo{journal}{Phys. Rev. D} \textbf{\bibinfo{volume}{56}},
  \bibinfo{pages}{1879} (\bibinfo{year}{1997}).

\bibitem[{\citenamefont{Bottino et~al.}(2001)\citenamefont{Bottino, Donato,
  Fornengo, and Scopel}}]{bottino}
\bibinfo{author}{\bibfnamefont{A.}~\bibnamefont{Bottino}},
  \bibinfo{author}{\bibfnamefont{F.}~\bibnamefont{Donato}},
  \bibinfo{author}{\bibfnamefont{N.}~\bibnamefont{Fornengo}}, \bibnamefont{and}
  \bibinfo{author}{\bibfnamefont{S.}~\bibnamefont{Scopel}},
  \bibinfo{journal}{Phys. Rev. D} \textbf{\bibinfo{volume}{63}},
  \bibinfo{pages}{125003} (\bibinfo{year}{2001}).

\bibitem[{\citenamefont{{Ellis} et~al.}(2000)\citenamefont{{Ellis}, {Falk},
  Gani, and {Olive}}}]{ellis00lep}
\bibinfo{author}{\bibfnamefont{J.}~\bibnamefont{{Ellis}}},
  \bibinfo{author}{\bibfnamefont{T.}~\bibnamefont{{Falk}}},
  \bibinfo{author}{\bibfnamefont{G.}~\bibnamefont{Gani}}, \bibnamefont{and}
  \bibinfo{author}{\bibfnamefont{K.~A.} \bibnamefont{{Olive}}},
  \bibinfo{journal}{Phys. Rev. D} \textbf{\bibinfo{volume}{62}},
  \bibinfo{pages}{075010} (\bibinfo{year}{2000}).

\bibitem[{\citenamefont{Lee and Weinberg}(1977)}]{lee}
\bibinfo{author}{\bibfnamefont{B.~W.} \bibnamefont{Lee}} \bibnamefont{and}
  \bibinfo{author}{\bibfnamefont{S.}~\bibnamefont{Weinberg}},
  \bibinfo{journal}{{Phys. Rev. Lett.}} \textbf{\bibinfo{volume}{39}},
  \bibinfo{pages}{165} (\bibinfo{year}{1977}).

\bibitem[{\citenamefont{Salucci and Persic}(1997)}]{salucci}
\bibinfo{author}{\bibfnamefont{P.}~\bibnamefont{Salucci}} \bibnamefont{and}
  \bibinfo{author}{\bibfnamefont{M.}~\bibnamefont{Persic}}, in
  \emph{\bibinfo{booktitle}{ASP Conf. Ser. 117: Dark and Visible Matter in
  Galaxies and Cosmological Implications}}, edited by
  \bibinfo{editor}{\bibfnamefont{P.}~\bibnamefont{Salucci}} \bibnamefont{and}
  \bibinfo{editor}{\bibfnamefont{M.}~\bibnamefont{Persic}}
  (\bibinfo{year}{1997}), pp. \bibinfo{pages}{1--27},
  \bibinfo{note}{asto-ph/9703027}.

\bibitem[{\citenamefont{{Goodman} and {Witten}}(1985)}]{goodman}
\bibinfo{author}{\bibfnamefont{M.~W.} \bibnamefont{{Goodman}}}
  \bibnamefont{and} \bibinfo{author}{\bibfnamefont{E.}~\bibnamefont{{Witten}}},
  \bibinfo{journal}{Phys. Rev. D} \textbf{\bibinfo{volume}{31}},
  \bibinfo{pages}{3059} (\bibinfo{year}{1985}).

\bibitem[{\citenamefont{{Primack} et~al.}(1988)\citenamefont{{Primack},
  {Seckel}, and {Sadoulet}}}]{primack}
\bibinfo{author}{\bibfnamefont{J.~R.} \bibnamefont{{Primack}}},
  \bibinfo{author}{\bibfnamefont{D.}~\bibnamefont{{Seckel}}}, \bibnamefont{and}
  \bibinfo{author}{\bibfnamefont{B.}~\bibnamefont{{Sadoulet}}},
  \bibinfo{journal}{Annu. Rev. Nucl. Part. Sci.} \textbf{\bibinfo{volume}{38}},
  \bibinfo{pages}{751} (\bibinfo{year}{1988}).

\bibitem[{\citenamefont{Lewin and Smith}(1996)}]{lewin}
\bibinfo{author}{\bibfnamefont{J.~D.} \bibnamefont{Lewin}} \bibnamefont{and}
  \bibinfo{author}{\bibfnamefont{P.~F.} \bibnamefont{Smith}},
  \bibinfo{journal}{Astropart. Phys.} \textbf{\bibinfo{volume}{6}},
  \bibinfo{pages}{87} (\bibinfo{year}{1996}).

\bibitem[{\citenamefont{Abusaidi et~al.}(2000)}]{r19prl}
\bibinfo{author}{\bibfnamefont{R.}~\bibnamefont{Abusaidi}}
  \bibnamefont{et~al.}, \bibinfo{journal}{Phys. Rev. Lett.}
  \textbf{\bibinfo{volume}{84}}, \bibinfo{pages}{5699} (\bibinfo{year}{2000}).

\bibitem[{\citenamefont{Shutt et~al.}(1992{\natexlab{a}})}]{tomprl2}
\bibinfo{author}{\bibfnamefont{T.}~\bibnamefont{Shutt}} \bibnamefont{et~al.},
  \bibinfo{journal}{Phys. Rev. Lett.} \textbf{\bibinfo{volume}{69}},
  \bibinfo{pages}{3425} (\bibinfo{year}{1992}{\natexlab{a}}).

\bibitem[{\citenamefont{Shutt et~al.}(1992{\natexlab{b}})}]{tomprl1}
\bibinfo{author}{\bibfnamefont{T.}~\bibnamefont{Shutt}} \bibnamefont{et~al.},
  \bibinfo{journal}{Phys. Rev. Lett.} \textbf{\bibinfo{volume}{69}},
  \bibinfo{pages}{3531} (\bibinfo{year}{1992}{\natexlab{b}}).

\bibitem[{\citenamefont{Gaitskell et~al.}(1997)}]{blipltd7}
\bibinfo{author}{\bibfnamefont{R.~J.} \bibnamefont{Gaitskell}}
  \bibnamefont{et~al.}, in \emph{\bibinfo{booktitle}{Proceedings of the Seventh
  International Workshop on Low Temperature Detectors}}, edited by
  \bibinfo{editor}{\bibfnamefont{S.}~\bibnamefont{{Cooper}}}
  (\bibinfo{organization}{Max Planck Institute of Physics},
  \bibinfo{address}{Munich}, \bibinfo{year}{1997}), pp.
  \bibinfo{pages}{221--223}.

\bibitem[{\citenamefont{{Irwin} et~al.}(1995)}]{irwin2}
\bibinfo{author}{\bibfnamefont{K.~D.} \bibnamefont{{Irwin}}}
  \bibnamefont{et~al.}, \bibinfo{journal}{Rev. Sci. Instr.}
  \textbf{\bibinfo{volume}{66}}, \bibinfo{pages}{5322} (\bibinfo{year}{1995}).

\bibitem[{\citenamefont{{Clarke} et~al.}(1997)}]{alexltd7}
\bibinfo{author}{\bibfnamefont{R.~M.} \bibnamefont{{Clarke}}}
  \bibnamefont{et~al.}, in \emph{\bibinfo{booktitle}{Proceedings of the Seventh
  International Workshop on Low Temperature Detectors}}, edited by
  \bibinfo{editor}{\bibfnamefont{S.}~\bibnamefont{{Cooper}}}
  (\bibinfo{organization}{Max Planck Institute of Physics},
  \bibinfo{address}{Munich}, \bibinfo{year}{1997}), pp.
  \bibinfo{pages}{229--231}, \bibinfo{note}{note that early designs of ZIPs are
  referred to as FLIPs in this and other references.}

\bibitem[{\citenamefont{{Clarke} et~al.}(1999)}]{alexsheffield}
\bibinfo{author}{\bibfnamefont{R.~M.} \bibnamefont{{Clarke}}}
  \bibnamefont{et~al.}, in \emph{\bibinfo{booktitle}{Proceedings of the Second
  International Workshop on the Identification of Dark Matter}}, edited by
  \bibinfo{editor}{\bibfnamefont{N.~J.~C.} \bibnamefont{{Spooner}}}
  \bibnamefont{and}
  \bibinfo{editor}{\bibfnamefont{V.}~\bibnamefont{{Kudryavtsev}}}
  (\bibinfo{publisher}{World Scientific}, \bibinfo{address}{Singapore},
  \bibinfo{year}{1999}), pp. \bibinfo{pages}{353--358}.

\bibitem[{\citenamefont{Clarke}(1999)}]{clarkethesis}
\bibinfo{author}{\bibfnamefont{R.~M.} \bibnamefont{Clarke}}, Ph.D. thesis,
  \bibinfo{school}{Stanford University} (\bibinfo{year}{1999}),
  \urlprefix\url{http://cosmology.berkeley.edu/preprints/cdms/
  Dissertations/rolandthesis.pdf}.

\bibitem[{\citenamefont{{Clarke} et~al.}(2000)}]{clarkeapl}
\bibinfo{author}{\bibfnamefont{R.~M.} \bibnamefont{{Clarke}}}
  \bibnamefont{et~al.}, \bibinfo{journal}{\apl} \textbf{\bibinfo{volume}{76}},
  \bibinfo{pages}{2958} (\bibinfo{year}{2000}).

\bibitem[{\citenamefont{{Sadoulet} et~al.}(1996)}]{hallereutectic}
\bibinfo{author}{\bibfnamefont{B.}~\bibnamefont{{Sadoulet}}}
  \bibnamefont{et~al.}, \bibinfo{journal}{Physica B}
  \textbf{\bibinfo{volume}{219}}, \bibinfo{pages}{741} (\bibinfo{year}{1996}).

\bibitem[{\citenamefont{Shutt et~al.}(2000)}]{tomltd8}
\bibinfo{author}{\bibfnamefont{T.}~\bibnamefont{Shutt}} \bibnamefont{et~al.},
  \bibinfo{journal}{Nucl. Instrum. Methods Phys. Res., Sect A}
  \textbf{\bibinfo{volume}{444}}, \bibinfo{pages}{340} (\bibinfo{year}{2000}).

\bibitem[{\citenamefont{Shutt}(1993)}]{shuttthesis}
\bibinfo{author}{\bibfnamefont{T.}~\bibnamefont{Shutt}}, Ph.D. thesis,
  \bibinfo{school}{University of California, Berkeley} (\bibinfo{year}{1993}).

\bibitem[{\citenamefont{Neganov and Trofimov}(1985)}]{neganov}
\bibinfo{author}{\bibfnamefont{B.}~\bibnamefont{Neganov}} \bibnamefont{and}
  \bibinfo{author}{\bibfnamefont{V.}~\bibnamefont{Trofimov}},
  \bibinfo{journal}{Otkrytia, Izobreteniya} \textbf{\bibinfo{volume}{146}},
  \bibinfo{pages}{215} (\bibinfo{year}{1985}).

\bibitem[{\citenamefont{Luke}(1988)}]{luke}
\bibinfo{author}{\bibfnamefont{P.~L.} \bibnamefont{Luke}}, \bibinfo{journal}{J.
  Appl. Phys.} \textbf{\bibinfo{volume}{64}}, \bibinfo{pages}{6858}
  (\bibinfo{year}{1988}).

\bibitem[{\citenamefont{{Shutt} et~al.}(2002)}]{tomltd9}
\bibinfo{author}{\bibfnamefont{T.}~\bibnamefont{{Shutt}}} \bibnamefont{et~al.},
  in \emph{\bibinfo{booktitle}{Low Temperature Detectors, AIP Conference
  Proceedings Vol. 605}}, edited by \bibinfo{editor}{\bibfnamefont{F.~S.}
  \bibnamefont{Porter}},
  \bibinfo{editor}{\bibfnamefont{D.}~\bibnamefont{McCammon}},
  \bibinfo{editor}{\bibfnamefont{M.}~\bibnamefont{Galeazzi}}, \bibnamefont{and}
  \bibinfo{editor}{\bibfnamefont{C.~K.} \bibnamefont{Stahle}}
  (\bibinfo{publisher}{AIP}, \bibinfo{address}{Melville, New York},
  \bibinfo{year}{2002}), pp. \bibinfo{pages}{513--516}.

\bibitem[{\citenamefont{Da~Silva}(1996)}]{dasilvathesis}
\bibinfo{author}{\bibfnamefont{A.}~\bibnamefont{Da~Silva}}, Ph.D. thesis,
  \bibinfo{school}{The University of British Columbia} (\bibinfo{year}{1996}).

\bibitem[{\citenamefont{Golwala}(2000)}]{golwalathesis}
\bibinfo{author}{\bibfnamefont{S.~R.} \bibnamefont{Golwala}}, Ph.D. thesis,
  \bibinfo{school}{University of California, Berkeley} (\bibinfo{year}{2000}),
  \urlprefix\url{http://cosmology.berkeley.edu/preprints/cdms/golwalathesis}.

\bibitem[{\citenamefont{{Akerib} et~al.}()}]{coldelect}
\bibinfo{author}{\bibfnamefont{D.~S.} \bibnamefont{{Akerib}}}
  \bibnamefont{et~al.}, \bibinfo{note}{in preparation}.

\bibitem[{\citenamefont{Wang}(1991)}]{nwthesis}
\bibinfo{author}{\bibfnamefont{N.}~\bibnamefont{Wang}}, Ph.D. thesis,
  \bibinfo{school}{University of California, Berkeley} (\bibinfo{year}{1991}).

\bibitem[{\citenamefont{Sonnenschein}(1999)}]{andrewthesis}
\bibinfo{author}{\bibfnamefont{A.~H.} \bibnamefont{Sonnenschein}}, Ph.D.
  thesis, \bibinfo{school}{University of California, Santa Barbara}
  (\bibinfo{year}{1999}),
  \urlprefix\url{http://cosmology.berkeley.edu/preprints/cdms/
  Dissertations/sonnenschein.ps}.

\bibitem[{\citenamefont{{Yvon} et~al.}(1996)}]{lockin}
\bibinfo{author}{\bibfnamefont{D.}~\bibnamefont{{Yvon}}} \bibnamefont{et~al.},
  \bibinfo{journal}{\nima} \textbf{\bibinfo{volume}{368}}, \bibinfo{pages}{778}
  (\bibinfo{year}{1996}).

\bibitem[{\citenamefont{Tamura}(1997)}]{tamura}
\bibinfo{author}{\bibfnamefont{S.}~\bibnamefont{Tamura}},
  \bibinfo{journal}{Phys. Rev. B} \textbf{\bibinfo{volume}{56}},
  \bibinfo{pages}{13630} (\bibinfo{year}{1997}).

\bibitem[{\citenamefont{Welty and Martinis}(1991)}]{squid1}
\bibinfo{author}{\bibfnamefont{R.~P.} \bibnamefont{Welty}} \bibnamefont{and}
  \bibinfo{author}{\bibfnamefont{J.~P.} \bibnamefont{Martinis}},
  \bibinfo{journal}{IEEE Trans. Magn.} \textbf{\bibinfo{volume}{27}},
  \bibinfo{pages}{2924} (\bibinfo{year}{1991}).

\bibitem[{\citenamefont{Welty and Martinis}(1993)}]{squid2}
\bibinfo{author}{\bibfnamefont{R.~P.} \bibnamefont{Welty}} \bibnamefont{and}
  \bibinfo{author}{\bibfnamefont{J.~P.} \bibnamefont{Martinis}},
  \bibinfo{journal}{IEEE Trans. Appl. Supercon.} \textbf{\bibinfo{volume}{3}},
  \bibinfo{pages}{2605} (\bibinfo{year}{1993}).

\bibitem[{\citenamefont{{Taylor} et~al.}(1996)}]{icebox}
\bibinfo{author}{\bibfnamefont{J.~D.} \bibnamefont{{Taylor}}}
  \bibnamefont{et~al.}, \bibinfo{journal}{Adv. Cryo. Eng.}
  \textbf{\bibinfo{volume}{41}}, \bibinfo{pages}{1971} (\bibinfo{year}{1996}).

\bibitem[{\citenamefont{Barnes}(1996)}]{pdbthesis}
\bibinfo{author}{\bibfnamefont{P.~D.} \bibnamefont{Barnes}, \bibfnamefont{Jr}},
  Ph.D. thesis, \bibinfo{school}{University of California, Berkeley}
  (\bibinfo{year}{1996}).

\bibitem[{\citenamefont{Akerib et~al.}(1997)}]{cdms_solder}
\bibinfo{author}{\bibfnamefont{D.~S.} \bibnamefont{Akerib}}
  \bibnamefont{et~al.}, \bibinfo{journal}{{Nucl. Instrum. Methods Phys. Res.,
  Sect. A}} \textbf{\bibinfo{volume}{400}}, \bibinfo{pages}{181}
  (\bibinfo{year}{1997}).

\bibitem[{\citenamefont{{Da~Silva} et~al.}(1995{\natexlab{a}})}]{dasilvapb210}
\bibinfo{author}{\bibfnamefont{A.}~\bibnamefont{{Da~Silva}}}
  \bibnamefont{et~al.}, \bibinfo{journal}{{Nucl. Instrum. Methods Phys. Res.,
  Sect. A}} \textbf{\bibinfo{volume}{364}}, \bibinfo{pages}{578}
  (\bibinfo{year}{1995}{\natexlab{a}}).

\bibitem[{nan()}]{nantes}
\bibinfo{note}{This lead was recovered from a sunken ship ballast near Nantes,
  France, and purchased from Lemer Pax, Protection Anti-X, 3 Rue de l'Europe,
  Zone Industrielle, F-44470 CARQUEFOU - FRANCE.}

\bibitem[{\citenamefont{{Da~Silva}
  et~al.}(1995{\natexlab{b}})}]{dasilvaneutron}
\bibinfo{author}{\bibfnamefont{A.}~\bibnamefont{{Da~Silva}}}
  \bibnamefont{et~al.}, \bibinfo{journal}{{Nucl. Instrum. Methods Phys. Res.,
  Sect. A}} \textbf{\bibinfo{volume}{354}}, \bibinfo{pages}{553}
  (\bibinfo{year}{1995}{\natexlab{b}}).

\bibitem[{\citenamefont{{Johnson} and {Gabriel}}(1988)}]{MICAP}
\bibinfo{author}{\bibfnamefont{J.~O.} \bibnamefont{{Johnson}}}
  \bibnamefont{and} \bibinfo{author}{\bibfnamefont{T.~A.}
  \bibnamefont{{Gabriel}}}, \bibinfo{type}{Tech. Rep.}
  \bibinfo{number}{TM-10340}, \bibinfo{institution}{ORNL}
  (\bibinfo{year}{1988}).

\bibitem[{\citenamefont{Fass\`o et~al.}(1993)}]{FLUKA}
\bibinfo{author}{\bibfnamefont{A.}~\bibnamefont{Fass\`o}} \bibnamefont{et~al.},
  in \emph{\bibinfo{booktitle}{Proceedings of the Workshop on Simulating
  Accelerator Radiation Environments, Santa Fe}} (\bibinfo{year}{1993}).

\bibitem[{\citenamefont{{Brun} and {Carminati}}(1993)}]{GEANT}
\bibinfo{author}{\bibfnamefont{R.}~\bibnamefont{{Brun}}} \bibnamefont{and}
  \bibinfo{author}{\bibfnamefont{F.}~\bibnamefont{{Carminati}}},
  \bibinfo{type}{CERN Program Library Long Writeup} \bibinfo{number}{W5013},
  \bibinfo{institution}{CERN} (\bibinfo{year}{1993}).

\bibitem[{\citenamefont{Khalchukov et~al.}(1983)}]{khalchukov1983}
\bibinfo{author}{\bibfnamefont{F.~F.} \bibnamefont{Khalchukov}}
  \bibnamefont{et~al.}, \bibinfo{journal}{{Nuovo Cimento}}
  \textbf{\bibinfo{volume}{6C}}, \bibinfo{pages}{3} (\bibinfo{year}{1983}).

\bibitem[{lab()}]{labview}
\bibinfo{note}{National Instruments Corporation; 11500 N Mopac Expwy; Austin,
  TX 78759-3504; (512) 794-0100.}

\bibitem[{\citenamefont{{Press} et~al.}(1992)\citenamefont{{Press},
  {Teukolsky}, {Vetterling}, and {Flannery}}}]{numrecipes}
\bibinfo{author}{\bibfnamefont{W.~H.} \bibnamefont{{Press}}},
  \bibinfo{author}{\bibfnamefont{S.~A.} \bibnamefont{{Teukolsky}}},
  \bibinfo{author}{\bibfnamefont{W.~T.} \bibnamefont{{Vetterling}}},
  \bibnamefont{and} \bibinfo{author}{\bibfnamefont{B.~P.}
  \bibnamefont{{Flannery}}}, \emph{\bibinfo{title}{{Numerical Recipes in C: The
  Art of Scientific Computing}}} (\bibinfo{publisher}{Cambridge University
  Press}, \bibinfo{address}{Cambridge}, \bibinfo{year}{1992}).

\bibitem[{\citenamefont{Knoll}(1989)}]{knoll}
\bibinfo{author}{\bibfnamefont{G.~F.} \bibnamefont{Knoll}},
  \emph{\bibinfo{title}{{Radiation Detection and Measurement, Second Edition}}}
  (\bibinfo{publisher}{John Wiley and Sons}, \bibinfo{address}{New York},
  \bibinfo{year}{1989}).

\bibitem[{\citenamefont{Perera}(2001)}]{TAPthesis}
\bibinfo{author}{\bibfnamefont{T.~A.} \bibnamefont{Perera}}, Ph.D. thesis,
  \bibinfo{school}{Case Western Reserve University} (\bibinfo{year}{2001}),
  \urlprefix\url{http://cosmology.berkeley.edu/preprints/cdms/
  Dissertations/tap_thesis.pdf}.

\bibitem[{\citenamefont{{Eadie} et~al.}(1971)\citenamefont{{Eadie}, {Drijard},
  {James}, {Roos}, and {Sadoulet}}}]{eadie}
\bibinfo{author}{\bibfnamefont{W.~T.} \bibnamefont{{Eadie}}},
  \bibinfo{author}{\bibfnamefont{D.}~\bibnamefont{{Drijard}}},
  \bibinfo{author}{\bibfnamefont{F.~E.} \bibnamefont{{James}}},
  \bibinfo{author}{\bibfnamefont{M.}~\bibnamefont{{Roos}}}, \bibnamefont{and}
  \bibinfo{author}{\bibfnamefont{B.}~\bibnamefont{{Sadoulet}}},
  \emph{\bibinfo{title}{{Statistical Methods in Experimental Physics}}}
  (\bibinfo{publisher}{North Holland}, \bibinfo{address}{Amsterdam},
  \bibinfo{year}{1971}).

\bibitem[{\citenamefont{Khalchukov et~al.}(1995)}]{khalchukov95}
\bibinfo{author}{\bibfnamefont{F.~F.} \bibnamefont{Khalchukov}}
  \bibnamefont{et~al.}, \bibinfo{journal}{{Nuovo Cimento}}
  \textbf{\bibinfo{volume}{18C}}, \bibinfo{pages}{5} (\bibinfo{year}{1995}).

\bibitem[{\citenamefont{Singer}(1974)}]{Singer}
\bibinfo{author}{\bibfnamefont{P.}~\bibnamefont{Singer}}, in
  \emph{\bibinfo{booktitle}{Springer Tracts in Modern Physics}}, edited by
  \bibinfo{editor}{\bibfnamefont{G.}~\bibnamefont{H\"{o}hler}}
  (\bibinfo{publisher}{Springer-Verlag}, \bibinfo{address}{Berlin},
  \bibinfo{year}{1974}), vol.~\bibinfo{volume}{71}, pp.
  \bibinfo{pages}{39--87}.

\bibitem[{\citenamefont{Cassiday et~al.}(1979)\citenamefont{Cassiday, Keuffel,
  and Thompson}}]{Cassiday}
\bibinfo{author}{\bibfnamefont{G.~L.} \bibnamefont{Cassiday}},
  \bibinfo{author}{\bibfnamefont{J.~W.} \bibnamefont{Keuffel}},
  \bibnamefont{and} \bibinfo{author}{\bibfnamefont{J.~A.}
  \bibnamefont{Thompson}}, \bibinfo{journal}{Phys. Rev. D}
  \textbf{\bibinfo{volume}{7}}, \bibinfo{pages}{2022} (\bibinfo{year}{1979}).

\bibitem[{\citenamefont{Lohmann et~al.}(1985)\citenamefont{Lohmann, Kopp, and
  Voss}}]{Lohmann}
\bibinfo{author}{\bibfnamefont{W.}~\bibnamefont{Lohmann}},
  \bibinfo{author}{\bibfnamefont{R.}~\bibnamefont{Kopp}}, \bibnamefont{and}
  \bibinfo{author}{\bibfnamefont{R.}~\bibnamefont{Voss}}, \bibinfo{type}{Tech.
  Rep.} \bibinfo{number}{85-03}, \bibinfo{institution}{CERN},
  \bibinfo{address}{Geneva} (\bibinfo{year}{1985}).

\bibitem[{\citenamefont{Wang et~al.}(2001)}]{YFWang}
\bibinfo{author}{\bibfnamefont{Y.-F.} \bibnamefont{Wang}} \bibnamefont{et~al.},
  \bibinfo{journal}{Phys. Rev. D} \textbf{\bibinfo{volume}{64}},
  \bibinfo{pages}{013012} (\bibinfo{year}{2001}).

\bibitem[{\citenamefont{{Saab} et~al.}(2002)}]{tarekltd9}
\bibinfo{author}{\bibfnamefont{T.}~\bibnamefont{{Saab}}} \bibnamefont{et~al.},
  in \emph{\bibinfo{booktitle}{Low Temperature Detectors, AIP Conference
  Proceedings Vol. 605}}, edited by \bibinfo{editor}{\bibfnamefont{F.~S.}
  \bibnamefont{Porter}},
  \bibinfo{editor}{\bibfnamefont{D.}~\bibnamefont{McCammon}},
  \bibinfo{editor}{\bibfnamefont{M.}~\bibnamefont{Galeazzi}}, \bibnamefont{and}
  \bibinfo{editor}{\bibfnamefont{C.~K.} \bibnamefont{Stahle}}
  (\bibinfo{publisher}{AIP}, \bibinfo{address}{Melville, New York},
  \bibinfo{year}{2002}), pp. \bibinfo{pages}{497--500}.

\bibitem[{\citenamefont{{Feldman} and {Cousins}}(1998)}]{feldmancousins}
\bibinfo{author}{\bibfnamefont{G.~J.} \bibnamefont{{Feldman}}}
  \bibnamefont{and}
  \bibinfo{author}{\bibfnamefont{R.}~\bibnamefont{{Cousins}}},
  \bibinfo{journal}{\prd} \textbf{\bibinfo{volume}{57}}, \bibinfo{pages}{3873}
  (\bibinfo{year}{1998}).

\bibitem[{\citenamefont{Yellin}(2002)}]{yellin}
\bibinfo{author}{\bibfnamefont{S.}~\bibnamefont{Yellin}}
  (\bibinfo{year}{2002}), \bibinfo{note}{to appear in Phys. Rev. D},
  \eprint{physics/0203002}.

\bibitem[{\citenamefont{Bernabei et~al.}(1996)}]{DAMApsa}
\bibinfo{author}{\bibfnamefont{R.}~\bibnamefont{Bernabei}}
  \bibnamefont{et~al.}, \bibinfo{journal}{Phys. Lett.}
  \textbf{\bibinfo{volume}{B389}}, \bibinfo{pages}{757} (\bibinfo{year}{1996}).

\bibitem[{\citenamefont{Bernabei et~al.}(2000)}]{DAMA2000}
\bibinfo{author}{\bibfnamefont{R.}~\bibnamefont{Bernabei}}
  \bibnamefont{et~al.}, \bibinfo{journal}{Phys. Lett.}
  \textbf{\bibinfo{volume}{B480}}, \bibinfo{pages}{23} (\bibinfo{year}{2000}).

\bibitem[{\citenamefont{Benoit et~al.}(2001)}]{edel2000}
\bibinfo{author}{\bibfnamefont{A.}~\bibnamefont{Benoit}} \bibnamefont{et~al.},
  \bibinfo{journal}{Phys. Lett.} \textbf{\bibinfo{volume}{B513}},
  \bibinfo{pages}{15} (\bibinfo{year}{2001}), \eprint{astro-ph/0106094}.

\bibitem[{\citenamefont{Morales et~al.}(2002)}]{igex2002}
\bibinfo{author}{\bibfnamefont{A.}~\bibnamefont{Morales}} \bibnamefont{et~al.},
  \bibinfo{journal}{Phys. Lett.} \textbf{\bibinfo{volume}{B532}},
  \bibinfo{pages}{8} (\bibinfo{year}{2002}).

\bibitem[{\citenamefont{Gaitskell and Mandic}()}]{dmplotter}
\bibinfo{author}{\bibfnamefont{R.~J.} \bibnamefont{Gaitskell}}
  \bibnamefont{and} \bibinfo{author}{\bibfnamefont{V.}~\bibnamefont{Mandic}},
  \bibinfo{note}{interactive sensitivity plots for direct detection of WIMP
  dark matter}, \urlprefix\url{http://dmtools.berkeley.edu/limitplots/}.

\bibitem[{\citenamefont{Baltz and Gondolo}(2001)}]{baltz}
\bibinfo{author}{\bibfnamefont{E.~A.} \bibnamefont{Baltz}} \bibnamefont{and}
  \bibinfo{author}{\bibfnamefont{P.}~\bibnamefont{Gondolo}},
  \bibinfo{journal}{{Phys. Rev. Lett.}} \textbf{\bibinfo{volume}{86}},
  \bibinfo{pages}{5004} (\bibinfo{year}{2001}).

\bibitem[{\citenamefont{{Ellis} et~al.}(2001)\citenamefont{{Ellis}, Ferstl, and
  {Olive}}}]{ellis01}
\bibinfo{author}{\bibfnamefont{J.}~\bibnamefont{{Ellis}}},
  \bibinfo{author}{\bibfnamefont{A.}~\bibnamefont{Ferstl}}, \bibnamefont{and}
  \bibinfo{author}{\bibfnamefont{K.~A.} \bibnamefont{{Olive}}},
  \bibinfo{journal}{Phys. Rev. D} \textbf{\bibinfo{volume}{63}},
  \bibinfo{pages}{065016} (\bibinfo{year}{2001}).

\bibitem[{\citenamefont{Copi and Krauss}(2002)}]{copikrauss}
\bibinfo{author}{\bibfnamefont{C.~J.} \bibnamefont{Copi}} \bibnamefont{and}
  \bibinfo{author}{\bibfnamefont{L.~M.} \bibnamefont{Krauss}}
  (\bibinfo{year}{2002}), \bibinfo{note}{submitted to Phys. Rev. Lett.},
  \eprint{astro-ph/0208010}.

\bibitem[{\citenamefont{Belli et~al.}(2002)\citenamefont{Belli, Cerulli,
  Fornengo, and Scopel}}]{DAMAgalaxy}
\bibinfo{author}{\bibfnamefont{P.}~\bibnamefont{Belli}},
  \bibinfo{author}{\bibfnamefont{R.}~\bibnamefont{Cerulli}},
  \bibinfo{author}{\bibfnamefont{N.}~\bibnamefont{Fornengo}}, \bibnamefont{and}
  \bibinfo{author}{\bibfnamefont{S.}~\bibnamefont{Scopel}},
  \bibinfo{journal}{Phys. Rev. D} \textbf{\bibinfo{volume}{66}},
  \bibinfo{pages}{043503} (\bibinfo{year}{2002}), \eprint{hep-ph/0203242}.

\bibitem[{\citenamefont{Smith and Weiner}(2001)}]{smit-wein-01}
\bibinfo{author}{\bibfnamefont{D.}~\bibnamefont{Smith}} \bibnamefont{and}
  \bibinfo{author}{\bibfnamefont{N.}~\bibnamefont{Weiner}},
  \bibinfo{journal}{Phys. Rev. D} \textbf{\bibinfo{volume}{64}},
  \bibinfo{pages}{043502} (\bibinfo{year}{2001}).

\bibitem[{\citenamefont{Bernabei et~al.}(2002)}]{DAMAinelastic}
\bibinfo{author}{\bibfnamefont{R.}~\bibnamefont{Bernabei}}
  \bibnamefont{et~al.}, \bibinfo{journal}{Eur. Phys. J. C}
  \textbf{\bibinfo{volume}{23}}, \bibinfo{pages}{61} (\bibinfo{year}{2002}).

\bibitem[{\citenamefont{Bernabei et~al.}(2001)}]{DAMAmixed}
\bibinfo{author}{\bibfnamefont{R.}~\bibnamefont{Bernabei}}
  \bibnamefont{et~al.}, \bibinfo{journal}{Phys. Lett.}
  \textbf{\bibinfo{volume}{B509}}, \bibinfo{pages}{197} (\bibinfo{year}{2001}).

\end{thebibliography}

\end{document}